\documentclass[aps,prc,showpacs,twocolumn,floatfix,nobibnotes,nofootinbib,amsmath,amssymb,longbibliography]{revtex4-2}
\usepackage{amsmath}
\usepackage{amssymb}
\usepackage{amsfonts}
\usepackage{mathtools}
\usepackage{bbm}
\usepackage{import}
\usepackage{graphicx}
\usepackage{fancyhdr}
\usepackage{tikz}
\usepackage{array}
\usepackage{units}
\usepackage{courier}
\usepackage{slashed}
\usepackage{bm}
\usepackage{dsfont}
\usepackage{epigraph}
\usepackage{verbatim}
\usepackage{listings}
\usepackage{comment}
\usepackage{multirow}
\usepackage{wrapfig} 
\usepackage{hyperref}
\usepackage{enumerate}

\usepackage[textsize=normalsize]{todonotes}

\newcounter{topic} 

\newcommand{\NNLOopt}{$\text{N2LO}_{\text{opt}}$}

\newcommand{\NijmI}{Nijmegen-I}
\newcommand{\INNNLO}{Idaho-N3LO}

\newcommand{\cm}{c.m.}

\newcommand{\wpcd}{WPCD}

\newcommand{\swp}{SWP}
\newcommand{\fwp}{FWP}

\newcommand{\NN}{$NN$}
\newcommand{\NNN}{$NNN$}
\newcommand{\Nd}{$Nd$}

\newcommand{\nd}{$nd$}
\newcommand{\pd}{$pd$}

\newcommand{\Elab}{E_\text{Lab}}
\newcommand{\Par}{\Pi}
\newcommand{\NWP}{N_{\text{WP}}}

\newcommand\inputpgf[2]{{
		\let\pgfimageWithoutPath\pgfimage
		\renewcommand{\pgfimage}[2][]{\pgfimageWithoutPath[##1]{#1/##2}}
		\input{#1/#2}
}}

\usetikzlibrary{shapes,arrows}	
\usetikzlibrary{arrows,decorations.markings}
\usetikzlibrary{positioning}
\definecolor{decisionColor}{HTML}{DEE1B2}
\definecolor{nodeColor}{HTML}{EBFAFF}
\tikzstyle{roundrect}=[rectangle, rounded corners, minimum width=0cm, minimum height=0.5cm, text centered, draw=black, fill=nodeColor, text width=2.2cm] 
\tikzstyle{decision} = [diamond, minimum width=3.7cm, minimum height=3.2cm, text centered, draw=black, fill=decisionColor, scale=1, text width=2.05cm]
\tikzstyle{arrow} = [decoration={markings,mark=at position 1 with
	{\arrow[scale=1.5,>=stealth]{>}}},postaction={decorate}]
\tikzset{
	desicion/.style={
		diamond,
		draw,
		text width=3em,
		text badly centered,
		inner sep=0pt
	},
	block/.style={
		rectangle,
		draw,
		text width=10em,
		text centered,
		rounded corners
	},
	cloud/.style={
		draw,
		ellipse,
		minimum height=2em
	},
	descr/.style={
		fill=white,
		inner sep=2.5pt
	},
	connector/.style={
		-latex
	},
	rectangle connector/.style={
		connector,
		to path={(\tikztostart) -- ++(#1,0pt) \tikztonodes |- (\tikztotarget) },
		pos=0.5
	},
	rectangle connector/.default=-2cm,
	straight connector/.style={
		connector,
		to path=--(\tikztotarget) \tikztonodes
	}
}

\begin{document}

\title{Neutron-deuteron scattering cross-sections with chiral \NN{} interactions using wave-packet continuum discretization}

\date{\today}
\author{Sean B. S. Miller}
\author{Andreas Ekstr\"om}
\affiliation{Department of Physics, Chalmers University of Technology, SE-412 96 Gothenburg, Sweden}
\author{Kai Hebeler}
\affiliation{Technische Universit\"at Darmstadt, Department of Physics, 64289 Darmstadt, Germany}
\affiliation{ExtreMe Matter Institute EMMI, GSI Helmholtzzentrum f\"ur Schwerionenforschung GmbH, 64291 Darmstadt, Germany}
\affiliation{Max-Planck-Institut f\"ur Kernphysik, Saupfercheckweg 1, 69117 Heidelberg, Germany}

\begin{abstract}
In this work we present a framework that allows to solve the Faddeev
equations for three-nucleon scattering using the wave-packet
continuum-discretization method. We perform systematic benchmarks
using results in the literature and study in detail the convergence of
this method with respect to the number of wave packets. We compute
several different elastic neutron-deuteron scattering cross-section
observables for a variety of energies using chiral nucleon-nucleon
interactions. For the interaction \NNLOopt~we find good agreement with
data for nucleon scattering-energies $\Elab \leq 70$ MeV and a
slightly larger maximum of the neutron analyzing power $A_y(n)$ at
$\Elab = 10$ MeV and 21 MeV compared with other interactions. This
work represents a first step towards a systematic inclusion of
three-nucleon scattering observables in the construction of
next-generation nuclear interactions.
\end{abstract}

\maketitle

\section{Introduction}
Nucleon-nucleon (\NN) and nucleon-deuteron (\Nd) scattering are
prototypical processes for analyzing \textit{ab initio} nuclear
Hamiltonians~\cite {Glockle:aYZcaRsU,Gloeckle:1995jg}. While \NN~cross
sections are straightforward to calculate and are nowadays routinely
being used to calibrate modern
\NN~interactions~\cite{Carlsson:2015vda,Reinert:2017usi,Piarulli:2014bda,Entem:2017gor},
the computation of three-nucleon (\NNN) scattering processes like
\Nd~scattering is much more demanding due to the presence of energy
poles in the underlying equations, contributions from \NNN
interactions, and the existence of multiple reaction channels. In
fact, it is a computationally challenging task to numerically solve
the Faddeev equations~\cite{Faddeev:1960su} and its
extensions~\cite{Weinberg:1964zza,Rosenberg:1965zz,Yakubovsky:1966ue,Alt:1967fx,Grassberger:1967zz}
in a reliable and accurate way. That is why it was only in the late
1980's that realistic quantum scattering
calculations~\cite{Witala:1988ty} began to emerge. Due to this
complexity, it has not yet been feasible to perform a simultaneous
statistical analysis of \NN~and \NNN~interactions using \Nd{} and
\NN{} cross section data.

In this paper we present an implementation of the wave-packet
continuum-discretization (\wpcd) method~\cite{Rubtsova:2015owa} to
solve the Faddeev equations with the chiral
\NNLOopt~\cite{Ekstrom:2013kea} interaction. The \wpcd~method is one
of many bound-state techniques~\cite {Carbonell:2013ywa} for solving
the multi-particle scattering problem. The main advantages of this
method are: i) coarse-graining the continuum using a square-integrable
basis smooths out all singularities and facilitates straightforward
numerical solutions of the Faddeev equations for the scattering
amplitude, ii) all on-shell energy dependence resides in a closed-form
expression of the channel resolvent, and iii) once the wave-packet
basis is antisymmetrized, which has to be done only once computationally, the
computational cost of predicting scattering observables scales sublinearly
with the number of scattering energies. This opens ways for efficient
computation of coarse-grained \Nd{} predictions for several scattering
energies. The accuracy of the solutions depends polynomially on the
number of wave-packets used for discretizing the continuum. To that
end we also study the convergence of the \wpcd{} results with respect
to the number of employed wave-packet basis states.

We benchmark the \wpcd{} results against published cross section
results for the traditional \NijmI{} \NN{}
interaction~\cite{Stoks:1994wp} and systematically compare and analyze
different cross section observables at a variety of energies using the
interactions \INNNLO{}~\cite{Entem:2003ft} and \NNLOopt. Both
interactions have a history of being routinely employed in \textit{ab
  initio} studies of nuclear structure and nucleon-nucleus
reactions. The latter one, \NNLOopt{}, is a next-to-next-to-leading
order chiral interaction optimized to reproduce \NN{} scattering phase
shifts, and yields an accurate description of low-energy \NN{}
scattering data up to 125 MeV scattering energy. More importantly, it
also reproduces key nuclear properties such as the location of the
oxygen neutron drip-line and calcium shell closures without having to
invoke \NNN~interactions. \NNLOopt{} also gives a rather good
description of selected nuclear structure physics, transitions, and
reaction data, see, e.g.,
Refs.~\cite{Dytrych:2018vkl,Burrows:2018ggt,Rotureau:2018pxk}. Of
course, a complete calculation requires \NNN~interactions and these
correlations can play a pivotal role for obtaining realistic
\textit{ab initio} predictions of bound and continuum nuclear
observables, see, e.g.,
Refs.~\cite{Witala:2000am,Pieper:2001mp,Epelbaum:2002vt,Navratil:2007we,Otsuka:2009cs,Kalantar-Nayestanaki:2011rzs,Calci:2016dfb,Hebeler:2020ocj}. From
these observations it is therefore interesting to predict
\Nd~scattering observables with the \NNLOopt{} interaction. In this
work we pay particular attention to the low-energy neutron ($n$)
analyzing power $A_y(n)$, where a long-standing
\textit{puzzle}~\cite{Huber:1998hu} resides\footnote{The so-called
  \Nd{} vector $A_y$-puzzle which is equally observed for low-energy
  $pd$- and $nd$-scattering. The same puzzle is observed for the
  deuteron vector analyzing power $iT_{11} = \frac{\sqrt{3}}{2}A_y(d)$
  whereas the deuteron tensor analyzing power is well understood.},
and the \nd{} differential cross section $d\sigma/d\Omega$ at nucleon
laboratory scattering energy $\Elab = 64.5$ MeV. The latter observable
is known to depend sensitively on
\NNN~interactions~\cite{Witala:1998ey, LENPIC:2018ewt}.

In Sec.~\ref{sec | elastic nd scattering} we present the formalism
that we implemented to solve the Faddeev equations for elastic \Nd{}
scattering and benchmark its convergence with respect to basis
dimension. In Sec.~\ref{sec | low-energy predicions of nd-scattering
  cross sections} we present predictions for \nd{} scattering cross
sections using the \NNLOopt{} interaction, and end with a summary and
outlook in Sec.~\ref{sec | summary and outlook}.

\section{Elastic \Nd{} scattering using the \wpcd{} method}
\label{sec | elastic nd scattering}

In this section, we present i) the \wpcd{} method for solving the
\Nd{} Faddeev equations in momentum space (Sec.~\ref{sec | faddeev in
  momentum space}), ii) how to construct a \wpcd-basis and its
partial-wave expansion (Sec.~\ref{sec | wpcd basis}), iii) our
computational implementation for solving the resulting matrix equation
(Sec.~\ref{sec | computational implementation}), and iv) a convergence
analysis of the \wpcd{} method (Sec.~\ref{sec | WP convergence}). All
detailed expressions are relegated to the appendices~\ref{app |
  permutation operator plane wave representation}-\ref{app | phase
  shifts}.

\subsection{The Faddeev equations in momentum space}
\label{sec | faddeev in momentum space}
The Faddeev equations can be reduced to the Alt-Grassberger-Sandhas
(AGS) equation~\cite{Alt:1967fx}, which for elastic \Nd{} scattering
and without a \NNN{} interaction can be written as
\begin{equation}
	\hat{U}_i(E) = \hat{P}\hat{v}_i + \hat{P}\hat{v}_i\hat{G}_{i}(E)\hat{U}_i(E) \:,
	\label{eq | faddeev equation | operator form}
\end{equation} where $E$ denotes the on-shell scattering energy and where we
 used the usual ``odd-man-out'' notation such that the index $i$ here refers
 to the incoming nucleon relative to an antisymmetric state of a nucleon pair
 $(jk)$, e.g., the deuteron, for unequal $i, j,
 k \in \{1,2,3 \}$. Our goal is to
 calculate elastic cross-sections via the elastic transition operator $\hat
 {U}_i$. The three operators $\hat{U}_i$ are related via the permutation
 operators $\hat{P}_{ijk}\equiv\hat{P}_{ij}\hat{P}_{jk}$:
\begin{align}
	\begin{split}
	\hat{U}_2 &= \hat{P}_{123}\hat{U}_1 = \hat{P}_{12}\hat{P}_{23}\hat{U}_1 \\
	\hat{U}_3 &= \hat{P}_{132}\hat{U}_1 = \hat{P}_{13}\hat{P}_{32}\hat{U}_1 \, ,
	\end{split}
\end{align}
where $\hat{P}_{12}$ permutes nucleons 1 and 2 etc. The operator
$\hat{P}\equiv 1 + \hat{P}_{123} + \hat{P}_{132}$ ensures full
antisymmetrization of the \Nd{} state. See App.~\ref{app | permutation
  operator plane wave representation} for the expressions we employ to
compute the partial-wave projected $\hat{P}_{123}$ operator. The two
remaining operators entering Eq.~\eqref{eq | faddeev equation |
  operator form} are the \NN{} potential $\hat{v}_i$ acting in the
pair-system $(jk)$, and the channel resolvent
$\hat{G}_i(E)\equiv\frac{1}{E-\hat{H}_i\pm i\epsilon}$, where
$\hat{H}_i\equiv \hat{h}_i \oplus \hat{h}_i^{0}$ is the full
Hamiltonian, $\hat{h}_i = \hat{h}_0 + \hat{v}_i$ is the \NN{}
Hamiltonian, $\hat{h}_0$ is the kinetic energy of the pair, and
$\hat{h}_i^{0}$ is the free Hamiltonian of the third nucleon relative
to the pair. Since
$\hat{U}_i$ for $i=1,2,3$ in Eq.~\eqref{eq | faddeev equation |
	operator form} are not independent it suffices to solve for only one
of them, e.g., $U_1$. For the most part we will also drop this
subscript. This will hopefully avoid possible confusion with respect
to subscripts denoting different basis states defined below.

In the \wpcd{} method, the channel resolvent $G(E)$ is diagonal and
straightforward to evaluate analytically using a \wpcd-basis of
scattering states. This has the advantage of removing all
complications from singularities that plague the Faddeev method
formulated in a plane-wave basis. Such points are essentially averaged
out when using wave packets to represent states in the
continuum. Furthermore, the entire $E$-dependence of the scattering
process resides in the channel resolvent $\hat{G}(E)$ and multiple
scattering energies can be accessed without inducing much
computational overhead.

We end this section by linking the form of the AGS equation used in
\wpcd{}, Eq.~\eqref{eq | faddeev equation | operator form}, to its
conventional formulation used as a starting point in standard Faddeev
methods. Using that $\hat{t}\hat{G}_0 \equiv \hat{v}\hat{G}$ and
$\hat{v}=\hat{G}_0^{-1}$ enables us to replace the interaction
$\hat{v}$ and the channel resolvent $\hat{G}$ with a fully off-shell
\NN{} $t-$matrix and the free resolvent $\hat{G}_0$ at the on-shell
energy $E$. This latter replacement is necessary since the channel
resolvent cannot be straightforwardly evaluated \textit{a priori}
using only a plane-wave basis. This also introduces an explicit
energy-dependence in the \NN{} $\hat{t}$-matrix and thereby in the
integral kernel of the AGS equation. In addition to this,
singularities arise in the representation of both these
operators~\cite{Gloeckle:1995jg}. This can be dealt with using
subtraction techniques. Such complications are avoided altogether in
the \wpcd{} method.

\subsection{Setting up the \wpcd{} basis}
\label{sec | wpcd basis}
We define a \emph{free} wave-packet (\fwp) for a pair of particles
with relative momenta $p$ within some interval (bin)
$\mathcal{D}_i\equiv[p_i,p_{i+1}]$ as
\begin{equation}
	|x_i\rangle = \frac{1}{N_i}\int_{\mathcal{D}_i} dp \: p \: f(p)|p\rangle \:,
\end{equation} 
 where $|p\rangle$ is a plane-wave state with momentum $p$ and
 normalization $\langle p'|p\rangle = \frac{\delta(p'-p)}{p'p}$. This
 normalization differs from the one used in Ref.~\cite
 {Rubtsova:2015owa}. Here, $N_i$ is a normalization constant. The function $f(p)$
 is a weighting function which allows us to define, for example,
 momentum wave-packets, $f(p)=1$, or energy wave-packets, $f(p)=\sqrt{\frac
 {p}{\mu_0}}$, where $\mu_0$ is the reduced mass. In this work $\mu_0$
 ($\mu_1$) denote the reduced mass for the two-body (three-body) system. The
 naming convention for the two kinds of wave packets indicate whether they
 correspond to eigenstates of the momentum operator $\hat{p}$, or the kinetic
 energy operator $\hat{h}_0$, of the two-body system. In this work we use
 both kinds of wave packets since it is simpler to use and derive operator
 projections onto momentum \fwp{}s, but the resolvent $\hat{G}$ is evaluated
 in an energy wave packet basis, see App.~\ref{app | WP projection}. In the
 three-body system we define momentum \fwp s as
\begin{equation}
	|X_{ij}\rangle = |x_i\rangle \otimes |\bar{x}_j\rangle,
	\label{eq | general wave-packet definition}
\end{equation}
where we use the bar-notation to denote wave packets with the momenta
$q \in \bar{\mathcal{D}}_j \equiv [q_j,q_{j+1}]$ of the third particle
relative to the centre of mass (\cm{}) of the pair. In this work, we
use the same number of wave packets, $\NWP$, when discretizing the
continuum of Jacobi momenta $p$ and $q$. We discretize the continuum
according to a Chebyshev grid, i.e.,
\begin{equation}
  p_i=q_i = \alpha \tan^t\left(\frac{2i-1}{4\NWP}\right)\:,\quad i=1,\ldots,\NWP,
\end{equation}
where we use $\alpha=200$ MeV and $t=1$, which yields wave packets
residing in momentum bins reaching momenta up to $\sim 10$ GeV. The
width of the momentum bins increases with $i$, such that the vast
majority of the wave packets reside below momenta of $\sim 500$ MeV,
which is where we typically have the most relevant contributions from
modern chiral \NN-potentials.

We work in a partial-wave representation of \NNN{} states and
introduce a spin-angular basis with total angular momentum $\mathcal{J}$ and isospin $\mathcal{T}$,
\begin{equation}
  | \alpha \rangle \equiv | (LS)J (l\:\nicefrac{1}{2})j(Jj)\mathcal{J}(T\:\nicefrac{1}{2})\mathcal{T}\rangle\:,
\end{equation}
where $L$, $S$, $J$, and $T$ denote the relative orbital angular
momentum, spin, total angular momentum, and isospin, respectively, for
the antisymmetric nucleon-pair system. The orbital angular momentum of the third
(spin-$1/2$) nucleon relative to the \cm{} of the
pair systems is denoted with $l$ and its total angular momentum is
denoted with $j$. Each $Jj$-coupled channel has a total angular
momentum $\mathcal{J}$. We can therefore construct \NNN{}
partial-waves as
\begin{align}
	\begin{split}
	|X_{ij}^{\alpha}\rangle &\equiv |x_i\rangle \otimes |\bar{x}_j\rangle \otimes |\alpha\rangle\\
	&= |x_i,\bar{x}_j;(LS)J(l\:\nicefrac{1}{2})j(Jj)\mathcal{J}(T\:\nicefrac{1}{2})\mathcal{T}\rangle.
	\end{split}
\end{align}
All \NNN{} partial-waves are equipped with a unique combination of good
quantum numbers $\mathcal{J}$ and parity $\Pi = (-1)^{L+l}$. In our
calculations we explicitly break isospin $\mathcal{T}$ by including
the charge dependence of the \NN{} interaction in the $^1S_0$
channel. The impact of this $\mathcal{T}=\frac{3}{2} - \frac{1}{2}$
isospin coupling on elastic \Nd{} scattering is very small~\cite
{Witala:2016inj}. On the other hand, the computational costs of
including it is negligible.

The \fwp{} states form a square-integrable basis, with appropriate
long-range behavior to approximate scattering
states~\cite{Pomerantsev:2008tw}. It is also straightforward to
represent matrix elements of the permutation operator $\hat{P}$ and
the \NN{} potential operator $\hat{v}_1$ in a basis of such
states. See App.~\ref{app | WP projection} for details regarding this
projection.

The elastic transition operator will be solved for in a basis of
\NNN{} \textit{scattering} wave packets (\swp) defined as
\begin{equation}
  |Z^{\alpha}_{ij}\rangle = |z_i\rangle \otimes |\bar{x}_j\rangle \otimes |\alpha\rangle, 
\end{equation}
where $|z_i\rangle$ are scattering wave-packets (eigenstates) of the \NN{}
Hamiltonian $\hat{h}$. The latter states can be approximated in a
finite \fwp-basis as
\begin{equation}
  |z_i\rangle \approx \sum_{j=1}^{\NWP} \langle x_j|z_i\rangle|x_j\rangle \equiv \sum_{j=1}^{\NWP} C_{ji}|x_j\rangle.
  \label{eq | transformation coefficients}
\end{equation}
where the (real) $C_{ji}$ coefficients are obtained via
straightforward diagonalization of the \NN{} Hamiltonian in a basis of
\fwp s $|x_i\rangle$. The coefficients allow for straightforward
transformation between \fwp{} and \swp{} partial-wave bases. From the
diagonalization we obtain eigenvectors and eigenvalues, i.e.,
scattering wave packets $|z_i\rangle$ with eigenenergies $\epsilon_i$
such that $\hat{h}|z_i\rangle = \epsilon_i|z_i\rangle$. The
eigenenergies define the bin boundaries
$\mathcal{D}_i\equiv\{\mathcal{E}_i,\mathcal{E}_{i+1}\}$ for the
scattering wave-packets $|z_i\rangle$ \cite{Rubtsova:2015owa}. We will
refer to the (negative) energy bin corresponding to the deuteron bound
state as $|z_{i_d}\rangle$ and the corresponding \NNN{} partial-waves
with a deuteron channel as $|\alpha_d\rangle$. Although the \fwp-basis
is sub-optimal for describing bound states, it yields a very good
approximation to scattering states which is more important here.

For all computations in this work we use a spin-angular basis of
positive and negative parity \NNN{} partial waves with
$\mathcal{J}\leq 17/2$ and $J\leq 3$. This leads to $\lesssim 60$
channels per \NNN{} partial wave. As we will discuss in Sec.~\ref{sec
  | WP convergence}, we find that using $\NWP\approx 125$ wave-packets
in both Jacobi momenta is more than sufficient for accurately
computing low-energy elastic scattering observables with $\Elab
\lesssim 100$ MeV~\cite{Gloeckle:1995jg}, which is the region we focus on in
this work. There are exceptions and we will discuss those below.

\subsection{Computational implementation}
\label{sec | computational implementation}
Naturally, we solve for the transition operator $\hat{U}$ for each
combination of \NNN{} total angular momentum $\mathcal{J}$ and parity
$\Pi$ separately. We represent Eq.~\eqref{eq | faddeev equation |
  operator form} in matrix form using a \swp-basis
\begin{equation}
	\mathbf{U}(E) = \mathbf{A} + \mathbf{A}\mathbf{G}(E)\mathbf{U} \, ,
	\label{eq | faddeev equation | WP matrix form}
\end{equation}
where $\mathbf{A} \equiv \mathbf{C}^T\mathbf{PVC}$.  Here, we defined
finite-dimensional matrices for the \NN{}-potential matrix
$\mathbf{V}$ and the permutation matrix $\mathbf{P}$ in a \NNN{}
momentum-\fwp-basis. They are obtained using the expressions in
App.~\ref{app | WP potential} and App.~\ref{app | WP permutation
  operator}, respectively. Note that $\mathbf{V}$ and $\mathbf{C}$ are
block diagonal for momenta $q$ in different bins $\bar{\mathcal{D}}_j$
and quantum numbers $l$ and $j$. Once we have diagonalized the \NN{}
Hamiltonian, we construct an approximate \swp-basis and setup the
(block-diagonal) matrix $\mathbf{C}$ of $C_{ij}$ coefficients in
Eq.~\eqref{eq | transformation coefficients}. The eigenvalues of
$\mathbf{G}$ are easily obtained in the \swp-basis, see App.~\ref{app
  | channel resolvent}. This is of key importance.

Formally, Eq.~\eqref{eq | faddeev equation | WP matrix form} is a
matrix-equation that can be solved via inversion. However,
straightforward inversion, or numerically stable equivalents, is
unviable for realistic nuclear potentials since the matrix
$\mathbf{A}$ is too large to be stored in memory for the basis sizes
we require for convergence. Fortunately it is possible to store the
matrices necessary to construct $\mathbf{A}$ in memory, i.e.,
$\mathbf{C}$, $\mathbf{V}$, and $\mathbf{P}$. Indeed, $\mathbf{P}$
only has to be computed once and is very sparse ($>99\%$). We see that
$\mathbf{A}$ is 100 times denser than $\mathbf{P}$.

We solve Eq.~\eqref{eq | faddeev equation | WP matrix form} for the
on-shell transition operator in the \swp-basis
$U_{i_dj}^{\alpha_d'\alpha_d} \equiv \langle Z_{i_dj}^{\alpha_d'} |
\hat{U} | Z_{i_dj}^{\alpha_d} \rangle$, i.e., the transition matrix
elements corresponding to an incoming nucleon with on-shell momentum
$q\in \bar{\mathcal{D}}_j$ scattering elastically off a deuteron. We
compute this amplitude by summing the first 20-30 terms of the
Neumann (or Born) series
\begin{equation}
	U_{i_dj}^{\alpha_d'\alpha_d} = \sum_{n=0}^{\infty} \left [ \mathbf{A}\mathbf{K}^n \right]_{i_dj}^{\alpha_d'\alpha_d} \:,
	\label{eq | WP on-shell U-matrix element}
\end{equation}
where we have defined $\mathbf{K}(E) \equiv \mathbf{G}(E)\mathbf{A}$.
Note that $\mathbf{G}$, and thereby also $\mathbf{K}$, depend on the on-shell scattering energy $E$,
and that since $\mathbf{G}$ is diagonal, $\mathbf{A}$ and $\mathbf{K}$ have identical densities.
Thus, $\mathbf{K}$ must be re-computed in segments and on-the-fly for the repeated matrix-vector multiplications
needed to generate the terms of the series above. We employ a Pad\'e
extrapolation~\cite{baker} to handle a divergent Neumann series and we
find that this rational approximant facilitates a convergent
resummation in our case, see App.~\ref{app | pade}.

\begin{figure*}[th]
	\centering
	\includegraphics[width=\textwidth]{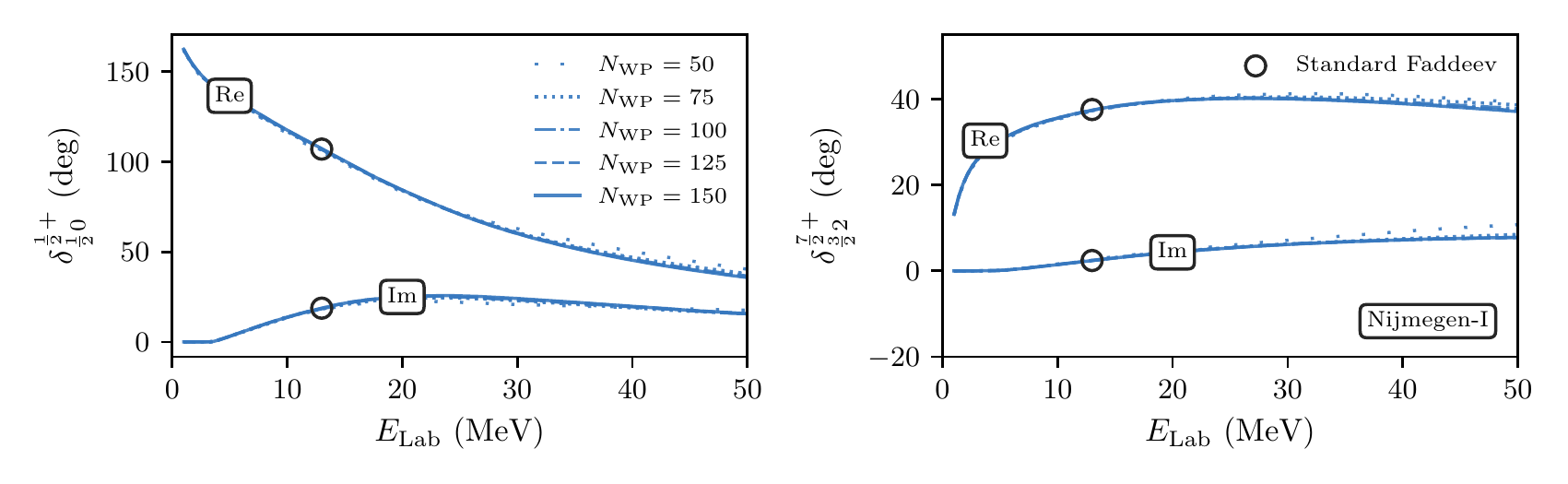}
	\caption{The \nd{} scattering phase shifts (real and imaginary
          parts) in the doublet and quartet spin channels of the
          $\frac{1}{2}^+$ (left panel) and $\frac{7}{2}^+$ (right
          panel) \NNN{} partial waves, respectively. Standard Faddeev
          results at $\Elab = 13$ MeV from
          Ref.~\cite{Gloeckle:1995jg}. See App.~\ref{app | phase
            shifts} for more on the notation.}
	\label{fig | nijmegen phases}
\end{figure*}

Next to the computational cost of initially constructing $\mathbf{P}$,
the cost of setting up the kernel $\mathbf{K}$ constitutes the
numerical bottleneck in our current implementation of the \wpcd{}
method as it must be repeated several times. The product $\mathbf{GA}$ is
trivial, which in turn makes it trivial to compute transition matrices
at several different energies $E$ with the \wpcd{} method. The product
$\mathbf{C}^T\mathbf{PVC}$ is a product of the sparse matrix
$\mathbf{P}$ with the block-diagonal matrices $\mathbf{C}^T$ and
$\mathbf{VC}$ on either side. Note also that $\mathbf{C}$ and
$\mathbf{V}$ have the same block-diagonal structure. For the product
$\mathbf{A}\mathbf{K}^n$ we re-use the on-shell row(s) of the
$\mathbf{A}\mathbf{K}^{n-1}$ matrix product computed for the $(n-1)$th
term. 

The (complex) on-shell transition amplitudes $U(E)$ for
spin$-\frac{1}{2}$--spin-$1$ scattering constitutes a $3\times 3$
matrix. Once this matrix is computed in all relevant \NNN{} partial
waves, i.e., for $\mathcal{J}\leq 17/2$ and $J \leq 3$ in our case, it is
straightforward to obtain the $6 \times 6$ spin-scattering matrix for
describing the elastic \Nd{} scattering cross sections at kinetic
energy $\Elab$ in the laboratory frame of reference, see App.~\ref{app
  | observables}-\ref{app | phase shifts}.

\subsection{Convergence with respect to $\NWP$}
\label{sec | WP convergence}
In the limit $\NWP \rightarrow \infty$, amplitudes computed using the
\wpcd{} method approaches results from the standard Faddeev method
utilizing a plane-wave basis. This infinite limit cannot be reached in
practice and all \wpcd{} predictions that we present are based on
solving Eq.~\eqref{eq | faddeev equation | operator form} in a finite
wave-packet basis. To analyze the convergence of predictions with
respect to increasing $\NWP$ we computed \nd{} scattering phase shifts,
shown in Fig.~\ref{fig | nijmegen phases} for the doublet and quartet
spin-channels in the $\mathcal{J}^{\Par}= \frac{1}{2}^{+}$ and
$\frac{7}{2}^{+}$ \NNN{} partial waves, respectively, using the
\NijmI{} \NN{} interaction~\cite{Stoks:1994wp}. For this potential
there exists published results~\cite{Gloeckle:1995jg} from a standard
Faddeev calculation at $\Elab = 13 $ MeV and this provides a valuable
benchmark to ensure the correctness of our implementation. Detailed
numerical inspection of the results reveal that we recover standard
Faddeev results for all imaginary and real parts of the \NNN{} phase
shifts for $\mathcal{J}^{\Par}\leq \frac{7}{2}^{\pm}$ within $\sim
1\%$ using $\NWP \gtrsim 125$ wave packets. We also observe that the
magnitude of the imaginary part of the phase shifts is
$|\text{Im}(\delta)|\lesssim 10^{-2}$ degrees for scattering energies
below the deuteron breakup threshold. The convergence with increasing
$\NWP$ is rather slow however, which is to be expected since a
packetized basis corresponds to a coarse grained continuum
representation across a wide range of energies simultaneously. In
Fig.~\ref{fig | nijmegen phases} it is nevertheless clear that the
\wpcd{} method yields highly accurate scattering phase shifts for
$\Elab\lesssim 50$ MeV already for $\NWP \gtrsim 75$. See
App.~\ref{app | phase shifts} for further information about how we
computed phase shifts from the partial-wave scattering amplitudes $U$.

\begin{figure}[t]
	\centering
	\includegraphics[width=\columnwidth]{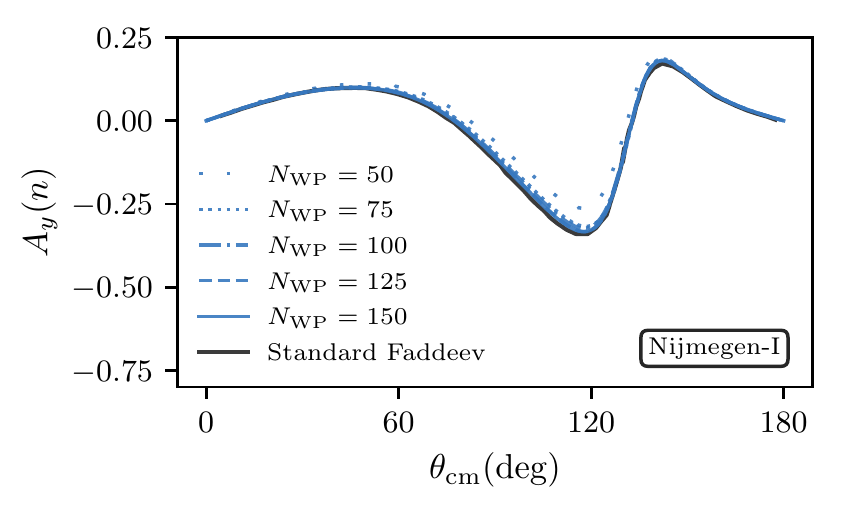}
	\caption{The neutron analyzing power $A_y(n)$ at
          $\Elab=35$ MeV versus \cm{} scattering
          angle $\theta_\text{cm}$ using the \NijmI{} \NN{}
          potential. The \wpcd{} prediction approaches the standard
          Faddeev result~\cite{Gloeckle:1995jg} with an increasing
          number of wave-packets $\NWP$ and they overlap for
          $\NWP \gtrsim 125$.}
	\label{fig | Ay 35 convergence}
\end{figure}

Predicting scattering observables is more interesting than
scattering phase shifts since they are directly comparable to
experimental data. To benchmark our \wpcd{} computation of observables
we compare with the results\footnote{Published results were traced
  from a figure in Ref.~\cite{Gloeckle:1995jg}. The calculations in
  that work are reported with $(1-2\%)$ accuracy.} from a standard
Faddeev calculation~\cite{Gloeckle:1995jg} of the neutron analyzing
power $A_{y}(n)$ at $\Elab=35$ MeV using the \NijmI{} \NN{} potential,
see Fig.~\ref{fig | Ay 35 convergence}. For the \wpcd-calculations we
varied the number of wave packets between $50 \leq \NWP \leq 150$. The
convergence pattern is very similar to the one we observed for the
phase shifts and for $\NWP \gtrsim 125$ we hence claim convergence for
this observable. Also in this calculation we included \NNN{}
partial-waves with $\mathcal{J}^{\Par} \leq \frac{17}{2}^{\pm}$ and
\NN{} channels with $J \leq 3$.

\begin{figure*}[t]
	\centering
	\includegraphics[width=\textwidth]{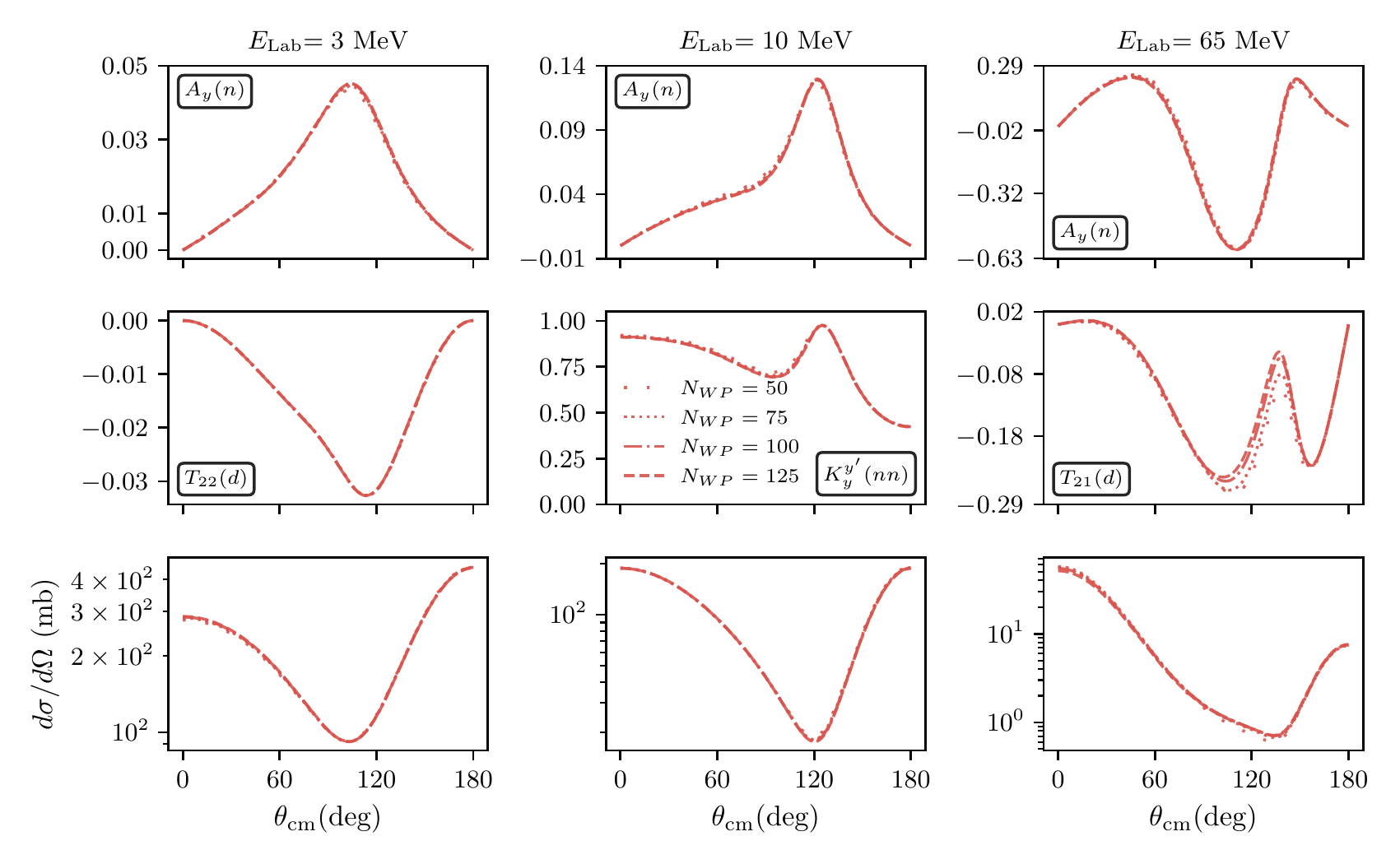}
	\caption{The convergence of \wpcd{} predictions for typical
          elastic \nd{} differential scattering cross sections and
          polarization observables for increasing neutron scattering
          energies with respect to an increasing number of wave
          packets using the \INNNLO{} \NN{}
          interaction~\cite{Entem:2003ft}.}
	\label{fig | wpcd convergence}
\end{figure*}

To further assess the convergence of the \wpcd{} method with
respect to $\NWP$, we study a range of vector ($A$) and spherical
tensor ($T$) analyzing powers, spin transfer coefficients ($K)$, and
differential cross sections for $50 \leq \NWP \leq 125$ at
$\Elab=3,10,65$ MeV using the well-known chiral \INNNLO{}
interaction~\cite{Entem:2003ft}, see Fig.~\ref{fig | wpcd
  convergence}.

With this result we can establish that for most elastic \nd{}
scattering observables it is indeed enough to employ $\NWP\gtrsim 75$
wave packets to obtain sufficiently accurate predictions for $\Elab
\lesssim 70$ MeV. If one can tolerate a \wpcd{} method-error
comparable to typical experimental errors of \Nd{} scattering data,
then even $\NWP \approx 50$ will be enough for most low-energy \Nd{}
predictions. Note that the number of wave packets dramatically impact the
computational cost of the \wpcd{} calculations since this scales as
$\sim \NWP^4$. Also, solving for all amplitudes with $\Elab \lesssim
100$ MeV is merely $\sim 2$ times slower than solving at a single
scattering energy.

In Fig.~\ref{fig | wpcd convergence} the wave-packet convergence of
the tensor analyzing power $T_{21}$ stands out and exhibits a
noticeable sensitivity to $\NWP$. This observable is known to also
depend more strongly on $J_\text{max}=4$ contributions of the \NN{}
interaction~\cite{Gloeckle:1995jg}. Observables that depend on finer
details of the nuclear interaction will exhibit a slower convergence
with respect to increasing $\NWP$, due to the coarse-graining of
\wpcd{}.  Fortunately, poorly converging predictions can be identified
straightforwardly. We explored various modifications to the
wave-packet distributions, e.g., increasing the density of
wave-packets in the vicinity of the scattering energy, but this did
not lead to any clear improvements.

\section{Predicting \nd{}-scattering cross sections using \NNLOopt{}}
\label{sec | low-energy predicions of nd-scattering cross sections}
In this section we present selected low-energy and elastic \nd{} cross
sections using the \NNLOopt{} \NN{} interaction and compare with
neutron-deuteron (\nd) as well as proton-deuteron (\pd) data. We can
neglect method uncertainties since we employ $\NWP=125$ wave-packets
for all predictions, unless otherwise stated.

The world database of \Nd{} scattering cross sections contains mostly
\pd{} data from experiments with either polarized or non-polarized
proton or deuteron beams. Indeed, \nd{} scattering is difficult to
perform. It is challenging to manipulate and focus electrically
neutral particles. The neutron itself is unstable and does not make
for a suitable target material on its own. Neutron detectors are also
less efficient compared charged-particle detectors. Theoretically, we
have the opposite situation. It is typically much easier to compute
\Nd{} scattering cross sections without a Coulomb interaction
~\cite{Deltuva:2005wx,Deltuva:2005cc}. Fortunately, Coulomb effects
are only significant at low energies, e.g., below the deuteron breakup
threshold, and for extremal scattering angles. As such, in most
kinematic regions \pd{} scattering data can be compared with
theoretical \Nd{} scattering results without any Coulomb
interaction. We will therefore use \pd{} data in case \nd{} data does
not exist or is very scarce. To be clear, we do not include any
Coulomb effects in our calculations. One can extend the \wpcd{} method
to incorporate such effects. Indeed, the challenge of treating a
long-range interaction for small momenta is alleviated when using a
square-integrable Coulomb wave-packet basis~\cite{Rubtsova:2015owa}.

\begin{figure}[t]
	\centering
	\includegraphics[width=\columnwidth]{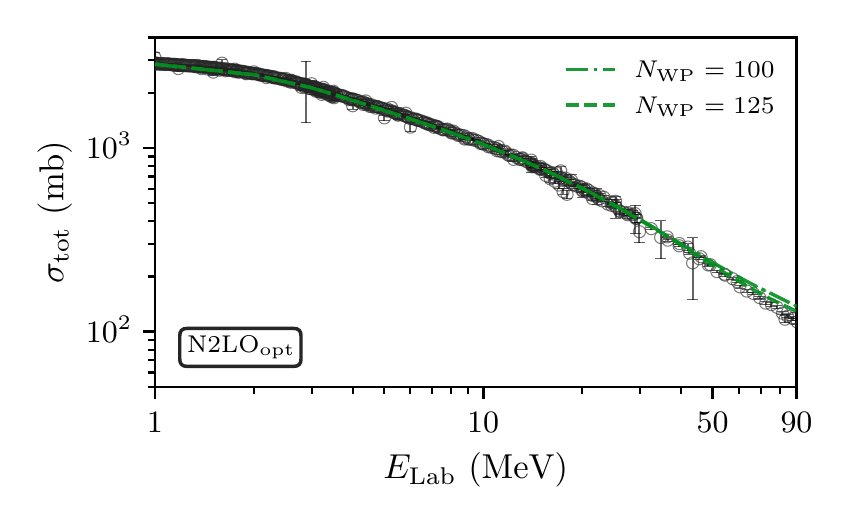}
	\caption{Total \nd{} cross sections computed using the \wpcd{}
          method and the optical theorem, see
          e.g. Ref.~\cite{Ishikawa:2000dq}. Experimental \nd{} data
          retrieved via EXFOR~\cite{Otuka:2014wzu}.}
	\label{fig | SGT}
\end{figure}

In Fig.~\ref{fig | SGT} we show our predictions for the total \nd{}
scattering cross section with the \NNLOopt{} interaction. The
reproduction of experimental \nd{} data is excellent up to $\Elab
\approx 70$ MeV. At this point we also begin to see a difference
between the $\NWP=100$ and $\NWP=125$ calculations. At $\Elab > 70$
MeV, the inclusion of $J>3$ \NN-channels will have a percent-level
effect on the predictions. We also note that $\Elab \approx 70$ MeV
corresponds to a relative momentum $q\approx 240$ MeV of the incident
neutron. This translates to a \NN{} scattering energy of 125 MeV in
the lab frame. The \NNLOopt{} goodness-of-fit measure , i.e., the
$\chi^2/N_\text{datum}$ with respect to \NN{} scattering data, is
$\approx 1$ up to 125 MeV scattering energy. As such, it is reasonable
to expect a gradual deterioration of the predictive power for
$\Elab>70$ MeV.

\begin{figure}[th]
	\centering
	\includegraphics[width=\columnwidth]{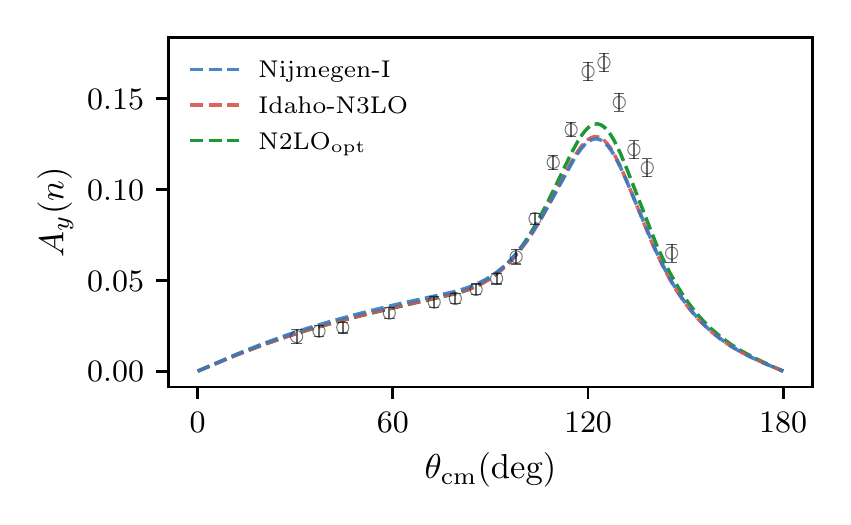}
	\caption{The neutron analyzing power $A_y(n)$ at $\Elab= 10$
          MeV computed using the \wpcd{} method with $\NWP=125$
          wave-packets. At the maximum, the top dashed line is the
          \NNLOopt{} result. Experimental \nd{} data from
          Ref.~\cite{Howell:1987uc}.}
	\label{fig | Ay 10 nnloopt}
\end{figure}

\begin{figure*}[th]
	\centering
	\includegraphics[width=\textwidth]{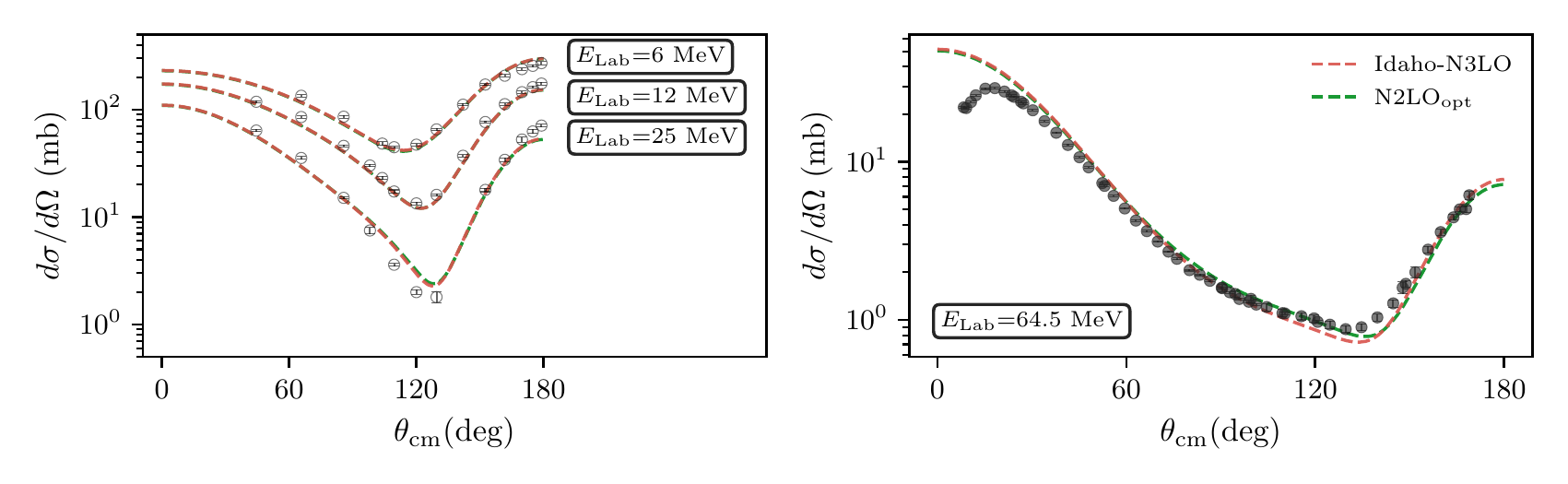}
	\caption{The differential cross section for elastic \nd{}
          scattering computed using the \wpcd{} method with $\NWP=125$
          wave-packets. The experimental \nd{} data (empty markers) at
          $\Elab=6,12,25$ MeV in the left panel are from
          Ref.~\cite{Schwarz:1983rlz} and the \pd{} data (filled
          markers) at $\Elab=64.5$ MeV in the right panel are from
          Ref.~\cite{Shimizu:1982lqa}. For $\Elab=64.5$ MeV (right
          panel) the \INNNLO{} predicts a slightly smaller cross
          section at the minimum.}
	\label{fig | DSG nnloopt}
\end{figure*}

At energies below $\Elab \approx 50-100$ MeV the effects of \NNN{}
interactions are typically
smaller~\cite{Witala:2000am,Kalantar-Nayestanaki:2011rzs,Epelbaum:2019zqc,LENPIC:2018ewt},
and the bulk of low-energy \Nd{} scattering observables can be
described quite well using only \NN{} interactions. However, there
exist a few scattering observables at these low energies that exhibit
discrepancies due to missing \NNN{} forces and (or) possibly
fine-tuning effects, e.g., low-energy analyzing powers, the
high-energy differential cross section minimum, and the \nd{} doublet
scattering length. The latter is known to correlate with the triton
binding energy via the well-known Phillips
line~\cite{Phillips:1968zze}. The \nd{} scattering length can be 
computed using a bound-state formulation of the Faddeev
equations~\cite{Witala:2003en} or via numerical extrapolation of the
scattering amplitude to $q\rightarrow 0$. Unfortunately, this limit is
challenging to reach in the \wpcd{} method with the Chebyshev
distribution we employ. Resorting to a basis with an increased number
wave-packets at small momenta will of course remedy this. However, to
simultaneously maintain accurate scattering amplitudes for higher
scattering energies will result in a needlessly large basis size.

The prediction of the neutron analyzing power $A_y(n)$ at $\Elab=10$
MeV with \NNLOopt{} is shown in Fig.~\ref{fig | Ay 10 nnloopt}. For
comparison we also include \wpcd{} results using the \INNNLO{} and
\NijmI{} potentials. All three potentials yield virtually the same
result and the discrepancy with respect to $^{2}H(n,n)^{2}H$
data~\cite{Howell:1987uc} at the \cm{} scattering angles $\theta_{cm}
\approx 120^{\circ}$, known as the
\textit{$A_y$-puzzle}~\cite{Huber:1998hu}, persists also with
\NNLOopt{}. There is some tendency of a slight increase using this
latter potential, but this is certainly not significant on an
absolute scale. This result reflects that the low-energy interaction
in the $^{3}P$-channels of \NNLOopt, to which we know that $A_y$ is
most sensitive~\cite{Huber:1998hu}, are similar to the ones in
\INNNLO{} and \NijmI{}. A detailed calculation~\cite{LENPIC:2018ewt}
to very high chiral orders suggests that the inclusion of leading
\NNN{} forces does not resolve the $A_y$-puzzle. Instead, there are
hints that the $A_y$-puzzle could be resolved with sub-leading \NNN{}
forces~\cite{Epelbaum:2019zqc}. Alternatively, the $A_y$ puzzle might
vanish in a simultaneous \NN+\NNN{} analysis conditioned on \NN{} and
\Nd{} scattering data and informed by model discrepancies such as the
truncation error in effective field theory.

Low-energy \nd{} differential cross-section data is very well
reproduced by \NNLOopt{}, see the left panel of Fig.~\ref{fig | DSG
  nnloopt}, and the results are identical to what is obtained using
the \INNNLO{} potential. At higher energies, however, there are some
discrepancies with respect to data and the two employed potentials
differ slightly in the vicinity of the cross section minimum. A
previous study~\cite{Witala:1998ey} concluded that the effects of
\NNN{} interactions are expected to be particularly noticeable in this
angular region. Although the \NNLOopt{} prediction lies marginally
closer to the experimental data at $\theta_\text{cm} \approx
120^{\circ}$ the shape of the differential cross section is not
correct.

\begin{figure*}[t]
	\centering
	\includegraphics[width=\textwidth]{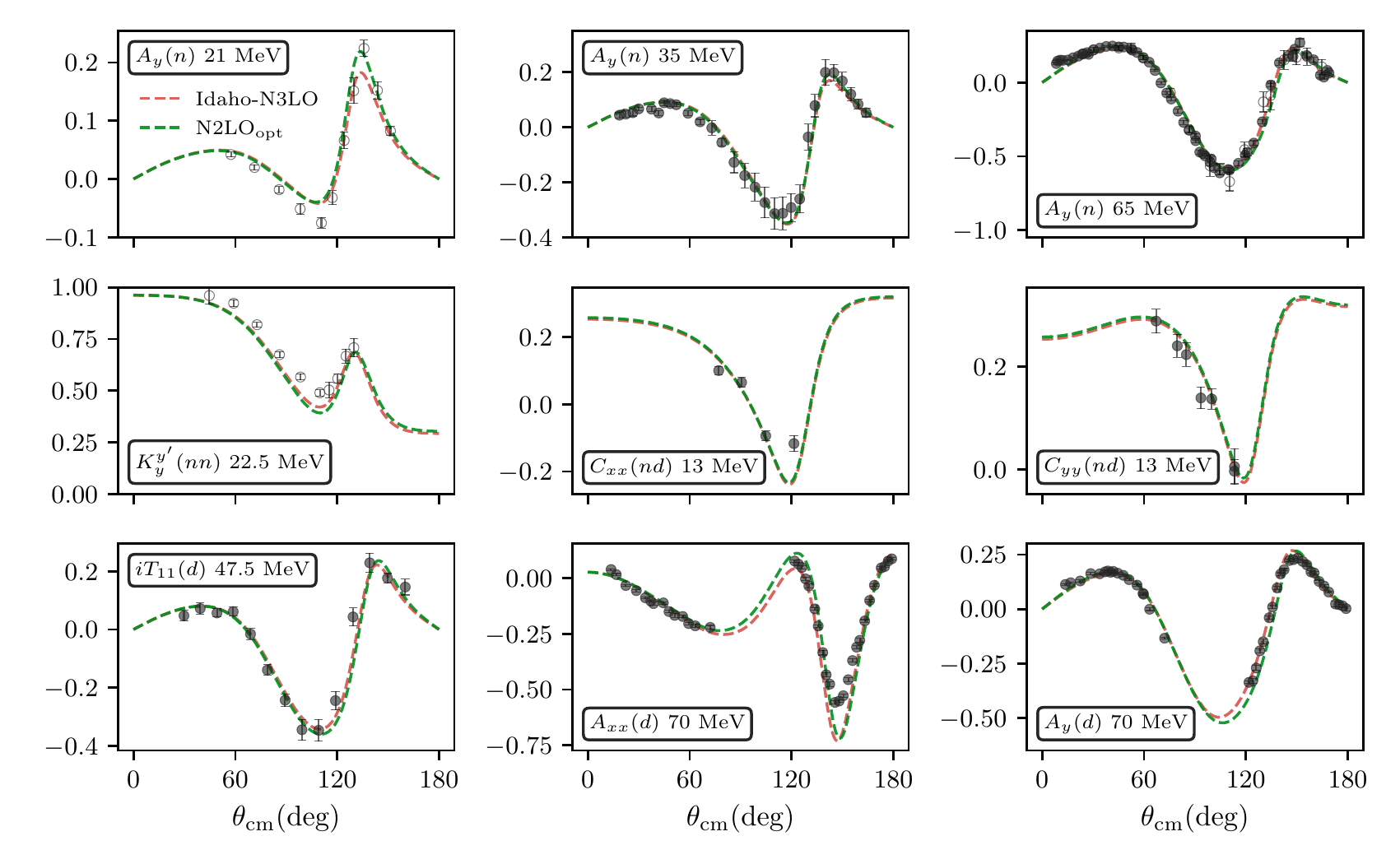}
	\caption{Spin observables for elastic \nd{} scattering
          computed using the \wpcd{} method with $\NWP=125$
          wave-packets and the \NNLOopt{} and the \INNNLO{}
          potentials. The experimental \nd{} data (empty markers) at
          $\Elab=21,22.5$ and 65 MeV are from
          Refs.~\cite{Weisel:2015min},~\cite{Clajus_1995},
          and~\cite{Ruhl:1991qop}, respectively. The \pd{} data
          (filled markers) at $\Elab=13,35,47.5,65$ and 70 MeV are
          from
          Refs.~\cite{Chauvin:1975fuc},~\cite{Bunker:1968omc},~\cite{Witaia:1993vz},~\cite{Shimizu:1982lqa},
          and~\cite{Sekiguchi:2004yb}, respectively. See the main text
          for detailed discussion.}
	\label{fig | nnloopt predictions}
\end{figure*}

In Fig.~\ref{fig | nnloopt predictions} we show a range of spin
observables for $\Elab=13-70$ MeV. Overall, \NNLOopt{} and \INNNLO{}
describe the data rather well in this energy region and the two
different potentials produce virtually identical results. In the top
row of Fig.~\ref{fig | nnloopt predictions} we present $A_y(n)$ for
increasing values of $\Elab$. It is well known that at energies below
$\sim 30$ MeV nearly all \NN{} interactions fail to describe the data
for this observable~\cite{Weisel:2015min}. As was discussed above, the
\NNLOopt{} interaction does not remedy the puzzle. Nevertheless,
careful inspection reveals that the predictions for $A_y$ at
$\Elab=21$ MeV fits the data slightly better at large scattering
angles when using \NNLOopt{}. Unfortunately, discrepancies with
respect to data persists for small scattering angles, i.e., at the
minimum value of $A_y$. In the second row of Fig.~\ref{fig | nnloopt
  predictions} we present low-energy neutron-to-neutron spin transfer
($K$) and neutron-to-deuteron correlation ($C$) observables. Previous
studies~\cite{Clajus:1990rzf} have found that the $K_y^{y'}$ spin
transfer is most sensitive to the structure of the \NN{} interaction
in the $^3S_1$--$^3D_1$ and $^1P_1$ channels. Since \INNNLO{} and
\NNLOopt{} have very similar \NN{} phase shifts below $\Elab=100$ MeV
in these channels it is not surprising to recover very similar results
also for these observables. Of course, the former potential
incorporates higher-order long- and short-range physics that modify
the off-shell structure of the potential, but this does not appear
alter the predictions much. Regarding the tensor analyzing powers
presented in the third row, the discrepancy between theory and data
for $A_{xx}(d)$ at $\theta_\text{cm} \approx 150 ^{\circ}$ persists
for both potentials. Inclusion of modern \NNN{} forces does not
resolve this~\cite{Epelbaum:2019zqc}.

\section{Summary and outlook}
\label{sec | summary and outlook}
In this work we presented a framework that allows to solve the Faddeev
equations for elastic \Nd{} scattering using a newly developed code
based on the \wpcd{} method. We analyzed the convergence of the
\wpcd{} method, applied to chiral potentials, with respect to the
number of basis wave packets $\NWP$. We find negligible method errors
when using $\NWP=125$ in the regime $\Elab \lesssim 70$ MeV.

We studied different \nd{} scattering observables up to $\Elab=70$ MeV with
the \NNLOopt{} \NN{} interaction and find a good overall reproduction of
the scattering data. However, the $A_y$-puzzle remains unsolved when applying
the \NNLOopt{} interaction. Compared to the \NijmI{} and
\INNNLO{} \NN{} interactions, we detect a minor increase in the
maximum of the theoretical predictions of this observable at low
energies. For other cross sections we find a good reproduction of
experimental data and the results are virtually indistinguishable from
the \INNNLO{} interaction. The \NNLOopt{} interaction prediction
for $d\sigma/d\Omega$ at $\Elab=64.5$ MeV is slightly closer to the
data at the differential cross section minimum. This observable is
typically associated with an increased sensitivity to
\NNN{} interactions.

Next, we will explore discretized bases with a different number of
wave-packets for the $p$ and $q$ momenta, make predictions for breakup
cross sections, and incorporate \NNN{} interactions in our
calculations. Although some observables, that depend sensitively on
finer details of the nuclear interaction, require more wave-packets to
be accurately resolved one can obtain sufficiently accurate
predictions for the vast majority of low-energy \Nd{} cross sections
with $N_\text{WP}\approx 50$ wave packets, which will help to reduce
the computational demands of the calculations to a level that allows a
study of \Nd{} scattering observables within a statistical
analysis. Specifically, in the near term we plan to employ \wpcd{}
predictions to sample Bayesian posterior predictive distributions for
\Nd{} scattering. Work in this direction, using frequentist methods,
was also initiated in~\cite {Skibinski:2018dot}. In the longer
perspective, emulator methods based on perturbation
theory~\cite{Witala:2021xqm} or eigenvector continuation~\cite
{Frame:2017fah} promise an efficient method for fast and accurate
emulation of scattering observables~\cite
{Furnstahl:2020abp,Bai:2021xok,Zhang:2021jmi,Drischler:2021qoy} and
will open new ways to systematically incorporate \Nd{} scattering
observables in the construction and fitting process of next-generation
\NN~and \NNN{} interactions.

\section*{Acknowledgments} This work was supported by the European Research
 Council (ERC) under the European Unions Horizon 2020 research and innovation
 programme (Grant agreement No. 758027), and in part by the  Deutsche
 Forschungsgemeinschaft (DFG,  German Research Foundation) -- Project-ID
 279384907 -- SFB 1245. The computations were enabled by resources provided
 by the Swedish National Infrastructure for Computing (SNIC) at Chalmers
 Centre for Computational Science and Engineering (C3SE) and the National
 Supercomputer Centre (NSC) partially funded by the Swedish Research
 Council.

\appendix
\section{Partial-wave decomposition of the permutation operator in a plane-wave basis}
\label{app | permutation operator plane wave representation}
In this section we present the permutation operator $\hat{P}_{123}$ in
a plane-wave partial-wave basis. The permutation operator
$\hat{P}_{ijk}=\hat{P}_{ij}\hat{P}_{jk}$ performs two pairwise
interchanges of particles: first $j\leftrightarrow k$ followed by
$i\leftrightarrow j$. Our derivation follows the steps presented in
\cite{Hebeler:2020ocj}, as well as the notation and convention for the
Jacobi momenta $\boldsymbol{p}$ and $\boldsymbol{q}$. For this section
we will use the indexing ($ij$) to denote the $ij$-pair system for the
sake of clarity, rather than the odd-man-out notation. To complement
\cite{Hebeler:2020ocj}, we use the $(12)$-subsystem as our initial
states upon which the permutation operator acts. One can show that all
representations of $\hat{P}_{123}$ are invariant under change of
reference system. Furthermore, one can show \cite{Hebeler:2020ocj,Glockle:aYZcaRsU}
that for a basis that is antisymmetric under exchange of particles 2 and 3, we have
$\langle\hat{P}_{123}\rangle=\langle\hat{P}_{132}\rangle$. This allows us
to express projections of $\hat{P}$ in Eq.~\eqref{eq | faddeev equation | operator form}
simply as $\langle \hat{P}\rangle = \langle 1+2\hat{P}_{123}\rangle$.

Our starting point is the following overlap
\begin{align}
	\begin{split}
	\prescript{}{12}{\langle}p'q';\alpha'|\hat{P}_{123}|pq;\alpha\rangle_{12}
	&= \prescript{}{12}{\langle} p'q';\alpha'|\hat{P}_{12}\hat{P}_{23}|pq;\alpha\rangle_{12} \\
	&= \prescript{}{12}{\langle} p'q';\alpha'|pq;\alpha\rangle_{23} \:.
	\end{split}
\end{align}
which, projected in a partial-wave basis, becomes (note that that
magnetic quantum numbers $m_{\mathcal{J}}$ and $m_{\mathcal{T}}$ are
implied in $|\alpha\rangle$)
\begin{align}
	\begin{split}
	\prescript{}{12}{\langle p'q';\alpha'}|pq;\alpha\rangle_{23}	
	&=
	\sum_{\mathcal{L}\mathcal{S}}
	\sum_{\mathcal{L}'\mathcal{S}'}
	\sqrt{\hat{J}'\hat{j}'\hat{\mathcal{L}}'\hat{\mathcal{S}}'}
	\sqrt{\hat{J}\hat{j}\hat{\mathcal{L}}\hat{\mathcal{S}}}\\
	&\hphantom{=}\times
	\begin{Bmatrix}
	L' & S' & J' \\
	l' & \frac{1}{2} & j' \\
	\mathcal{L}' & \mathcal{S}' & \mathcal{J}'
	\end{Bmatrix}
	\begin{Bmatrix}
	L & S & J \\
	l & \frac{1}{2} & j \\
	\mathcal{L} & \mathcal{S} & \mathcal{J}
	\end{Bmatrix}\\
	&\hphantom{=}\times
	\sum_{m_{\mathcal{L}}m_{\mathcal{S}}}
	\sum_{m_{\mathcal{L}'}m_{\mathcal{S}'}} C_{\mathcal{L}'m_{\mathcal{L}'},\mathcal{S}'m_{\mathcal{S}}'}^{\mathcal{J}'m_{\mathcal{J}'}}
	C_{\mathcal{L}m_{\mathcal{L}},\mathcal{S}m_{\mathcal{S}}}^{\mathcal{J}m_{\mathcal{J}}}\\
	&\hphantom{=}\times
	\prescript{}{12}{\langle} p'q';L'l'\mathcal{L}'m_{\mathcal{L}'}|pq;Ll\mathcal{L}m_{\mathcal{L}}\rangle_{23} \\
	&\hphantom{=}\times
	\prescript{}{12}{\langle}(S'\frac{1}{2})\mathcal{S}'m_{\mathcal{S}'}|(S\frac{1}{2})\mathcal{S}m_{\mathcal{S}}\rangle_{23} \\
	&\hphantom{=}\times \prescript{}{12}{\langle}(T'\frac{1}{2})\mathcal{T}'m_{\mathcal{T}'}|(T\frac{1}{2})\mathcal{T}m_{\mathcal{T}}\rangle_{23}\:, \\
	\end{split}
	\label{eq | permutation matrix element first step}
\end{align}
where the spin and isospin recouplings are given by the Wigner-6j
symbols,
\begin{align}
	\begin{split}
	&\prescript{}{12}{\langle}(S'\frac{1}{2})\mathcal{S}'m_{\mathcal{S}'}|(S\frac{1}{2})\mathcal{S}m_{\mathcal{S}}\rangle_{23} \\
	=& \delta_{\mathcal{S}'\mathcal{S}}\delta_{m_{\mathcal{S}'}m_{\mathcal{S}}}(-1)^{S}\sqrt{\hat{S}'\hat{S}}
	\begin{Bmatrix}
	\frac{1}{2} & \frac{1}{2} & S' \\
	\frac{1}{2} & \mathcal{S} & S
	\end{Bmatrix}\:.
	\end{split}
\end{align}
Here, $C_{l_1m_1,l_2m_2}^{l_3m_3}$ denote Clebsch-Gordan
coefficients and we use the notation $\hat{n}\equiv \sqrt{2n+1}$
The recoupling of orbital angular momenta are calculated
using momentum-space projection in the pair-systems $(12)$ and $(23)$,
\begin{align}
	\begin{split}
	&\prescript{}{12}{\langle} p'q';L'l'\mathcal{L}'m_{\mathcal{L}'}|pq;Ll\mathcal{L}m_{\mathcal{L}}\rangle_{23}\\
	&= \int_0^\infty \:d\bm{p}_{12}'''\:d\bm{q}_{12}'''d\bm{p}_{23}''\:d\bm{q}_{23}''\: \mathcal{Y}_{L'l'}^{*\mathcal{L}'m_{\mathcal{L}'}}(\hat{\bm{p}}_{12}''',\hat{\bm{q}}_{12}''')\\
	&\hphantom{=}\times \prescript{}{12}{\langle} \bm{p}'''\bm{q}'''|\bm{p}''\bm{q}''\rangle_{23} \: \mathcal{Y}_{Ll}^{\mathcal{L}m_{\mathcal{L}}}(\hat{\bm{p}}_{23}'',\hat{\bm{q}}_{23}'') \\
	&\hphantom{=}\times \left[\frac{\delta(p'-p''')}{p'p'''}\frac{\delta(q'-q''')}{q'q'''}\right]_{12} \\
	&\hphantom{=}\times \left[\frac{\delta(p''-p)}{p''p}\frac{\delta(q''-q)}{q''q}\right]_{23}\:,\\
	\end{split}
	\label{eq | permutation matrix element calculation step 1}
\end{align}
where the hat-notation on vectors indicate unit vectors, and we introduced the coupled spherical harmonics,
\begin{equation}
	\mathcal{Y}_{l_1l_2}^{l_3m_3}(\hat{\bm{a}},\hat{\bm{b}}) = \sum_{m_1m_2} C_{l_1m_1,l_2m_2}^{l_3m_3} Y_{l_1m_1}(\hat{\bm{a}})Y_{l_2m_2}(\hat{\bm{b}})\:,
	\label{eq | definition of coupled spherical harmonics}
\end{equation}
and where $Y_{lm}(\hat{\bm{a}})$ are the spherical harmonics. Note that we
use the Condon-Shortley phase factor. Furthermore we use proper
normalization of the spherical harmonics, such that numerical
evaluation is performed in terms of the associated Legendre
polynomials as $Y_{lm}(\hat{\bm{a}}) = e^{im\theta} P_l^m(\cos(\phi))$
where $\hat{\bm{a}}=(\theta,\phi)$.

The inner product of Jacobi momenta,
\begin{equation}
	\prescript{}{12}{\langle} \bm{p}'''\bm{q}'''|\bm{p}''\bm{q}''\rangle_{23} = \delta(\bm{p}'''_{12}-\bm{p}''_{12})\delta(\bm{q}'''_{12}-\bm{q}''_{12}) \:,
\end{equation}
can be resolved using the identities $\bm{p}''_{12} =
-\frac{1}{2}\bm{p}_{23}''+\frac{3}{4}\bm{q}_{23}''$ and $\bm{q}''_{12}
= -\bm{p}_{23}''-\frac{1}{2}\bm{q}_{23}''$, from which we can define
four variables that fulfil momentum conservation,
\begin{align}
	\begin{split}
	\prescript{}{12}{\langle} \bm{p}'''\bm{q}'''|\bm{p}''\bm{q}''\rangle_{23}
	&= \delta(\bm{p}_{23}''-\bm{\bar{p}})\delta(\bm{q}_{23}''-\bm{\bar{q}}) \\
	&= \delta(\bm{p}_{12}'''-\bm{\tilde{p}})\delta(\bm{q}_{12}'''-\bm{\tilde{q}}) \\
	&= \delta(\bm{p}_{12}'''-\bm{\pi}')\delta(\bm{p}_{23}''-\bm{\pi}) \\
	&= \delta(\bm{q}_{12}'''-\bm{\kappa}')\delta(\bm{q}_{23}''-\bm{\kappa}) \:.
	\end{split}
	\label{eq | permutation matrix dirac delta choice}
\end{align}
Here, we also defined (in the pair-system $(12)$)
\begin{align}
	\begin{split}
	&\left.\begin{cases}
	\bm{\bar{p}}  &\equiv -\frac{1}{2}\bm{p}_{12}''' - \frac{3}{4}\bm{q}_{12}''' \\
	\bm{\bar{q}}  &\equiv \bm{p}_{12}''' - \frac{1}{2}\bm{q}_{12}'''
	\end{cases}\right\}
	\:,\quad
	\left.\begin{cases}
	\bm{\tilde{p}}  &\equiv -\frac{1}{2}\bm{p}_{23}'' + \frac{3}{4}\bm{q}_{23}'' \\
	\bm{\tilde{q}}  &\equiv -\bm{p}_{23}'' - \frac{1}{2}\bm{q}_{23}''
	\end{cases}\right\} \:,
	\\	
	&\left.\begin{cases}
	\bm{\pi}  &\equiv  -\frac{1}{2}\bm{q}_{23}''' - \bm{q}_{12}''\\
	\bm{\pi}' &\equiv \frac{1}{2}\bm{q}_{12}'' + \bm{q}_{23}'''
	\end{cases}\right\}
	\:,\quad
	\left.\begin{cases}
	\bm{\kappa}  &\equiv  \frac{2}{3}\bm{p}_{23}''' + \frac{4}{3}\bm{p}_{12}''\\
	\bm{\kappa}' &\equiv -\frac{4}{3}\bm{p}_{23}''' - \frac{2}{3}\bm{p}_{12}''
	\end{cases}\right\}\:.
	\end{split}
	\label{eq | pi bar tilde momenta}
\end{align}
Note that the exact form of these relations depend on the choice of
pair-system. See \cite{Hebeler:2020ocj} for a summary of three-body
kinematics.

Choosing to conserve $(\bm{\bar{p}},\bm{\bar{q}})$ will
restrict the bra-momenta of an operator to the right of the
permutation operator in the Faddeev equation due to the ensuing
delta-functions (used in
e.g. \cite{Hebeler:2020ocj,Gloeckle:1995jg}). Likewise,
$(\bm{\tilde{p}},\bm{\tilde{q}})$ will restrict the ket-momenta of an
operator to the left, while $(\bm{\kappa},\bm{\kappa}')$ will restrict
$p$ of operators on both sides, and lastly $(\bm{\pi},\bm{\pi}')$ will
restrict $q$ on both sides (used in
e.g. \cite{glockle1983quantum,Pomerantsev:2014ija}).
In this work we have followed \cite{Hebeler:2020ocj}
and conserve $(\bm{\bar{p}},\bm{\bar{q}})$. From this point on we drop
the $(12)$ subscript on momenta and get
\begin{align}
	\begin{split}
	\prescript{}{12}{\langle} p'q';L'l'\mathcal{L}'m_{\mathcal{L}'}|pq;Ll\mathcal{L}m_{\mathcal{L}}\rangle_{23} &= \int_0^\infty d\bm{p}'''\:d\bm{q}'''\\
	&\hphantom{=}\times \mathcal{Y}_{L'l'}^{*\mathcal{L}'m_{\mathcal{L}'}}(\hat{\bm{p}}''',\hat{\bm{q}}''')\\
	&\hphantom{=}\times \mathcal{Y}_{Ll}^{\mathcal{L}m_{\mathcal{L}}}(\hat{\bm{\bar{p}}},\hat{\bm{\bar{q}}}) \\
	&\hphantom{=}\times \frac{\delta(\bar{p}-p)}{\bar{p}p}\frac{\delta(\bar{q}-q)}{\bar{q}q} \:,
	\end{split}
	\label{eq | permutation matrix element calculation step 2}
\end{align}
where $\bar{p} = |\bar{\bm{p}}|$ and $\bar{q} = |\bar{\bm{q}}|$. The
integral is invariant under rotations, making it proportional to
$\delta_{\mathcal{L}'\mathcal{L}}\delta_{m_{\mathcal{L}'}m_{\mathcal{L}}}$.
This invariance allows us to simply average
out $m_\mathcal{L}$, giving a factor $\frac{1}{2\mathcal{L}+1}$. We now
have freedom in the choice of axes. Choosing
$\hat{z}\parallel\bm{\hat{p}}'''$ and the polar angle of
$\hat{\bm{q}}'''$ to zero, we can simplify a spherical harmonic:
$Y_{L'm_{L'}}(\hat{\bm{p}}''')=\sqrt{\frac{\hat{L}'}{4\pi}}\delta_{m_{L'}0}$. By
solving the remaining angular integrals we are left with
\begin{align}
	\begin{split}
	\prescript{}{12}{\langle} p'q';L'l'\mathcal{L}'|pq;Ll\mathcal{L}\rangle_{23}
	&= 8\pi^2\frac{\delta_{\mathcal{L}'\mathcal{L}}}{\hat{\mathcal{L}}}\int_{-1}^{+1} dx \\
	&\hphantom{=}\times \frac{\delta(\bar{p}-p)}{\bar{p}p}\frac{\delta(\bar{q}-q)}{\bar{q}q} \\
	&\hphantom{=}\times \sum_{m_{\mathcal{L}}}\mathcal{Y}_{L'l'}^{*\mathcal{L}'m_{\mathcal{L}'}}(\hat{\bm{p}}''',\hat{\bm{q}}''')\\
	&\hphantom{=}\times \mathcal{Y}_{Ll}^{\mathcal{L}m_{\mathcal{L}}}(\hat{\bar{\bm{p}}},\hat{\bar{\bm{q}}}) \:,
	\end{split}
        \label{eq | angle}
\end{align}
where vectors are functions of $(p',q',x)$ and $x=\cos(\phi)$ (the
angle from $\bm{q}'''$ to $\bm{p}'''$). Notice the change in notation
of states on the left-hand side as we averaged with respect to $m_{\mathcal{L}}$. Inserting
Eq.~\eqref{eq | angle} and Eq.~\eqref{eq | definition of coupled
  spherical harmonics} back into Eq.~\eqref{eq | permutation matrix
  element first step} gives
\begin{widetext}
\begin{equation}
	\prescript{}{12}{\langle}p'q';\alpha'|\hat{P}_{123}|pq;\alpha\rangle_{12} = \int_{-1}^1 dx\: G_{\alpha\alpha'}(p',q',x) \frac{\delta(\bar{p}-p)}{\bar{p}p}\frac{\delta(\bar{q}-q)}{\bar{q}q} \:,
	\label{eq | permutation matrix analytical expression}
\end{equation}
with the geometrical function $G_{\alpha\alpha'}(p',q',x)$ in a form that is straightforward to implement algorithmically,
\begin{align}
	\begin{split}
	 G_{\alpha\alpha'}(p',q',x)= \delta_{\mathcal{J}'\mathcal{J}}\delta_{\mathcal{T}'\mathcal{T}}&\sum_{\mathcal{L}\mathcal{S}}\sqrt{\hat{J}'\hat{J}}\sqrt{\hat{j}'\hat{j}}
	\hat{\mathcal{S}}\begin{Bmatrix}
	L' & S' & J' \\
	l' & \frac{1}{2} & j' \\
	\mathcal{L} & \mathcal{S} & \mathcal{J}
	\end{Bmatrix}
	\begin{Bmatrix}
	L & S & J \\
	l & \frac{1}{2} & j \\
	\mathcal{L} & \mathcal{S} & \mathcal{J}
	\end{Bmatrix}
	(-1)^{S}\sqrt{\hat{S}'\hat{S}}
	\begin{Bmatrix}
	\frac{1}{2} & \frac{1}{2} & S' \\
	\frac{1}{2} & \mathcal{S} & S
	\end{Bmatrix}(-1)^{T}\sqrt{\hat{T}'\hat{T}}
	\begin{Bmatrix}
	\frac{1}{2} & \frac{1}{2} & T' \\
	\frac{1}{2} & \mathcal{T} & T
	\end{Bmatrix} \\
	\times\frac{8\pi^2}{\sqrt{4\pi}}
	&\sum_{m_{\mathcal{L}}m_Lm_l} C_{L'0,lm_{\mathcal{L}}}^{\mathcal{L}m_{\mathcal{L}}}C_{Lm_L,lm_l}^{\mathcal{L}m_{\mathcal{L}}} P_L^{m_L}(\cos(\theta_1))P_l^{m_l}(\cos(\theta_2))(-1)^{m_{\mathcal{L}}}P_{l'}^{m_{\mathcal{L}}}(x) \:,\quad m_{\mathcal{L}}\equiv m_L+m_l \:,\\
	\end{split}
	\label{eq | geometric function}
\end{align}
\end{widetext}
and where we defined
\begin{align}
	\begin{split}
	\cos(\theta_1) &\equiv \frac{|\bar{\bm{p}}_z|}{\bar{p}} = \frac{-\frac{1}{2}p' - \frac{3}{4}q'x}{\bar{p}} \:,\\
	\cos(\theta_2) &\equiv \frac{|\bar{\bm{q}}_z|}{\bar{q}} = \frac{p' - \frac{1}{2}q'x}{\bar{q}} \:.
	\end{split}
\end{align}
and where we used $m_{L'}=0$. Given that we have $\mathcal{S}=\mathcal{S}'$, $m_{\mathcal{S}}=m_{\mathcal{S}'}$, $\mathcal{L}=\mathcal{L}'$, and $m_{\mathcal{L}}=m_{\mathcal{L}'}$, we used the orthogonality of Clebsch-Gordan coefficients,
\begin{equation}
	\sum_{m_1m_2} C_{l_1m_1,l_2m_2}^{l_3'm_3'}C_{l_1m_1,l_2m_2}^{l_3m_3} = \delta_{l_3'l_3}\delta_{m_3'm_3}\:.
\end{equation}
to set $\mathcal{J}'=\mathcal{J}$ and $m_{\mathcal{J}}=m_{\mathcal{J}'}$.

Previously, in~\cite{glockle1983quantum} the angular dependence in
$G_{\alpha\alpha'}(p',q',x)$ was evaluated separately from the
recoupling terms. This allows pre-calculation of the geometric
recouplings before doing the angular integration. However, it turns
out that keeping the angular dependence as above is both more
numerically efficient and stabler with higher $l$ and $L$
\cite{Hebeler:2020ocj}. As this function is the most computationally
costly part of evaluating the integral of Eq.~\eqref{eq | permutation matrix analytical expression},
we mention some key optimizations one can use.

The simplest and most effective optimization is to calculate $G_{\alpha\alpha'}(p',q',x)$ and store it in the computer memory in its entirety. From a computational viewpoint the function is 5-dimensional, which is still storable in the computer memory for the basis sizes and number of quadrature points we typically require.

Regardless of whether prestorage of $G_{\alpha\alpha'}(p',q',x)$ is possible,
we still wish to speed up the calculation of $G_{\alpha\alpha'}(p',q',x)$.
To this end, everything before the second sum in
Eq.~\eqref{eq | geometric function} (i.e. all geometric recoupling)
can easily be precalculated to improve computational performance.
The second summation can be sped up by prestoring the three Legendre polynomials individually,
which is usually still manageable and quite fast.

\section{Projecting operators to the wave-packet basis}
\label{app | WP projection}
In this section we present the expressions we employ for projecting
operators to a wave-packet basis. Although Eq.~\eqref{eq | general
  wave-packet definition} provides a definition of a three-body wave
packet, it is written explicitly for plane-wave states. Identically,
and in general, any three-body wave packet $|Y_{ij}\rangle$ (where we
omit partial-wave indexing) can be defined from continuum states
$|\omega_p\rangle$ and $|\omega_q\rangle$ of some operator
(e.g. $\hat{p}$, $\hat{h}_0$, or $\hat{h}_1$) with Jacobi momenta
$p\in\mathcal{D}_i$ and $q\in\bar{\mathcal{D}}_j$, 
\begin{equation}
	Y_{ij} = \frac{1}{N_{ij}}\int_{\mathcal{D}_i,\bar{\mathcal{D}}_j} dp\:p\:dq\:q\: f(p)\bar{f}(q)|\omega_{p},\omega_{q}\rangle \:,
\end{equation}
where $N_{ij}$ is the wave-packet normalization constant, and $f(p)$ and
$\bar{f}(q)$ are the weighting functions defined in Sec.~\ref{sec | wpcd basis}.
The wave packet can be
projected onto the continuum basis straightforwardly,
\begin{align}
	\begin{split}
	\langle \omega_p,\omega_q|Y_{ij}\rangle &= \frac{1}{N_{ij}}\int_{\mathcal{D}_i,\bar{\mathcal{D}}_j} dp'p'dq'q'\: f(p')\bar{f}(q')\\
	&\hspace{2cm}\times \langle\omega_p,\omega_q|\omega_{p'},\omega_{q'}\rangle \\
	&= \frac{1}{N_{ij}}\frac{f(p)\bar{f}(q)}{pq}\mathds{1}_{\mathcal{D}_i}(p)\mathds{1}_{\bar{\mathcal{D}}_j}(q) \:,
	\end{split}
	\label{eq | continuum-WP inner product}
\end{align}
where $\mathds{1}_{\mathcal{D}_i}(p)$ is the indicator function. From this
it is easy to show that $N_{ij}=\sqrt{D_i\bar{D}_j}$, where
$D_i$($\bar{D}_j$) is the width of
$\mathcal{D}_i$($\bar{\mathcal{D}}_j$). Note that for energy wave packets the
width is expressed in energy as, for example, $D_i=E_{i+1}-E_i$ for boundaries $E_i$.
The general form of Eq.~\eqref{app | WP free hamiltonian} comes in
handy for analytical derivation as it applies to both free and
scattering wave packets alike.

Going back to a \fwp{} basis, a projection of a general \NNN{} operator will
look as follows,
\begin{align}
	\begin{split}
	\langle X_{i'j'}^{\alpha'}|\hat{O}|X_{ij}^\alpha\rangle = \frac{1}{N_{i'j'}N_{ij}}&\int_{\mathcal{D}_{i'j'}} dp'\:p'\:dq'\:q' \:f^*(p')\bar{f}^*(q')\\
	\times&\int_{\mathcal{D}_{ij}} dp\:p\:dq\:q\: f(p)\bar{f}(q)\\
	\times &\langle p'q';\alpha'|\hat{O}|pq;\alpha\rangle \:.
	\end{split}
\label{eq | general operator in fWP basis}
\end{align}

\subsection{Two-body free Hamiltonian}
\label{app | WP free hamiltonian}
For our chosen normalization, the free \NN{} Hamiltonian is given by
\begin{equation}
	\langle p'q';\alpha'|\hat{h}_0|pq;\alpha\rangle = \delta_{\alpha'\alpha}\frac{\delta(p'-p)}{p^2}\frac{\delta(q'-q)}{q^2}\langle p|\hat{h}_{0}|p\rangle \:.
\end{equation}
Clearly, the free Hamiltonian is also diagonal in the \fwp{} basis. Depending on the choice of wave packet, we get for the \NN{} free Hamiltonian,
\begin{equation}
	\langle X_{i'j'}^{\alpha'}|\hat{h}_{0}|X_{ij}^\alpha\rangle = \begin{cases}
	\delta_{\alpha'\alpha}\delta_{i'i}\delta_{j'j}\frac{p_{i+1}^2 + p_{i}^2}{2\mu_0} ,& f(p)=\sqrt{\frac{p}{\mu_0}} \\
	\delta_{\alpha'\alpha}\delta_{i'i}\delta_{j'j}\frac{p_{i+1}^2 + p_{i+1}p_{i} + p_{i}^2}{6\mu_0} ,&f(p)=1 \, .
	\end{cases}
\end{equation}

\subsection{The \NN{} potential}
\label{app | WP potential}
The \NN{} potential $\hat{v}$ in a \NNN{} partial-wave basis reduces to
\begin{equation}
	\langle p'q';\alpha'|\hat{v}|pq;\alpha\rangle = \delta_{\gamma'\gamma}\delta_{\Gamma'\Gamma}\frac{\delta(q'-q)}{q^2}\langle p'|\hat{v}_{n'n}|p\rangle \:,
\end{equation}
where $\gamma$ denotes all the quantum numbers for the third nucleon
relative to the pair system, i.e., $\gamma=\{l,j\}$,
$\Gamma=\{\mathcal{J},\mathcal{T}\}$ denotes the coupled \NNN{}
quantum numbers, and the pair-system quantum numbers are jointly
referred to as $n=\{L,S,J,T\}$. For our predictions we break total
$\mathcal{T}$ isospin conservation, and the expressions below must be modified in
an obvious way. We obtain the \NN{} interaction in the \NNN{} \fwp-basis via Eq.~\eqref{eq | general operator
  in fWP basis}, and easily resolving the $q$-integral,
\begin{align}
	\begin{split}
	\langle X_{i'j'}^{\alpha'}|\hat{v}|X_{ij}^\alpha\rangle = &\frac{\delta_{\gamma'\gamma}\delta_{\Gamma \Gamma'}\delta_{j'j}}{D_{i'}D_i}\int_{\mathcal{D}_{i'}} dp'\:p'
	\int_{\mathcal{D}_{i}} dp\:p \\
	&\times f^*(p')f(p)\langle p'|\hat{v}_{n'n}|p\rangle \:.
	\end{split}
\end{align}
This expression is straightforward to evaluate numerically using quadrature.

\subsection{The permutation operator}
\label{app | WP permutation operator}
The permutation operator $\hat{P}_{123}$ in a partial-wave basis,
Eq.~\eqref{eq | permutation matrix analytical expression}, can be
inserted into Eq.~\eqref{eq | general operator in fWP basis}. The
delta-functions are only non-zero when the Jacobi momenta $\bar{p}$
and $\bar{q}$ fall within the bins $\mathcal{D}_i$ and
$\bar{\mathcal{D}}_j$, which we express using the indicator
function. Choosing momentum wave-packets gives
\begin{align}
	\begin{split}
	\langle X_{i'j'}^{\alpha'}|\hat{P}_{123}|X_{ij}^\alpha\rangle = \frac{1}{N_{i'j'}N_{ij}}&\int_{\mathcal{D}_{i'j'}} dp'\:p'\:dq'\:q'\\
	\times&\int_{-1}^1\:dx\: G_{\alpha\alpha'}(p',q',x)\\
	\times&\frac{\mathds{1}_{\mathcal{D}_j}(\bar{p})}{\bar{p}}\frac{\mathds{1}_{\bar{\mathcal{D}}_j}(\bar{q})}{\bar{q}}\:.
	\end{split}
\label{eq | app permutation operator temp step}
\end{align}
The indicator function is discontinuous and an evaluation of
Eq.\eqref{eq | app permutation operator temp step} using Gaussian
quadrature in the $p$ and $q$ momenta yields poor convergence with an
increasing number of quadrature points. Therefore, we transform the
integral over $p'$ and $q'$ to polar coordinates using the procedure
presented in Ref.~\cite{Pomerantsev:2014ija}:
\begin{equation}
\left.\begin{cases}
q' &= k\cos(\phi) \\
p' &= k\sin(\phi)
\end{cases}\right\} \:,\quad
\left.\begin{cases}
\phi &= \arctan\left(\frac{p'}{q'}\right) \\
k &= p'^2 + q'^2
\end{cases}\right\}\:,
\label{eq | parametrise p' and q'}
\end{equation}
Note that the integral-boundaries of $k$ and $\phi$ depend on each
other. With this parametrization, Eq.~\eqref{eq | app permutation
  operator temp step} can be expressed as
\begin{widetext}
\begin{align}
	\begin{split}
	\langle X_{i'j'}^{\alpha'}|\hat{P}_{123}|X_{ij}^\alpha\rangle =\frac{1}{N_{i'j'}N_{ij}}\int_{-1}^1\:dx\:\int_{\phi_{\text{min}}}^{\phi_{\text{max}}}d\phi\: \frac{\cos(\phi)\sin(\phi)}{\zeta_1\zeta_2}G_{\alpha\alpha'}(\cos(\phi),\sin(\phi),x)\frac{k_{\textrm{max}}'^2(\phi)-k_{\min}'^2(\phi)}{2} \:,
	\end{split}
	\label{eq | P123 in WPCD basis}
\end{align}
\end{widetext}
where we have used the following equality
$G_{\alpha\alpha'}(\cos(\phi),\sin(\phi),x) =
G_{\alpha\alpha'}(p',q',x)$ (see App.~\ref{app | permutation operator
  plane wave representation}), and where the momenta $\bar{p}$ and
$\bar{q}$ are replaced by $\zeta_1$ and $\zeta_2$,
\begin{align}
\begin{split}
\zeta_1 &\equiv \frac{\bar{p}}{k} = \sqrt{\frac{1}{4}\sin^2(\phi) + \frac{9}{16}\cos^2(\phi) + \frac{3}{4}x\cos(\phi)\sin(\phi)} \:, \\
\zeta_2 &\equiv \frac{\bar{q}}{k} = \sqrt{\sin^2(\phi) + \frac{1}{4}\cos^2(\phi) - x\cos(\phi)\sin(\phi)} \:.
\end{split}
\end{align}
Furthermore we have defined
\begin{align}
\begin{split}
k'_{\textrm{min}}(\phi) &\equiv \max\left[\frac{p_{i'}}{\sin(\phi)}, \frac{q_{j'}}{\cos(\phi)}, \frac{p_i}{\zeta_1}, \frac{q_j}{\zeta_2}\right] \:, \\
k'_{\textrm{max}}(\phi) &\equiv \min\left[\frac{p_{i'+1}}{\sin(\phi)}, \frac{q_{j'+1}}{\cos(\phi)}, \frac{p_{i+1}}{\zeta_1}, \frac{q_{j+1}}{\zeta_2}\right] \:,
\end{split}
\end{align}
which incorporates all integration limits imposed on $k$ by the
wave-packet bin boundaries and by $\phi$. To evaluate this expression
we must construct a quadrature mesh for $\phi$ which depends on the
bin indices $i'$, $j'$, $i$, and $j$. An important step in optimizing
the numerical evaluation of this integral is to first verify that
$\phi_{\text{min}}\leq\phi_{\text{max}}$ and $k_{\textrm{min}}\leq
k_{\textrm{max}}$. We find that the
$P_{123}$-matrix in a \fwp{}-basis is less than $0.1\%$ dense due to
momentum conservation.

The optimization steps discussed at the end of App.~\ref{app |
  permutation operator plane wave representation} are not all viable
in the \wpcd{} method. Since $\phi$ parametrizes $p'$ and $q'$, but
depends on 4 bin indices, $G_{\alpha\alpha'}(p',q',x)$ is essentially
7-dimensional, incurring a massive memory cost compared to the
continuum representation.  This can leave the precalculation of
individual Legendre polynomials as the only remaining viable
optimization step, provided enough computer memory to store them.  The
calculation of Eq.~\eqref{eq | P123 in WPCD basis} is somewhat costly,
but the resulting matrix is independent of the interaction and can be
stored to disk in a sparse format and reused.

\subsection{The channel resolvent}
\label{app | channel resolvent}
The channel resolvent $\hat{G}_1$ can be evaluated in closed form in a
\swp{}-basis. The relevant operator is defined as
\begin{equation}
	\hat{G}_a(E) = (E - \hat{h}_a)^{-1} \:,\quad a=(1,2,3) \:.
\end{equation}
This can also be expressed as a convolution
\cite{bianchi1964convolution} of the two-body resolvents $g_a^{(+)}$
and $g_0^{(+)}$. These depend on the two-body Hamiltonians $\hat{h}_a$
and $\hat{h}_0$, respectively, where $\hat{h}_a$ is the pair-system
Hamiltonian and $\hat{h}_0$ is the kinetic Hamiltonian of the third
particle relative to the pair-system. The result is
\begin{equation}
	\hat{G}_a(E) = \frac{1}{2\pi i}\int_{-\infty}^{\infty} d\epsilon\:\hat{g}_a^{(+)}(E-\epsilon)\hat{g}_0^{(+)}(\epsilon) \:.
\end{equation}
Following \cite{kukulin2007wave}, this can be expressed as the sum of two terms,
\begin{equation}
	\hat{G}_a(E) = \hat{R}_a(E) + \hat{Q}_a(E) \:,
	\label{eq | resovlent post-convolution}
\end{equation}
where
\begin{equation}
	\hat{R}_a(E) = \sum_{\alpha}\sum_{n=0}^{N_b} \int_0^\infty dE_q \frac{|\psi_{p,n}^\alpha, \psi_q^\alpha\rangle\langle \psi_{p,n}^\alpha, \psi_q^\alpha|}{E - \epsilon_n - E_q \pm i\epsilon} \:,
\end{equation}
and
\begin{equation}
	\hat{Q}_a(E) = \sum_{\alpha} \int_0^\infty dE_p\:dE_q \frac{|\psi_p^\alpha, \psi_q^\alpha\rangle\langle \psi_p^\alpha, \psi_q^\alpha|}{E - E_p - E_q \pm i\epsilon} \:,
\end{equation}
are the bound-continuum (BC) and continuum-continuum (CC) parts of the
channel resolvent, respectively. Here we have defined eigenstates
$\{|\psi_{p,n}^\alpha\rangle, |\psi_p^\alpha\rangle\}$ of $\hat{h}_a$
with eigenenergies $\epsilon_n<0$ for $n\leq N_b$ and $E_p>0$,
respectively, and eigenstates $\{|\psi_q^\alpha\rangle\}$ of
$\hat{h}_0$ with eigenenergies $E_q$.

Equation~\eqref{eq | resovlent post-convolution} can be projected onto
a \swp{} basis $\{|Z_{ij}^\alpha\rangle\}$, as the
inner-products $\langle \psi_{p,n}^\alpha,
\psi_q^\alpha|Z_{ij}^\alpha\rangle$ and $\langle \psi_p^\alpha,
\psi_q^\alpha|Z_{ij}^\alpha\rangle$ are known analytically
(Eq.~\eqref{eq | continuum-WP inner product}), such
that 
\begin{equation}
	\langle Z_{i'j'}^{\alpha'}|\hat{G}(E)|Z_{ij}^{\alpha}\rangle = \delta_{i'i}\delta_{j'j}\delta_{\alpha'\alpha}\left[R_{ij}^\alpha(E) + Q_{ij}^\alpha(E)\right] \:,
\end{equation}
where $R_{ij}^\alpha(E)\equiv\langle Z_{ij}^{\alpha}|\hat{R}(E)|Z_{ij}^{\alpha}\rangle$ is given by
\begin{equation}
	R_{ij}^\alpha(E) = \frac{1}{\bar{D}_j}\int_{\bar{\mathcal{D}_j}}dq\:\frac{|\bar{f}(q)|^2}{E - \epsilon_i - \frac{q^2}{2\mu_1} \pm i\epsilon}\:,
\end{equation}
and $Q_{ij}^\alpha(E)\equiv\langle Z_{ij}^{\alpha}|\hat{Q}(E)|Z_{ij}^{\alpha}\rangle$ is given by
\begin{equation}
	Q_{ij}^\alpha(E) = \frac{1}{D_i\bar{D}_j}\int_{\mathcal{D}_i, \bar{\mathcal{D}_j}}dp\:dq\:\frac{|f(p)|^2\bar{f}(q)|^2}{E - \frac{p^2}{2\mu_0} - \frac{q^2}{2\mu_1} \pm i\epsilon}\:.
\end{equation}
These integrals can be solved analytically and in the case of energy \swp{}s we get
\begin{align}
	\begin{split}
	\text{Re}(R_{ij}^\alpha(E)) &= \frac{1}{\bar{D}_j}\ln\left|\frac{\mathcal{E}_{j-1}+\epsilon_i - E}{\mathcal{E}_j + \epsilon_i - E}\right| \:,\\
	\text{Im}(R_{ij}^\alpha(E)) &= -\frac{\pi}{\bar{D}_j}\big[\Theta\left(\mathcal{E}_j+\epsilon_i-E\right) \\
	&\hspace{1.cm}- \Theta\left(\mathcal{E}_{j-1}+\epsilon_i-E\right)\big] \:,
	\end{split}
\end{align}
and
\begin{align}
	\begin{split}
	\text{Re}(Q_{ij}^\alpha(E)) &=
	\frac{1}{D_k\bar{D}_j}\big[
	                 \left(\Delta+\Delta_-\right)\ln\left|\Delta+\Delta_-\right|\\
	&\hspace{1.24cm}+\left(\Delta-\Delta_-\right)\ln\left|\Delta-\Delta_-\right|\\
	&\hspace{1.24cm}-\left(\Delta+\Delta_+\right)\ln\left|\Delta+\Delta_+\right|\\
	&\hspace{1.24cm}-\left(\Delta-\Delta_+\right)\ln\left|\Delta-\Delta_+\right|\big] \:,\\
	\text{Im}(Q_{ij}^\alpha(E)) &= -\frac{\pi}{D_k\bar{D}_j}\big[
	                \left(\Delta+\Delta_+\right)\Theta\left(\Delta+\Delta_+\right)\\
	&\hspace{1.5cm}+\left(\Delta-\Delta_+\right)\Theta\left(\Delta-\Delta_+\right)\\
	&\hspace{1.5cm}-\left(\Delta+\Delta_-\right)\Theta\left(\Delta+\Delta_-\right)\\
	&\hspace{1.5cm}-\left(\Delta-\Delta_-\right)\Theta\left(\Delta-\Delta_-\right)\big] \:,
	\end{split}
\end{align}
where
\begin{equation}
	\Delta \equiv \epsilon_i + \bar{\epsilon}_j - E \:,\quad
	\Delta_\pm \equiv \frac{D_i\pm\bar{D}_j}{2} \:,
\end{equation}
and where
$\hat{h}_1^0|\bar{x}_j\rangle=\bar{\epsilon}_j|\bar{x}_j\rangle$. We
also denoted the Heaviside step function with $\Theta$. We do not
distinguish between $\epsilon_i\lessgtr0$ since this follows
automatically from the operator being calculated, i.e. $R_{ij}^\alpha$
or $Q_{ij}^\alpha$. In \Nd{} scattering there is only one
\NN{} bound state, the deuteron, such that there should only be one
index $i=i_d$ where $R_{ij}^\alpha\neq0$, but here we have kept the
expressions above general.

\section{Neumann series and Pad\'e extrapolant}
\label{app | pade}
The Faddeev equation, just as the Lippmann-Schwinger and
Faddeev-Yakubovsky equations, are Fredholm type II equations (integral
equations), generally written as
\begin{equation}
	f(x) = \varphi(x) + \int K(x,y)f(y)\:dy\:,
\end{equation}
for any-dimensional variables $x$ and $y$. The Neumann series of this equation is written,
\begin{equation}
	f(x) = \sum_{n=0}^\infty K^n\varphi \:.
\end{equation}
This series only converges if all so-called Weinberg eigenvalues
$\eta_i$ of $K$ satisfy $|\eta_i|<1$, and this is by no means
guaranteed in nuclear physics. Indeed, analyzing the Weinberg
eigenvalues for nuclear interactions reveals the non-perturbative
character in many partial-waves, e.g., where we have bound
states~\cite{Hoppe:2017lok}. Using Pad\'e approximants is a convenient
method for resumming the terms of the Neumann series and extrapolating
beyond its radius of convergence. See
Refs.~\cite{baker,kukulin2013theory} for more details.

In brief, a Pad\'e approximant of a meromorphic function $f(z)$, which
is analytic near $z=0$, amounts to formulating the ratio of two
polynomial functions $P_N(z)$ and $Q_M(z)$ of degrees $N$ and $M$,
respectively, such that
\begin{equation}
	f(z) = a_0 + a_1z + a_2z^2 + a_3z^3 + \ldots = \frac{P_{N}(z)}{Q_{M}(	z)} + \mathcal{O}(z^{N+M+1}) \:,
	\label{eq | general neumann series}
\end{equation}
The advantage of this Pad\'e approximant is that, contrary to a simple polynomial
approximation which would only converge within some radius $|z|<R$, we
can now approximate singularities in $f(z)$. Finding the (unique)
coefficients of the polynomials $P_N$ and $Q_N$ amounts to solving a
system of polynomial equations. The solutions are effectively obtained
by evaluating the following determinants built from the terms in
the Neumann series, $\{a_n\}_{n=0}^{N+M}$,
\begin{equation}
	P_N(z) = \begin{vmatrix}
	a_{N-M+1} & a_{N-M+2} & \ldots & a_{N+1} \\
	\vdots & \vdots & \ddots & \vdots \\
	a_{N} & a_{N+1} & \ldots & a_{N+M} \\
	\sum\limits_{j=M}^N a_{j-M}z^j & \sum\limits_{j=M-1}^N a_{j-M+1}z^j & \ldots & \sum\limits_{j=0}^N a_{j}z^j \\
	\end{vmatrix} \:,
	\label{eq | Pade P-polynomial}
\end{equation}
and
\begin{equation}
	Q_M(z) = \begin{vmatrix}
	a_{N-M+1} & a_{N-M+2} & \ldots & a_{N+1} \\
	\vdots & \vdots & \ddots & \vdots \\
	a_{N} & a_{N+1} & \ldots & a_{N+M} \\
	z^M & z^{M-1} & \ldots & 1 \\
	\end{vmatrix} \:.
	\label{eq | Pade Q-polynomial}
\end{equation}
For our studies we have only used ``diagonal'' Pad\'e approximants
where we use $M=N\approx 15$ to ensure a convergent scattering
amplitude.

\section{Elastic scattering cross sections, polarizations observables, and the channel-spin scattering matrix}
\label{app | observables}
All elastic \Nd{} observables (total and differential cross sections
and spin observables) were calculated using expressions presented in
\cite{Ohlsen:1972zz}, which are straightforward to evaluate once the
spin-scattering matrix $M$ in a ``channel spin'' basis representation
has been obtained. For explicit forms of spin-projection operators in
such a basis we refer the reader to e.g. \cite{Seyler:1969sii}.

We define the channel spin $\bm{\Sigma}$ as the coupling of the
pair-system total angular momentum $\bm{J}$ and the spin of the
third nucleon $\bm{s}$,
\begin{equation}
	 \bm{\Sigma} \equiv \bm{J} + \bm{s}\:.
\end{equation}
In our conventions, the elastic spin-scattering matrix
$M(\theta)$ at some energy $E$ is represented as a
$6\times6$ matrix with elements given by
\begin{align}
	\begin{split}
	M_{\Sigma' m_{\Sigma'},\Sigma m_{\Sigma}}(\theta) &= \frac{\sqrt{\pi}}{ik} \sum_{\mathcal{J}l'l} i^{l'-l}\sqrt{2l+1} \\
	&\hspace{1.5cm}\times C_{\Sigma m_{\Sigma}, l 0}^{\mathcal{J} m_{\Sigma}} \\
	&\hspace{1.5cm}\times C_{\Sigma' m_{\Sigma'}, l' (m_{\Sigma} - m_{\Sigma'})}^{\mathcal{J} m_{\Sigma}} \\
	&\hspace{1.5cm}\times \left(S^{\mathcal{J}}_{l'\Sigma',l\Sigma} - \delta_{\Sigma'\Sigma}\delta_{l'l}\right) \\
	&\hspace{1.5cm}\times Y_{l'}^{(m_{\Sigma} - m_{\Sigma'}) }(\theta,0) \, ,
	\end{split}
\end{align}
where the $S$-matrix is given by
\begin{equation}
	S^{\mathcal{J}}_{l'\Sigma',l\Sigma} = \delta_{l'l}\delta_{\Sigma'\Sigma} - 2\pi iq m_N i^{l'-l}U^{\mathcal{J}}_{l'\Sigma',l\Sigma}\:,
\end{equation}
and where $m_N\equiv\frac{2m_pm_n}{m_p+m_n}$ is the nucleon
mass. Note that $m_N\equiv\frac{m_p+m_n}{2}$ is also commonly
  used. The difference between the two expressions occurs at the 7th
  significant digit and is not observed to be of any importance in our
  work. The channel-spin $U$-matrix of on-shell transition elements
are obtained by recoupling the $Jj$-coupled elements via
\begin{align}
	\begin{split}
	U_{l'\Sigma',l\Sigma}^{\mathcal{J}} = \sum_{j'j}&\sqrt{\hat{j}'\hat{\Sigma}'}(-1)^{\mathcal{J} + j'}
	\begin{Bmatrix}
	l' & \frac{1}{2} & j' \\
	J_d & \mathcal{J} & \Sigma'
	\end{Bmatrix}\\
	\times&
	\sqrt{\hat{j}\hat{\Sigma}}(-1)^{\mathcal{J} + j}
	\begin{Bmatrix}
	l & \frac{1}{2} & j \\
	J_d & \mathcal{J} & \Sigma
	\end{Bmatrix}
	U_{l'j',lj}^{\mathcal{J}}
	\end{split} \:,
\end{align}
where $J_d\equiv1=|\bm{J}|$ is the total angular momentum of the deuteron. The on-shell U-matrix in a plane-wave 
representation is extracted from a wave-packet representation $U_{i_dj}^{\alpha_d'\alpha_d}$, calculated through Eq.~\eqref{eq | WP on-shell U-matrix element},
 using Eq.~\eqref{eq | continuum-WP inner product},
\begin{equation}
	U_{l'j',lj}^{\mathcal{J}} = \frac{|\bar{f}(q)|^2}{q^2\bar{D}_j}U_{i_dj}^{\alpha_d'\alpha_d}\mathds{1}_{\bar{\mathcal{D}}_j}(q) \:,
\end{equation}
Usually, we find it best to let $q$ fall on bin midpoints and then interpolate $U_{l'j',lj}^{\mathcal{J}}$ to do predictions at arbitrary energies $E$. This approach works quite well and we see no noticeable difference in observables in going from linear to higher-order polynomial interpolation.

\section{Phase shifts and mixing angles}
\label{app | phase shifts}
Phase shifts and mixing angles are obtained by diagonalizing the channel-spin $S$-matrix in the \NNN{} partial wave $\mathcal{J}^{\Par}$ given by
\begin{equation}
	S^{\mathcal{J}} =
	\begin{pmatrix}
	S_{\mathcal{J}\mp\frac{3}{2}\:\frac{3}{2}, \mathcal{J}\mp\frac{3}{2}\:\frac{3}{2}}^{\mathcal{J}} &
	S_{\mathcal{J}\mp\frac{3}{2}\:\frac{3}{2}, \mathcal{J}\pm\frac{1}{2}\:\frac{1}{2}}^{\mathcal{J}} &
	S_{\mathcal{J}\mp\frac{3}{2}\:\frac{3}{2}, \mathcal{J}\pm\frac{1}{2}\:\frac{3}{2}}^{\mathcal{J}} \\
	S_{\mathcal{J}\pm\frac{1}{2}\:\frac{1}{2}, \mathcal{J}\mp\frac{3}{2}\:\frac{3}{2}}^{\mathcal{J}} &
	S_{\mathcal{J}\pm\frac{1}{2}\:\frac{1}{2}, \mathcal{J}\pm\frac{1}{2}\:\frac{1}{2}}^{\mathcal{J}} &
	S_{\mathcal{J}\pm\frac{1}{2}\:\frac{1}{2}, \mathcal{J}\pm\frac{1}{2}\:\frac{3}{2}}^{\mathcal{J}} \\
	S_{\mathcal{J}\pm\frac{1}{2}\:\frac{3}{2}, \mathcal{J}\mp\frac{3}{2}\:\frac{3}{2}}^{\mathcal{J}} &
	S_{\mathcal{J}\pm\frac{1}{2}\:\frac{3}{2}, \mathcal{J}\pm\frac{1}{2}\:\frac{1}{2}}^{\mathcal{J}} &
	S_{\mathcal{J}\pm\frac{1}{2}\:\frac{3}{2}, \mathcal{J}\pm\frac{1}{2}\:\frac{3}{2}}^{\mathcal{J}} \\
	\end{pmatrix} \:.
\end{equation}
where the upper and lower sign correspond to parities $\Par =
(-1)^{\mathcal{J}\pm \frac{1}{2}}$. We define
\begin{equation}
	S^{\mathcal{J}} = U^Te^{2i\delta}U,
\end{equation}
where $\delta$ represents three phase shifts and where $U$ are the
eigenvectors of $S$. The three mixing angles are derived using a
generalization~\cite{Seyler:1969sii} of the Blatt-Biedenharn method
\cite{Blatt:1952zz} for \NN{} phase-shift parametrization,
\begin{equation}
	U = uwx \:,
\end{equation}
where $u$, $w$, and $x$ are rotation matrices in the {$yz$-}, {$xz$-,} and
$xy$-planes, respectively, according to the Madison
convention~\cite{osti_4726823} for the scattering plane:
\begin{align}
	\begin{split}
	u &=
	\begin{pmatrix}
	1 & 0 & 0 \\
	0 & \cos(\epsilon) & \sin(\epsilon) \\
	0 & -\sin(\epsilon) & \cos(\epsilon)
	\end{pmatrix} \:, \\
	w &=
	\begin{pmatrix}
	\cos(\xi) & 0 & \sin(\xi)) \\
	0 & 1 & 0 \\
	-\sin(\xi) & 0 & \cos(\xi)
	\end{pmatrix} \:, \\
	x &=
	\begin{pmatrix}
	\cos(\eta) & \sin(\eta) & 0 \\
	-\sin(\eta) & \cos(\eta) & 0 \\
	0 & 0 & 1
	\end{pmatrix} \:.
	\end{split}
\end{align}
Uniquely identifying the phase shifts and mixing angles requires a
convention for the ordering of eigenvectors. Below the deuteron
breakup threshold we order the eigenvectors (which can be chosen to be
real) to have a dominant and positive diagonal
\cite{Huber:1995zza}. Above the threshold we will start getting
imaginary components and it becomes necessary to use, for example, the
continuity of eigenvectors to arrange $U$ correctly to identify phase
shifts \cite{Huber1995}.

\bibliography{bibliography}

\begin{thebibliography}{72}%
\makeatletter
\providecommand \@ifxundefined [1]{%
 \@ifx{#1\undefined}
}%
\providecommand \@ifnum [1]{%
 \ifnum #1\expandafter \@firstoftwo
 \else \expandafter \@secondoftwo
 \fi
}%
\providecommand \@ifx [1]{%
 \ifx #1\expandafter \@firstoftwo
 \else \expandafter \@secondoftwo
 \fi
}%
\providecommand \natexlab [1]{#1}%
\providecommand \enquote  [1]{``#1''}%
\providecommand \bibnamefont  [1]{#1}%
\providecommand \bibfnamefont [1]{#1}%
\providecommand \citenamefont [1]{#1}%
\providecommand \href@noop [0]{\@secondoftwo}%
\providecommand \href [0]{\begingroup \@sanitize@url \@href}%
\providecommand \@href[1]{\@@startlink{#1}\@@href}%
\providecommand \@@href[1]{\endgroup#1\@@endlink}%
\providecommand \@sanitize@url [0]{\catcode `\\12\catcode `\$12\catcode
  `\&12\catcode `\#12\catcode `\^12\catcode `\_12\catcode `\%12\relax}%
\providecommand \@@startlink[1]{}%
\providecommand \@@endlink[0]{}%
\providecommand \url  [0]{\begingroup\@sanitize@url \@url }%
\providecommand \@url [1]{\endgroup\@href {#1}{\urlprefix }}%
\providecommand \urlprefix  [0]{URL }%
\providecommand \Eprint [0]{\href }%
\providecommand \doibase [0]{https://doi.org/}%
\providecommand \selectlanguage [0]{\@gobble}%
\providecommand \bibinfo  [0]{\@secondoftwo}%
\providecommand \bibfield  [0]{\@secondoftwo}%
\providecommand \translation [1]{[#1]}%
\providecommand \BibitemOpen [0]{}%
\providecommand \bibitemStop [0]{}%
\providecommand \bibitemNoStop [0]{.\EOS\space}%
\providecommand \EOS [0]{\spacefactor3000\relax}%
\providecommand \BibitemShut  [1]{\csname bibitem#1\endcsname}%
\let\auto@bib@innerbib\@empty
\bibitem [{\citenamefont {Gl{\"o}ckle}(1983{\natexlab{a}})}]{Glockle:aYZcaRsU}%
  \BibitemOpen
  \bibfield  {author} {\bibinfo {author} {\bibfnamefont {W.}~\bibnamefont
  {Gl{\"o}ckle}},\ }\href@noop {} {\emph {\bibinfo {title} {{The quantum
  mechanical few-body problem}}}}\ (\bibinfo  {publisher} {Springer-Verlag},\
  \bibinfo {year} {1983})\BibitemShut {NoStop}%
\bibitem [{\citenamefont {Gl{\"o}ckle}\ \emph {et~al.}(1996)\citenamefont
  {Gl{\"o}ckle}, \citenamefont {Witala}, \citenamefont {Huber}, \citenamefont
  {Kamada},\ and\ \citenamefont {Golak}}]{Gloeckle:1995jg}%
  \BibitemOpen
  \bibfield  {author} {\bibinfo {author} {\bibfnamefont {W.}~\bibnamefont
  {Gl{\"o}ckle}}, \bibinfo {author} {\bibfnamefont {H.}~\bibnamefont {Witala}},
  \bibinfo {author} {\bibfnamefont {D.}~\bibnamefont {Huber}}, \bibinfo
  {author} {\bibfnamefont {H.}~\bibnamefont {Kamada}},\ and\ \bibinfo {author}
  {\bibfnamefont {J.}~\bibnamefont {Golak}},\ }\bibfield  {title} {\bibinfo
  {title} {{The Three nucleon continuum: Achievements, challenges and
  applications}},\ }\href {https://doi.org/10.1016/0370-1573(95)00085-2}
  {\bibfield  {journal} {\bibinfo  {journal} {Phys. Rept.}\ }\textbf {\bibinfo
  {volume} {274}},\ \bibinfo {pages} {107} (\bibinfo {year}
  {1996})}\BibitemShut {NoStop}%
\bibitem [{\citenamefont {Carlsson}\ \emph {et~al.}(2016)\citenamefont
  {Carlsson}, \citenamefont {Ekstr\"om}, \citenamefont {Forss\'en},
  \citenamefont {Str\"omberg}, \citenamefont {Jansen}, \citenamefont {Lilja},
  \citenamefont {Lindby}, \citenamefont {Mattsson},\ and\ \citenamefont
  {Wendt}}]{Carlsson:2015vda}%
  \BibitemOpen
  \bibfield  {author} {\bibinfo {author} {\bibfnamefont {B.~D.}\ \bibnamefont
  {Carlsson}}, \bibinfo {author} {\bibfnamefont {A.}~\bibnamefont {Ekstr\"om}},
  \bibinfo {author} {\bibfnamefont {C.}~\bibnamefont {Forss\'en}}, \bibinfo
  {author} {\bibfnamefont {D.~F.}\ \bibnamefont {Str\"omberg}}, \bibinfo
  {author} {\bibfnamefont {G.~R.}\ \bibnamefont {Jansen}}, \bibinfo {author}
  {\bibfnamefont {O.}~\bibnamefont {Lilja}}, \bibinfo {author} {\bibfnamefont
  {M.}~\bibnamefont {Lindby}}, \bibinfo {author} {\bibfnamefont {B.~A.}\
  \bibnamefont {Mattsson}},\ and\ \bibinfo {author} {\bibfnamefont {K.~A.}\
  \bibnamefont {Wendt}},\ }\bibfield  {title} {\bibinfo {title} {{Uncertainty
  analysis and order-by-order optimization of chiral nuclear interactions}},\
  }\href {https://doi.org/10.1103/PhysRevX.6.011019} {\bibfield  {journal}
  {\bibinfo  {journal} {Phys. Rev. X}\ }\textbf {\bibinfo {volume} {6}},\
  \bibinfo {pages} {011019} (\bibinfo {year} {2016})},\ \Eprint
  {https://arxiv.org/abs/1506.02466} {arXiv:1506.02466 [nucl-th]} \BibitemShut
  {NoStop}%
\bibitem [{\citenamefont {Reinert}\ \emph {et~al.}(2018)\citenamefont
  {Reinert}, \citenamefont {Krebs},\ and\ \citenamefont
  {Epelbaum}}]{Reinert:2017usi}%
  \BibitemOpen
  \bibfield  {author} {\bibinfo {author} {\bibfnamefont {P.}~\bibnamefont
  {Reinert}}, \bibinfo {author} {\bibfnamefont {H.}~\bibnamefont {Krebs}},\
  and\ \bibinfo {author} {\bibfnamefont {E.}~\bibnamefont {Epelbaum}},\
  }\bibfield  {title} {\bibinfo {title} {{Semilocal momentum-space regularized
  chiral two-nucleon potentials up to fifth order}},\ }\href
  {https://doi.org/10.1140/epja/i2018-12516-4} {\bibfield  {journal} {\bibinfo
  {journal} {Eur. Phys. J. A}\ }\textbf {\bibinfo {volume} {54}},\ \bibinfo
  {pages} {86} (\bibinfo {year} {2018})},\ \Eprint
  {https://arxiv.org/abs/1711.08821} {arXiv:1711.08821 [nucl-th]} \BibitemShut
  {NoStop}%
\bibitem [{\citenamefont {Piarulli}\ \emph {et~al.}(2015)\citenamefont
  {Piarulli}, \citenamefont {Girlanda}, \citenamefont {Schiavilla},
  \citenamefont {Navarro~P\'erez}, \citenamefont {Amaro},\ and\ \citenamefont
  {Ruiz~Arriola}}]{Piarulli:2014bda}%
  \BibitemOpen
  \bibfield  {author} {\bibinfo {author} {\bibfnamefont {M.}~\bibnamefont
  {Piarulli}}, \bibinfo {author} {\bibfnamefont {L.}~\bibnamefont {Girlanda}},
  \bibinfo {author} {\bibfnamefont {R.}~\bibnamefont {Schiavilla}}, \bibinfo
  {author} {\bibfnamefont {R.}~\bibnamefont {Navarro~P\'erez}}, \bibinfo
  {author} {\bibfnamefont {J.~E.}\ \bibnamefont {Amaro}},\ and\ \bibinfo
  {author} {\bibfnamefont {E.}~\bibnamefont {Ruiz~Arriola}},\ }\bibfield
  {title} {\bibinfo {title} {{Minimally nonlocal nucleon-nucleon potentials
  with chiral two-pion exchange including $\Delta$ resonances}},\ }\href
  {https://doi.org/10.1103/PhysRevC.91.024003} {\bibfield  {journal} {\bibinfo
  {journal} {Phys. Rev. C}\ }\textbf {\bibinfo {volume} {91}},\ \bibinfo
  {pages} {024003} (\bibinfo {year} {2015})},\ \Eprint
  {https://arxiv.org/abs/1412.6446} {arXiv:1412.6446 [nucl-th]} \BibitemShut
  {NoStop}%
\bibitem [{\citenamefont {Entem}\ \emph {et~al.}(2017)\citenamefont {Entem},
  \citenamefont {Machleidt},\ and\ \citenamefont {Nosyk}}]{Entem:2017gor}%
  \BibitemOpen
  \bibfield  {author} {\bibinfo {author} {\bibfnamefont {D.~R.}\ \bibnamefont
  {Entem}}, \bibinfo {author} {\bibfnamefont {R.}~\bibnamefont {Machleidt}},\
  and\ \bibinfo {author} {\bibfnamefont {Y.}~\bibnamefont {Nosyk}},\ }\bibfield
   {title} {\bibinfo {title} {{High-quality two-nucleon potentials up to fifth
  order of the chiral expansion}},\ }\href
  {https://doi.org/10.1103/PhysRevC.96.024004} {\bibfield  {journal} {\bibinfo
  {journal} {Phys. Rev. C}\ }\textbf {\bibinfo {volume} {96}},\ \bibinfo
  {pages} {024004} (\bibinfo {year} {2017})},\ \Eprint
  {https://arxiv.org/abs/1703.05454} {arXiv:1703.05454 [nucl-th]} \BibitemShut
  {NoStop}%
\bibitem [{\citenamefont {Faddeev}(1960)}]{Faddeev:1960su}%
  \BibitemOpen
  \bibfield  {author} {\bibinfo {author} {\bibfnamefont {L.~D.}\ \bibnamefont
  {Faddeev}},\ }\bibfield  {title} {\bibinfo {title} {{Scattering theory for a
  three particle system}},\ }\href@noop {} {\bibfield  {journal} {\bibinfo
  {journal} {Zh. Eksp. Teor. Fiz.}\ }\textbf {\bibinfo {volume} {39}},\
  \bibinfo {pages} {1459} (\bibinfo {year} {1960})}\BibitemShut {NoStop}%
\bibitem [{\citenamefont {Weinberg}(1964)}]{Weinberg:1964zza}%
  \BibitemOpen
  \bibfield  {author} {\bibinfo {author} {\bibfnamefont {S.}~\bibnamefont
  {Weinberg}},\ }\bibfield  {title} {\bibinfo {title} {{Systematic Solution of
  Multiparticle Scattering Problems}},\ }\href
  {https://doi.org/10.1103/PhysRev.133.B232} {\bibfield  {journal} {\bibinfo
  {journal} {Phys. Rev.}\ }\textbf {\bibinfo {volume} {133}},\ \bibinfo {pages}
  {B232} (\bibinfo {year} {1964})}\BibitemShut {NoStop}%
\bibitem [{\citenamefont {Rosenberg}(1965)}]{Rosenberg:1965zz}%
  \BibitemOpen
  \bibfield  {author} {\bibinfo {author} {\bibfnamefont {L.}~\bibnamefont
  {Rosenberg}},\ }\bibfield  {title} {\bibinfo {title} {{Generalized Faddeev
  Integral Equations for Multiparticle Scattering Amplitudes}},\ }\href
  {https://doi.org/10.1103/PhysRev.140.B217} {\bibfield  {journal} {\bibinfo
  {journal} {Phys. Rev.}\ }\textbf {\bibinfo {volume} {140}},\ \bibinfo {pages}
  {B217} (\bibinfo {year} {1965})}\BibitemShut {NoStop}%
\bibitem [{\citenamefont {Yakubovsky}(1967)}]{Yakubovsky:1966ue}%
  \BibitemOpen
  \bibfield  {author} {\bibinfo {author} {\bibfnamefont {O.~A.}\ \bibnamefont
  {Yakubovsky}},\ }\bibfield  {title} {\bibinfo {title} {{On the Integral
  equations in the theory o N particle scattering}},\ }\href@noop {} {\bibfield
   {journal} {\bibinfo  {journal} {Sov. J. Nucl. Phys.}\ }\textbf {\bibinfo
  {volume} {5}},\ \bibinfo {pages} {937} (\bibinfo {year} {1967})}\BibitemShut
  {NoStop}%
\bibitem [{\citenamefont {Alt}\ \emph {et~al.}(1967)\citenamefont {Alt},
  \citenamefont {Grassberger},\ and\ \citenamefont {Sandhas}}]{Alt:1967fx}%
  \BibitemOpen
  \bibfield  {author} {\bibinfo {author} {\bibfnamefont {E.~O.}\ \bibnamefont
  {Alt}}, \bibinfo {author} {\bibfnamefont {P.}~\bibnamefont {Grassberger}},\
  and\ \bibinfo {author} {\bibfnamefont {W.}~\bibnamefont {Sandhas}},\
  }\bibfield  {title} {\bibinfo {title} {{Reduction of the three - particle
  collision problem to multichannel two - particle Lippmann-Schwinger
  equations}},\ }\href {https://doi.org/10.1016/0550-3213(67)90016-8}
  {\bibfield  {journal} {\bibinfo  {journal} {Nucl. Phys. B}\ }\textbf
  {\bibinfo {volume} {2}},\ \bibinfo {pages} {167} (\bibinfo {year}
  {1967})}\BibitemShut {NoStop}%
\bibitem [{\citenamefont {Grassberger}\ and\ \citenamefont
  {Sandhas}(1967)}]{Grassberger:1967zz}%
  \BibitemOpen
  \bibfield  {author} {\bibinfo {author} {\bibfnamefont {P.}~\bibnamefont
  {Grassberger}}\ and\ \bibinfo {author} {\bibfnamefont {W.}~\bibnamefont
  {Sandhas}},\ }\bibfield  {title} {\bibinfo {title} {{Systematical treatment
  of the non-relativistic n-particle scattering problem}},\ }\href
  {https://doi.org/10.1016/0550-3213(67)90017-X} {\bibfield  {journal}
  {\bibinfo  {journal} {Nucl. Phys. B}\ }\textbf {\bibinfo {volume} {2}},\
  \bibinfo {pages} {181} (\bibinfo {year} {1967})}\BibitemShut {NoStop}%
\bibitem [{\citenamefont {Witala}\ \emph {et~al.}(1988)\citenamefont {Witala},
  \citenamefont {Cornelius},\ and\ \citenamefont
  {Gl{\"o}ckle}}]{Witala:1988ty}%
  \BibitemOpen
  \bibfield  {author} {\bibinfo {author} {\bibfnamefont {H.}~\bibnamefont
  {Witala}}, \bibinfo {author} {\bibfnamefont {T.}~\bibnamefont {Cornelius}},\
  and\ \bibinfo {author} {\bibfnamefont {W.}~\bibnamefont {Gl{\"o}ckle}},\
  }\bibfield  {title} {\bibinfo {title} {{Elastic scattering and break-up
  processes in then-d system}},\ }\href {https://doi.org/10.1007/BF01086331}
  {\bibfield  {journal} {\bibinfo  {journal} {Few Body Systems}\ }\textbf
  {\bibinfo {volume} {3}},\ \bibinfo {pages} {123} (\bibinfo {year}
  {1988})}\BibitemShut {NoStop}%
\bibitem [{\citenamefont {Rubtsova}\ \emph {et~al.}(2015)\citenamefont
  {Rubtsova}, \citenamefont {Kukulin},\ and\ \citenamefont
  {Pomerantsev}}]{Rubtsova:2015owa}%
  \BibitemOpen
  \bibfield  {author} {\bibinfo {author} {\bibfnamefont {O.~A.}\ \bibnamefont
  {Rubtsova}}, \bibinfo {author} {\bibfnamefont {V.~I.}\ \bibnamefont
  {Kukulin}},\ and\ \bibinfo {author} {\bibfnamefont {V.~N.}\ \bibnamefont
  {Pomerantsev}},\ }\bibfield  {title} {\bibinfo {title} {{Wave-packet
  continuum discretization for quantum scattering}},\ }\href
  {https://doi.org/10.1016/j.aop.2015.04.028} {\bibfield  {journal} {\bibinfo
  {journal} {Annals Phys.}\ }\textbf {\bibinfo {volume} {360}},\ \bibinfo
  {pages} {613} (\bibinfo {year} {2015})},\ \Eprint
  {https://arxiv.org/abs/1501.02531} {arXiv:1501.02531 [nucl-th]} \BibitemShut
  {NoStop}%
\bibitem [{\citenamefont {Ekstr\"om}\ \emph {et~al.}(2013)\citenamefont
  {Ekstr\"om} \emph {et~al.}}]{Ekstrom:2013kea}%
  \BibitemOpen
  \bibfield  {author} {\bibinfo {author} {\bibfnamefont {A.}~\bibnamefont
  {Ekstr\"om}} \emph {et~al.},\ }\bibfield  {title} {\bibinfo {title}
  {{Optimized Chiral Nucleon-Nucleon Interaction at Next-to-Next-to-Leading
  Order}},\ }\href {https://doi.org/10.1103/PhysRevLett.110.192502} {\bibfield
  {journal} {\bibinfo  {journal} {Phys. Rev. Lett.}\ }\textbf {\bibinfo
  {volume} {110}},\ \bibinfo {pages} {192502} (\bibinfo {year} {2013})},\
  \Eprint {https://arxiv.org/abs/1303.4674} {arXiv:1303.4674 [nucl-th]}
  \BibitemShut {NoStop}%
\bibitem [{\citenamefont {Carbonell}\ \emph {et~al.}(2014)\citenamefont
  {Carbonell}, \citenamefont {Deltuva}, \citenamefont {Fonseca},\ and\
  \citenamefont {Lazauskas}}]{Carbonell:2013ywa}%
  \BibitemOpen
  \bibfield  {author} {\bibinfo {author} {\bibfnamefont {J.}~\bibnamefont
  {Carbonell}}, \bibinfo {author} {\bibfnamefont {A.}~\bibnamefont {Deltuva}},
  \bibinfo {author} {\bibfnamefont {A.~C.}\ \bibnamefont {Fonseca}},\ and\
  \bibinfo {author} {\bibfnamefont {R.}~\bibnamefont {Lazauskas}},\ }\bibfield
  {title} {\bibinfo {title} {{Bound state techniques to solve the multiparticle
  scattering problem}},\ }\href {https://doi.org/10.1016/j.ppnp.2013.10.003}
  {\bibfield  {journal} {\bibinfo  {journal} {Prog. Part. Nucl. Phys.}\
  }\textbf {\bibinfo {volume} {74}},\ \bibinfo {pages} {55} (\bibinfo {year}
  {2014})},\ \Eprint {https://arxiv.org/abs/1310.6631} {arXiv:1310.6631
  [nucl-th]} \BibitemShut {NoStop}%
\bibitem [{\citenamefont {Stoks}\ \emph {et~al.}(1994)\citenamefont {Stoks},
  \citenamefont {Klomp}, \citenamefont {Terheggen},\ and\ \citenamefont
  {de~Swart}}]{Stoks:1994wp}%
  \BibitemOpen
  \bibfield  {author} {\bibinfo {author} {\bibfnamefont {V.~G.~J.}\
  \bibnamefont {Stoks}}, \bibinfo {author} {\bibfnamefont {R.~A.~M.}\
  \bibnamefont {Klomp}}, \bibinfo {author} {\bibfnamefont {C.~P.~F.}\
  \bibnamefont {Terheggen}},\ and\ \bibinfo {author} {\bibfnamefont {J.~J.}\
  \bibnamefont {de~Swart}},\ }\bibfield  {title} {\bibinfo {title}
  {{Construction of high quality N N potential models}},\ }\href
  {https://doi.org/10.1103/PhysRevC.49.2950} {\bibfield  {journal} {\bibinfo
  {journal} {Phys. Rev. C}\ }\textbf {\bibinfo {volume} {49}},\ \bibinfo
  {pages} {2950} (\bibinfo {year} {1994})},\ \Eprint
  {https://arxiv.org/abs/nucl-th/9406039} {arXiv:nucl-th/9406039} \BibitemShut
  {NoStop}%
\bibitem [{\citenamefont {Entem}\ and\ \citenamefont
  {Machleidt}(2003)}]{Entem:2003ft}%
  \BibitemOpen
  \bibfield  {author} {\bibinfo {author} {\bibfnamefont {D.~R.}\ \bibnamefont
  {Entem}}\ and\ \bibinfo {author} {\bibfnamefont {R.}~\bibnamefont
  {Machleidt}},\ }\bibfield  {title} {\bibinfo {title} {{Accurate charge
  dependent nucleon nucleon potential at fourth order of chiral perturbation
  theory}},\ }\href {https://doi.org/10.1103/PhysRevC.68.041001} {\bibfield
  {journal} {\bibinfo  {journal} {Phys. Rev. C}\ }\textbf {\bibinfo {volume}
  {68}},\ \bibinfo {pages} {041001} (\bibinfo {year} {2003})},\ \Eprint
  {https://arxiv.org/abs/nucl-th/0304018} {arXiv:nucl-th/0304018} \BibitemShut
  {NoStop}%
\bibitem [{\citenamefont {Dytrych}\ \emph {et~al.}(2020)\citenamefont
  {Dytrych}, \citenamefont {Launey}, \citenamefont {Draayer}, \citenamefont
  {Rowe}, \citenamefont {Wood}, \citenamefont {Rosensteel}, \citenamefont
  {Bahri}, \citenamefont {Langr},\ and\ \citenamefont
  {Baker}}]{Dytrych:2018vkl}%
  \BibitemOpen
  \bibfield  {author} {\bibinfo {author} {\bibfnamefont {T.}~\bibnamefont
  {Dytrych}}, \bibinfo {author} {\bibfnamefont {K.~D.}\ \bibnamefont {Launey}},
  \bibinfo {author} {\bibfnamefont {J.~P.}\ \bibnamefont {Draayer}}, \bibinfo
  {author} {\bibfnamefont {D.}~\bibnamefont {Rowe}}, \bibinfo {author}
  {\bibfnamefont {J.}~\bibnamefont {Wood}}, \bibinfo {author} {\bibfnamefont
  {G.}~\bibnamefont {Rosensteel}}, \bibinfo {author} {\bibfnamefont
  {C.}~\bibnamefont {Bahri}}, \bibinfo {author} {\bibfnamefont
  {D.}~\bibnamefont {Langr}},\ and\ \bibinfo {author} {\bibfnamefont {R.~B.}\
  \bibnamefont {Baker}},\ }\bibfield  {title} {\bibinfo {title} {{Physics of
  nuclei: Key role of an emergent symmetry}},\ }\href
  {https://doi.org/10.1103/PhysRevLett.124.042501} {\bibfield  {journal}
  {\bibinfo  {journal} {Phys. Rev. Lett.}\ }\textbf {\bibinfo {volume} {124}},\
  \bibinfo {pages} {042501} (\bibinfo {year} {2020})},\ \Eprint
  {https://arxiv.org/abs/1810.05757} {arXiv:1810.05757 [nucl-th]} \BibitemShut
  {NoStop}%
\bibitem [{\citenamefont {Burrows}\ \emph {et~al.}(2019)\citenamefont
  {Burrows}, \citenamefont {Elster}, \citenamefont {Weppner}, \citenamefont
  {Launey}, \citenamefont {Maris}, \citenamefont {Nogga},\ and\ \citenamefont
  {Popa}}]{Burrows:2018ggt}%
  \BibitemOpen
  \bibfield  {author} {\bibinfo {author} {\bibfnamefont {M.}~\bibnamefont
  {Burrows}}, \bibinfo {author} {\bibfnamefont {C.}~\bibnamefont {Elster}},
  \bibinfo {author} {\bibfnamefont {S.~P.}\ \bibnamefont {Weppner}}, \bibinfo
  {author} {\bibfnamefont {K.~D.}\ \bibnamefont {Launey}}, \bibinfo {author}
  {\bibfnamefont {P.}~\bibnamefont {Maris}}, \bibinfo {author} {\bibfnamefont
  {A.}~\bibnamefont {Nogga}},\ and\ \bibinfo {author} {\bibfnamefont
  {G.}~\bibnamefont {Popa}},\ }\bibfield  {title} {\bibinfo {title} {{Ab initio
  folding potentials for nucleon-nucleus scattering based on no-core
  shell-model one-body densities}},\ }\href
  {https://doi.org/10.1103/PhysRevC.99.044603} {\bibfield  {journal} {\bibinfo
  {journal} {Phys. Rev. C}\ }\textbf {\bibinfo {volume} {99}},\ \bibinfo
  {pages} {044603} (\bibinfo {year} {2019})},\ \Eprint
  {https://arxiv.org/abs/1810.06442} {arXiv:1810.06442 [nucl-th]} \BibitemShut
  {NoStop}%
\bibitem [{\citenamefont {Rotureau}\ \emph {et~al.}(2018)\citenamefont
  {Rotureau}, \citenamefont {Danielewicz}, \citenamefont {Hagen}, \citenamefont
  {Jansen},\ and\ \citenamefont {Nunes}}]{Rotureau:2018pxk}%
  \BibitemOpen
  \bibfield  {author} {\bibinfo {author} {\bibfnamefont {J.}~\bibnamefont
  {Rotureau}}, \bibinfo {author} {\bibfnamefont {P.}~\bibnamefont
  {Danielewicz}}, \bibinfo {author} {\bibfnamefont {G.}~\bibnamefont {Hagen}},
  \bibinfo {author} {\bibfnamefont {G.~R.}\ \bibnamefont {Jansen}},\ and\
  \bibinfo {author} {\bibfnamefont {F.~M.}\ \bibnamefont {Nunes}},\ }\bibfield
  {title} {\bibinfo {title} {{Microscopic optical potentials for calcium
  isotopes}},\ }\href {https://doi.org/10.1103/PhysRevC.98.044625} {\bibfield
  {journal} {\bibinfo  {journal} {Phys. Rev. C}\ }\textbf {\bibinfo {volume}
  {98}},\ \bibinfo {pages} {044625} (\bibinfo {year} {2018})},\ \Eprint
  {https://arxiv.org/abs/1808.04535} {arXiv:1808.04535 [nucl-th]} \BibitemShut
  {NoStop}%
\bibitem [{\citenamefont {Witala}\ \emph {et~al.}(2001)\citenamefont {Witala},
  \citenamefont {Gl{\"o}ckle}, \citenamefont {Golak}, \citenamefont {Kamada},
  \citenamefont {Kuros-Zolnierczuk}, \citenamefont {Nogga},\ and\ \citenamefont
  {Skibinski}}]{Witala:2000am}%
  \BibitemOpen
  \bibfield  {author} {\bibinfo {author} {\bibfnamefont {H.}~\bibnamefont
  {Witala}}, \bibinfo {author} {\bibfnamefont {W.}~\bibnamefont {Gl{\"o}ckle}},
  \bibinfo {author} {\bibfnamefont {J.}~\bibnamefont {Golak}}, \bibinfo
  {author} {\bibfnamefont {H.}~\bibnamefont {Kamada}}, \bibinfo {author}
  {\bibfnamefont {J.}~\bibnamefont {Kuros-Zolnierczuk}}, \bibinfo {author}
  {\bibfnamefont {A.}~\bibnamefont {Nogga}},\ and\ \bibinfo {author}
  {\bibfnamefont {R.}~\bibnamefont {Skibinski}},\ }\bibfield  {title} {\bibinfo
  {title} {{Nd elastic scattering as a tool to probe properties of three
  nucleon forces}},\ }\href {https://doi.org/10.1103/PhysRevC.63.024007}
  {\bibfield  {journal} {\bibinfo  {journal} {Phys. Rev. C}\ }\textbf {\bibinfo
  {volume} {63}},\ \bibinfo {pages} {024007} (\bibinfo {year} {2001})},\
  \Eprint {https://arxiv.org/abs/nucl-th/0010013} {arXiv:nucl-th/0010013}
  \BibitemShut {NoStop}%
\bibitem [{\citenamefont {Pieper}\ and\ \citenamefont
  {Wiringa}(2001)}]{Pieper:2001mp}%
  \BibitemOpen
  \bibfield  {author} {\bibinfo {author} {\bibfnamefont {S.~C.}\ \bibnamefont
  {Pieper}}\ and\ \bibinfo {author} {\bibfnamefont {R.~B.}\ \bibnamefont
  {Wiringa}},\ }\bibfield  {title} {\bibinfo {title} {{Quantum Monte Carlo
  calculations of light nuclei}},\ }\href
  {https://doi.org/10.1146/annurev.nucl.51.101701.132506} {\bibfield  {journal}
  {\bibinfo  {journal} {Ann. Rev. Nucl. Part. Sci.}\ }\textbf {\bibinfo
  {volume} {51}},\ \bibinfo {pages} {53} (\bibinfo {year} {2001})},\ \Eprint
  {https://arxiv.org/abs/nucl-th/0103005} {arXiv:nucl-th/0103005} \BibitemShut
  {NoStop}%
\bibitem [{\citenamefont {Epelbaum}\ \emph {et~al.}(2002)\citenamefont
  {Epelbaum}, \citenamefont {Nogga}, \citenamefont {Gl{\"o}ckle}, \citenamefont
  {Kamada}, \citenamefont {Meissner},\ and\ \citenamefont
  {Witala}}]{Epelbaum:2002vt}%
  \BibitemOpen
  \bibfield  {author} {\bibinfo {author} {\bibfnamefont {E.}~\bibnamefont
  {Epelbaum}}, \bibinfo {author} {\bibfnamefont {A.}~\bibnamefont {Nogga}},
  \bibinfo {author} {\bibfnamefont {W.}~\bibnamefont {Gl{\"o}ckle}}, \bibinfo
  {author} {\bibfnamefont {H.}~\bibnamefont {Kamada}}, \bibinfo {author}
  {\bibfnamefont {U.~G.}\ \bibnamefont {Meissner}},\ and\ \bibinfo {author}
  {\bibfnamefont {H.}~\bibnamefont {Witala}},\ }\bibfield  {title} {\bibinfo
  {title} {{Three nucleon forces from chiral effective field theory}},\ }\href
  {https://doi.org/10.1103/PhysRevC.66.064001} {\bibfield  {journal} {\bibinfo
  {journal} {Phys. Rev. C}\ }\textbf {\bibinfo {volume} {66}},\ \bibinfo
  {pages} {064001} (\bibinfo {year} {2002})},\ \Eprint
  {https://arxiv.org/abs/nucl-th/0208023} {arXiv:nucl-th/0208023} \BibitemShut
  {NoStop}%
\bibitem [{\citenamefont {Navratil}\ \emph {et~al.}(2007)\citenamefont
  {Navratil}, \citenamefont {Gueorguiev}, \citenamefont {Vary}, \citenamefont
  {Ormand},\ and\ \citenamefont {Nogga}}]{Navratil:2007we}%
  \BibitemOpen
  \bibfield  {author} {\bibinfo {author} {\bibfnamefont {P.}~\bibnamefont
  {Navratil}}, \bibinfo {author} {\bibfnamefont {V.~G.}\ \bibnamefont
  {Gueorguiev}}, \bibinfo {author} {\bibfnamefont {J.~P.}\ \bibnamefont
  {Vary}}, \bibinfo {author} {\bibfnamefont {W.~E.}\ \bibnamefont {Ormand}},\
  and\ \bibinfo {author} {\bibfnamefont {A.}~\bibnamefont {Nogga}},\ }\bibfield
   {title} {\bibinfo {title} {{Structure of A=10-13 nuclei with two plus
  three-nucleon interactions from chiral effective field theory}},\ }\href
  {https://doi.org/10.1103/PhysRevLett.99.042501} {\bibfield  {journal}
  {\bibinfo  {journal} {Phys. Rev. Lett.}\ }\textbf {\bibinfo {volume} {99}},\
  \bibinfo {pages} {042501} (\bibinfo {year} {2007})},\ \Eprint
  {https://arxiv.org/abs/nucl-th/0701038} {arXiv:nucl-th/0701038} \BibitemShut
  {NoStop}%
\bibitem [{\citenamefont {Otsuka}\ \emph {et~al.}(2010)\citenamefont {Otsuka},
  \citenamefont {Suzuki}, \citenamefont {Holt}, \citenamefont {Schwenk},\ and\
  \citenamefont {Akaishi}}]{Otsuka:2009cs}%
  \BibitemOpen
  \bibfield  {author} {\bibinfo {author} {\bibfnamefont {T.}~\bibnamefont
  {Otsuka}}, \bibinfo {author} {\bibfnamefont {T.}~\bibnamefont {Suzuki}},
  \bibinfo {author} {\bibfnamefont {J.~D.}\ \bibnamefont {Holt}}, \bibinfo
  {author} {\bibfnamefont {A.}~\bibnamefont {Schwenk}},\ and\ \bibinfo {author}
  {\bibfnamefont {Y.}~\bibnamefont {Akaishi}},\ }\bibfield  {title} {\bibinfo
  {title} {{Three-body forces and the limit of oxygen isotopes}},\ }\href
  {https://doi.org/10.1103/PhysRevLett.105.032501} {\bibfield  {journal}
  {\bibinfo  {journal} {Phys. Rev. Lett.}\ }\textbf {\bibinfo {volume} {105}},\
  \bibinfo {pages} {032501} (\bibinfo {year} {2010})},\ \Eprint
  {https://arxiv.org/abs/0908.2607} {arXiv:0908.2607 [nucl-th]} \BibitemShut
  {NoStop}%
\bibitem [{\citenamefont {Kalantar-Nayestanaki}\ \emph
  {et~al.}(2012)\citenamefont {Kalantar-Nayestanaki}, \citenamefont {Epelbaum},
  \citenamefont {Messchendorp},\ and\ \citenamefont
  {Nogga}}]{Kalantar-Nayestanaki:2011rzs}%
  \BibitemOpen
  \bibfield  {author} {\bibinfo {author} {\bibfnamefont {N.}~\bibnamefont
  {Kalantar-Nayestanaki}}, \bibinfo {author} {\bibfnamefont {E.}~\bibnamefont
  {Epelbaum}}, \bibinfo {author} {\bibfnamefont {J.~G.}\ \bibnamefont
  {Messchendorp}},\ and\ \bibinfo {author} {\bibfnamefont {A.}~\bibnamefont
  {Nogga}},\ }\bibfield  {title} {\bibinfo {title} {{Signatures of
  three-nucleon interactions in few-nucleon systems}},\ }\href
  {https://doi.org/10.1088/0034-4885/75/1/016301} {\bibfield  {journal}
  {\bibinfo  {journal} {Rept. Prog. Phys.}\ }\textbf {\bibinfo {volume} {75}},\
  \bibinfo {pages} {016301} (\bibinfo {year} {2012})},\ \Eprint
  {https://arxiv.org/abs/1108.1227} {arXiv:1108.1227 [nucl-th]} \BibitemShut
  {NoStop}%
\bibitem [{\citenamefont {Calci}\ \emph {et~al.}(2016)\citenamefont {Calci},
  \citenamefont {Navr\'atil}, \citenamefont {Roth}, \citenamefont
  {Dohet-Eraly}, \citenamefont {Quaglioni},\ and\ \citenamefont
  {Hupin}}]{Calci:2016dfb}%
  \BibitemOpen
  \bibfield  {author} {\bibinfo {author} {\bibfnamefont {A.}~\bibnamefont
  {Calci}}, \bibinfo {author} {\bibfnamefont {P.}~\bibnamefont {Navr\'atil}},
  \bibinfo {author} {\bibfnamefont {R.}~\bibnamefont {Roth}}, \bibinfo {author}
  {\bibfnamefont {J.}~\bibnamefont {Dohet-Eraly}}, \bibinfo {author}
  {\bibfnamefont {S.}~\bibnamefont {Quaglioni}},\ and\ \bibinfo {author}
  {\bibfnamefont {G.}~\bibnamefont {Hupin}},\ }\bibfield  {title} {\bibinfo
  {title} {{Can Ab Initio Theory Explain the Phenomenon of Parity Inversion in
  $^{11}$Be?}},\ }\href {https://doi.org/10.1103/PhysRevLett.117.242501}
  {\bibfield  {journal} {\bibinfo  {journal} {Phys. Rev. Lett.}\ }\textbf
  {\bibinfo {volume} {117}},\ \bibinfo {pages} {242501} (\bibinfo {year}
  {2016})},\ \Eprint {https://arxiv.org/abs/1608.03318} {arXiv:1608.03318
  [nucl-th]} \BibitemShut {NoStop}%
\bibitem [{\citenamefont {Hebeler}(2021)}]{Hebeler:2020ocj}%
  \BibitemOpen
  \bibfield  {author} {\bibinfo {author} {\bibfnamefont {K.}~\bibnamefont
  {Hebeler}},\ }\bibfield  {title} {\bibinfo {title} {{Three-nucleon forces:
  Implementation and applications to atomic nuclei and dense matter}},\ }\href
  {https://doi.org/10.1016/j.physrep.2020.08.009} {\bibfield  {journal}
  {\bibinfo  {journal} {Phys. Rept.}\ }\textbf {\bibinfo {volume} {890}},\
  \bibinfo {pages} {1} (\bibinfo {year} {2021})},\ \Eprint
  {https://arxiv.org/abs/2002.09548} {arXiv:2002.09548 [nucl-th]} \BibitemShut
  {NoStop}%
\bibitem [{\citenamefont {Huber}\ and\ \citenamefont
  {Friar}(1998)}]{Huber:1998hu}%
  \BibitemOpen
  \bibfield  {author} {\bibinfo {author} {\bibfnamefont {D.}~\bibnamefont
  {Huber}}\ and\ \bibinfo {author} {\bibfnamefont {J.~L.}\ \bibnamefont
  {Friar}},\ }\bibfield  {title} {\bibinfo {title} {{The $A_y$ puzzle and the
  nuclear force}},\ }\href {https://doi.org/10.1103/PhysRevC.58.674} {\bibfield
   {journal} {\bibinfo  {journal} {Phys. Rev. C}\ }\textbf {\bibinfo {volume}
  {58}},\ \bibinfo {pages} {674} (\bibinfo {year} {1998})},\ \Eprint
  {https://arxiv.org/abs/nucl-th/9803038} {arXiv:nucl-th/9803038} \BibitemShut
  {NoStop}%
\bibitem [{\citenamefont {Witala}\ \emph {et~al.}(1998)\citenamefont {Witala},
  \citenamefont {Gl{\"o}ckle}, \citenamefont {Huber}, \citenamefont {Golak},\
  and\ \citenamefont {Kamada}}]{Witala:1998ey}%
  \BibitemOpen
  \bibfield  {author} {\bibinfo {author} {\bibfnamefont {H.}~\bibnamefont
  {Witala}}, \bibinfo {author} {\bibfnamefont {W.}~\bibnamefont {Gl{\"o}ckle}},
  \bibinfo {author} {\bibfnamefont {D.}~\bibnamefont {Huber}}, \bibinfo
  {author} {\bibfnamefont {J.}~\bibnamefont {Golak}},\ and\ \bibinfo {author}
  {\bibfnamefont {H.}~\bibnamefont {Kamada}},\ }\bibfield  {title} {\bibinfo
  {title} {{The Cross-section minima in elastic Nd scattering: A `Smoking gun'
  for three nucleon force effects}},\ }\href
  {https://doi.org/10.1103/PhysRevLett.81.1183} {\bibfield  {journal} {\bibinfo
   {journal} {Phys. Rev. Lett.}\ }\textbf {\bibinfo {volume} {81}},\ \bibinfo
  {pages} {1183} (\bibinfo {year} {1998})},\ \Eprint
  {https://arxiv.org/abs/nucl-th/9801018} {arXiv:nucl-th/9801018} \BibitemShut
  {NoStop}%
\bibitem [{\citenamefont {Epelbaum}\ \emph {et~al.}(2019)\citenamefont
  {Epelbaum} \emph {et~al.}}]{LENPIC:2018ewt}%
  \BibitemOpen
  \bibfield  {author} {\bibinfo {author} {\bibfnamefont {E.}~\bibnamefont
  {Epelbaum}} \emph {et~al.} (\bibinfo {collaboration} {LENPIC}),\ }\bibfield
  {title} {\bibinfo {title} {{Few- and many-nucleon systems with semilocal
  coordinate-space regularized chiral two- and three-body forces}},\ }\href
  {https://doi.org/10.1103/PhysRevC.99.024313} {\bibfield  {journal} {\bibinfo
  {journal} {Phys. Rev. C}\ }\textbf {\bibinfo {volume} {99}},\ \bibinfo
  {pages} {024313} (\bibinfo {year} {2019})},\ \Eprint
  {https://arxiv.org/abs/1807.02848} {arXiv:1807.02848 [nucl-th]} \BibitemShut
  {NoStop}%
\bibitem [{\citenamefont {Wita\l{}a}\ \emph {et~al.}(2016)\citenamefont
  {Wita\l{}a}, \citenamefont {Golak}, \citenamefont {Skibi\'nski},
  \citenamefont {Topolnicki}, \citenamefont {Epelbaum}, \citenamefont
  {Hebeler}, \citenamefont {Kamada}, \citenamefont {Krebs}, \citenamefont
  {Mei\ss{}ner},\ and\ \citenamefont {Nogga}}]{Witala:2016inj}%
  \BibitemOpen
  \bibfield  {author} {\bibinfo {author} {\bibfnamefont {H.}~\bibnamefont
  {Wita\l{}a}}, \bibinfo {author} {\bibfnamefont {J.}~\bibnamefont {Golak}},
  \bibinfo {author} {\bibfnamefont {R.}~\bibnamefont {Skibi\'nski}}, \bibinfo
  {author} {\bibfnamefont {K.}~\bibnamefont {Topolnicki}}, \bibinfo {author}
  {\bibfnamefont {E.}~\bibnamefont {Epelbaum}}, \bibinfo {author}
  {\bibfnamefont {K.}~\bibnamefont {Hebeler}}, \bibinfo {author} {\bibfnamefont
  {H.}~\bibnamefont {Kamada}}, \bibinfo {author} {\bibfnamefont
  {H.}~\bibnamefont {Krebs}}, \bibinfo {author} {\bibfnamefont {U.~G.}\
  \bibnamefont {Mei\ss{}ner}},\ and\ \bibinfo {author} {\bibfnamefont
  {A.}~\bibnamefont {Nogga}},\ }\bibfield  {title} {\bibinfo {title} {{Role of
  the total isospin 3/2 component in three-nucleon reactions}},\ }\href
  {https://doi.org/10.1007/s00601-016-1156-3} {\bibfield  {journal} {\bibinfo
  {journal} {Few Body Syst.}\ }\textbf {\bibinfo {volume} {57}},\ \bibinfo
  {pages} {1213} (\bibinfo {year} {2016})},\ \Eprint
  {https://arxiv.org/abs/1605.02011} {arXiv:1605.02011 [nucl-th]} \BibitemShut
  {NoStop}%
\bibitem [{\citenamefont {Pomerantsev}\ \emph {et~al.}(2009)\citenamefont
  {Pomerantsev}, \citenamefont {Kukulin},\ and\ \citenamefont
  {Rubtsova}}]{Pomerantsev:2008tw}%
  \BibitemOpen
  \bibfield  {author} {\bibinfo {author} {\bibfnamefont {V.~N.}\ \bibnamefont
  {Pomerantsev}}, \bibinfo {author} {\bibfnamefont {V.~I.}\ \bibnamefont
  {Kukulin}},\ and\ \bibinfo {author} {\bibfnamefont {O.~A.}\ \bibnamefont
  {Rubtsova}},\ }\bibfield  {title} {\bibinfo {title} {{Solving three-body
  scattering problem in the momentum lattice representation}},\ }\href
  {https://doi.org/10.1103/PhysRevC.79.034001} {\bibfield  {journal} {\bibinfo
  {journal} {Phys. Rev. C}\ }\textbf {\bibinfo {volume} {79}},\ \bibinfo
  {pages} {034001} (\bibinfo {year} {2009})},\ \Eprint
  {https://arxiv.org/abs/0812.0572} {arXiv:0812.0572 [nucl-th]} \BibitemShut
  {NoStop}%
\bibitem [{\citenamefont {Baker}(1975)}]{baker}%
  \BibitemOpen
  \bibfield  {author} {\bibinfo {author} {\bibfnamefont {G.~A.}\ \bibnamefont
  {Baker}},\ }\href@noop {} {\emph {\bibinfo {title} {Essentials of Pade
  approximants}}}\ (\bibinfo  {publisher} {Academic Press New York},\ \bibinfo
  {year} {1975})\BibitemShut {NoStop}%
\bibitem [{\citenamefont {Deltuva}\ \emph
  {et~al.}(2005{\natexlab{a}})\citenamefont {Deltuva}, \citenamefont
  {Fonseca},\ and\ \citenamefont {Sauer}}]{Deltuva:2005wx}%
  \BibitemOpen
  \bibfield  {author} {\bibinfo {author} {\bibfnamefont {A.}~\bibnamefont
  {Deltuva}}, \bibinfo {author} {\bibfnamefont {A.~C.}\ \bibnamefont
  {Fonseca}},\ and\ \bibinfo {author} {\bibfnamefont {P.~U.}\ \bibnamefont
  {Sauer}},\ }\bibfield  {title} {\bibinfo {title} {{Momentum-space treatment
  of Coulomb interaction in three-nucleon reactions with two protons}},\ }\href
  {https://doi.org/10.1103/PhysRevC.71.054005} {\bibfield  {journal} {\bibinfo
  {journal} {Phys. Rev. C}\ }\textbf {\bibinfo {volume} {71}},\ \bibinfo
  {pages} {054005} (\bibinfo {year} {2005}{\natexlab{a}})},\ \Eprint
  {https://arxiv.org/abs/nucl-th/0503012} {arXiv:nucl-th/0503012} \BibitemShut
  {NoStop}%
\bibitem [{\citenamefont {Deltuva}\ \emph
  {et~al.}(2005{\natexlab{b}})\citenamefont {Deltuva}, \citenamefont
  {Fonseca},\ and\ \citenamefont {Sauer}}]{Deltuva:2005cc}%
  \BibitemOpen
  \bibfield  {author} {\bibinfo {author} {\bibfnamefont {A.}~\bibnamefont
  {Deltuva}}, \bibinfo {author} {\bibfnamefont {A.~C.}\ \bibnamefont
  {Fonseca}},\ and\ \bibinfo {author} {\bibfnamefont {P.~U.}\ \bibnamefont
  {Sauer}},\ }\bibfield  {title} {\bibinfo {title} {{Momentum-space description
  of three-nucleon breakup reactions including the Coulomb interaction}},\
  }\href {https://doi.org/10.1103/PhysRevC.72.059903} {\bibfield  {journal}
  {\bibinfo  {journal} {Phys. Rev. C}\ }\textbf {\bibinfo {volume} {72}},\
  \bibinfo {pages} {054004} (\bibinfo {year} {2005}{\natexlab{b}})},\ \bibinfo
  {note} {[Erratum: Phys.Rev.C 72, 059903 (2005)]},\ \Eprint
  {https://arxiv.org/abs/nucl-th/0509034} {arXiv:nucl-th/0509034} \BibitemShut
  {NoStop}%
\bibitem [{\citenamefont {Ishikawa}\ \emph {et~al.}(2001)\citenamefont
  {Ishikawa}, \citenamefont {Tanifuji},\ and\ \citenamefont
  {Iseri}}]{Ishikawa:2000dq}%
  \BibitemOpen
  \bibfield  {author} {\bibinfo {author} {\bibfnamefont {S.}~\bibnamefont
  {Ishikawa}}, \bibinfo {author} {\bibfnamefont {M.}~\bibnamefont {Tanifuji}},\
  and\ \bibinfo {author} {\bibfnamefont {Y.}~\bibnamefont {Iseri}},\ }\bibfield
   {title} {\bibinfo {title} {{A Complete set of total cross-sections for
  imaginary parts of n d forward scattering amplitudes and three nucleon force
  effects}},\ }\href {https://doi.org/10.1103/PhysRevC.64.024001} {\bibfield
  {journal} {\bibinfo  {journal} {Phys. Rev. C}\ }\textbf {\bibinfo {volume}
  {64}},\ \bibinfo {pages} {024001} (\bibinfo {year} {2001})},\ \Eprint
  {https://arxiv.org/abs/nucl-th/0011030} {arXiv:nucl-th/0011030} \BibitemShut
  {NoStop}%
\bibitem [{\citenamefont {Otuka}\ \emph {et~al.}(2014)\citenamefont {Otuka}
  \emph {et~al.}}]{Otuka:2014wzu}%
  \BibitemOpen
  \bibfield  {author} {\bibinfo {author} {\bibfnamefont {N.}~\bibnamefont
  {Otuka}} \emph {et~al.},\ }\bibfield  {title} {\bibinfo {title} {{Towards a
  More Complete and Accurate Experimental Nuclear Reaction Data Library
  (EXFOR): International Collaboration Between Nuclear Reaction Data Centres
  (NRDC)}},\ }\href {https://doi.org/10.1016/j.nds.2014.07.065} {\bibfield
  {journal} {\bibinfo  {journal} {Nucl. Data Sheets}\ }\textbf {\bibinfo
  {volume} {120}},\ \bibinfo {pages} {272} (\bibinfo {year} {2014})},\ \Eprint
  {https://arxiv.org/abs/2002.07114} {arXiv:2002.07114 [nucl-ex]} \BibitemShut
  {NoStop}%
\bibitem [{\citenamefont {Howell}\ \emph {et~al.}(1987)\citenamefont {Howell},
  \citenamefont {Tornow}, \citenamefont {Murphy}, \citenamefont {Pf{\"u}tzner},
  \citenamefont {Roberts}, \citenamefont {Li}, \citenamefont {Felsher},
  \citenamefont {Walter}, \citenamefont {Slaus}, \citenamefont {Treado},\ and\
  \citenamefont {Koike}}]{Howell:1987uc}%
  \BibitemOpen
  \bibfield  {author} {\bibinfo {author} {\bibfnamefont {C.~R.}\ \bibnamefont
  {Howell}}, \bibinfo {author} {\bibfnamefont {W.}~\bibnamefont {Tornow}},
  \bibinfo {author} {\bibfnamefont {K.}~\bibnamefont {Murphy}}, \bibinfo
  {author} {\bibfnamefont {H.~G.}\ \bibnamefont {Pf{\"u}tzner}}, \bibinfo
  {author} {\bibfnamefont {M.~L.}\ \bibnamefont {Roberts}}, \bibinfo {author}
  {\bibfnamefont {A.}~\bibnamefont {Li}}, \bibinfo {author} {\bibfnamefont
  {P.~D.}\ \bibnamefont {Felsher}}, \bibinfo {author} {\bibfnamefont {R.~L.}\
  \bibnamefont {Walter}}, \bibinfo {author} {\bibfnamefont {I.}~\bibnamefont
  {Slaus}}, \bibinfo {author} {\bibfnamefont {P.~A.}\ \bibnamefont {Treado}},\
  and\ \bibinfo {author} {\bibfnamefont {Y.}~\bibnamefont {Koike}},\ }\bibfield
   {title} {\bibinfo {title} {{Comparisons of vector analyzing-power data and
  calculations for neutron-deuteron elastic scattering from 10 to 14 MeV}},\
  }\href {https://doi.org/10.1007/BF01078989} {\bibfield  {journal} {\bibinfo
  {journal} {Few Body Systems}\ }\textbf {\bibinfo {volume} {2}},\ \bibinfo
  {pages} {19} (\bibinfo {year} {1987})}\BibitemShut {NoStop}%
\bibitem [{\citenamefont {Schwarz}\ \emph {et~al.}(1983)\citenamefont
  {Schwarz}, \citenamefont {Klages}, \citenamefont {Doll}, \citenamefont
  {Haesner}, \citenamefont {Wilczynski}, \citenamefont {Zeitnitz},\ and\
  \citenamefont {Kecskemeti}}]{Schwarz:1983rlz}%
  \BibitemOpen
  \bibfield  {author} {\bibinfo {author} {\bibfnamefont {P.}~\bibnamefont
  {Schwarz}}, \bibinfo {author} {\bibfnamefont {H.~O.}\ \bibnamefont {Klages}},
  \bibinfo {author} {\bibfnamefont {P.}~\bibnamefont {Doll}}, \bibinfo {author}
  {\bibfnamefont {B.}~\bibnamefont {Haesner}}, \bibinfo {author} {\bibfnamefont
  {J.}~\bibnamefont {Wilczynski}}, \bibinfo {author} {\bibfnamefont
  {B.}~\bibnamefont {Zeitnitz}},\ and\ \bibinfo {author} {\bibfnamefont
  {J.}~\bibnamefont {Kecskemeti}},\ }\bibfield  {title} {\bibinfo {title}
  {{Elastic neutron-deuteron scattering in the energy range from 2.5 MeV to 30
  MeV}},\ }\href {https://doi.org/10.1016/0375-9474(83)90645-0} {\bibfield
  {journal} {\bibinfo  {journal} {Nucl. Phys. A}\ }\textbf {\bibinfo {volume}
  {398}},\ \bibinfo {pages} {1} (\bibinfo {year} {1983})}\BibitemShut {NoStop}%
\bibitem [{\citenamefont {Shimizu}\ \emph {et~al.}(1982)\citenamefont
  {Shimizu}, \citenamefont {Imai}, \citenamefont {Tamura}, \citenamefont
  {Nisimura}, \citenamefont {Hatanaka}, \citenamefont {Saito}, \citenamefont
  {Koike},\ and\ \citenamefont {Taniguchi}}]{Shimizu:1982lqa}%
  \BibitemOpen
  \bibfield  {author} {\bibinfo {author} {\bibfnamefont {H.}~\bibnamefont
  {Shimizu}}, \bibinfo {author} {\bibfnamefont {K.}~\bibnamefont {Imai}},
  \bibinfo {author} {\bibfnamefont {N.}~\bibnamefont {Tamura}}, \bibinfo
  {author} {\bibfnamefont {K.}~\bibnamefont {Nisimura}}, \bibinfo {author}
  {\bibfnamefont {K.}~\bibnamefont {Hatanaka}}, \bibinfo {author}
  {\bibfnamefont {T.}~\bibnamefont {Saito}}, \bibinfo {author} {\bibfnamefont
  {Y.}~\bibnamefont {Koike}},\ and\ \bibinfo {author} {\bibfnamefont
  {Y.}~\bibnamefont {Taniguchi}},\ }\bibfield  {title} {\bibinfo {title}
  {{Analyzing powers and cross sections in elastic p - d scattering at 65
  MeV}},\ }\href {https://doi.org/10.1016/0375-9474(82)90134-8} {\bibfield
  {journal} {\bibinfo  {journal} {Nucl. Phys. A}\ }\textbf {\bibinfo {volume}
  {382}},\ \bibinfo {pages} {242} (\bibinfo {year} {1982})}\BibitemShut
  {NoStop}%
\bibitem [{\citenamefont {Epelbaum}\ \emph {et~al.}(2020)\citenamefont
  {Epelbaum} \emph {et~al.}}]{Epelbaum:2019zqc}%
  \BibitemOpen
  \bibfield  {author} {\bibinfo {author} {\bibfnamefont {E.}~\bibnamefont
  {Epelbaum}} \emph {et~al.},\ }\bibfield  {title} {\bibinfo {title} {{Towards
  high-order calculations of three-nucleon scattering in chiral effective field
  theory}},\ }\href {https://doi.org/10.1140/epja/s10050-020-00102-2}
  {\bibfield  {journal} {\bibinfo  {journal} {Eur. Phys. J. A}\ }\textbf
  {\bibinfo {volume} {56}},\ \bibinfo {pages} {92} (\bibinfo {year} {2020})},\
  \Eprint {https://arxiv.org/abs/1907.03608} {arXiv:1907.03608 [nucl-th]}
  \BibitemShut {NoStop}%
\bibitem [{\citenamefont {Phillips}(1968)}]{Phillips:1968zze}%
  \BibitemOpen
  \bibfield  {author} {\bibinfo {author} {\bibfnamefont {A.~C.}\ \bibnamefont
  {Phillips}},\ }\bibfield  {title} {\bibinfo {title} {{Consistency of the
  low-energy three-nucleon observables and the separable interaction model}},\
  }\href {https://doi.org/10.1016/0375-9474(68)90737-9} {\bibfield  {journal}
  {\bibinfo  {journal} {Nucl. Phys. A}\ }\textbf {\bibinfo {volume} {107}},\
  \bibinfo {pages} {209} (\bibinfo {year} {1968})}\BibitemShut {NoStop}%
\bibitem [{\citenamefont {Witala}\ \emph {et~al.}(2003)\citenamefont {Witala},
  \citenamefont {Nogga}, \citenamefont {Kamada}, \citenamefont {Gl{\"o}ckle},
  \citenamefont {Golak},\ and\ \citenamefont {Skibinski}}]{Witala:2003en}%
  \BibitemOpen
  \bibfield  {author} {\bibinfo {author} {\bibfnamefont {H.}~\bibnamefont
  {Witala}}, \bibinfo {author} {\bibfnamefont {A.}~\bibnamefont {Nogga}},
  \bibinfo {author} {\bibfnamefont {H.}~\bibnamefont {Kamada}}, \bibinfo
  {author} {\bibfnamefont {W.}~\bibnamefont {Gl{\"o}ckle}}, \bibinfo {author}
  {\bibfnamefont {J.}~\bibnamefont {Golak}},\ and\ \bibinfo {author}
  {\bibfnamefont {R.}~\bibnamefont {Skibinski}},\ }\bibfield  {title} {\bibinfo
  {title} {{Modern nuclear force predictions for the neutron deuteron
  scattering lengths}},\ }\href {https://doi.org/10.1103/PhysRevC.68.034002}
  {\bibfield  {journal} {\bibinfo  {journal} {Phys. Rev. C}\ }\textbf {\bibinfo
  {volume} {68}},\ \bibinfo {pages} {034002} (\bibinfo {year}
  {2003})}\BibitemShut {NoStop}%
\bibitem [{\citenamefont {Weisel}\ \emph {et~al.}(2015)\citenamefont {Weisel},
  \citenamefont {Tornow},\ and\ \citenamefont {Esterline}}]{Weisel:2015min}%
  \BibitemOpen
  \bibfield  {author} {\bibinfo {author} {\bibfnamefont {G.~J.}\ \bibnamefont
  {Weisel}}, \bibinfo {author} {\bibfnamefont {W.}~\bibnamefont {Tornow}},\
  and\ \bibinfo {author} {\bibfnamefont {J.~H.}\ \bibnamefont {Esterline}},\
  }\bibfield  {title} {\bibinfo {title} {{Neutron\textendash{}deuteron
  analyzing power data at En = 21 MeV and the energy dependence of the
  three-nucleon analyzing power puzzle}},\ }\href
  {https://doi.org/10.1088/0954-3899/42/8/085106} {\bibfield  {journal}
  {\bibinfo  {journal} {J. Phys. G}\ }\textbf {\bibinfo {volume} {42}},\
  \bibinfo {pages} {085106} (\bibinfo {year} {2015})}\BibitemShut {NoStop}%
\bibitem [{\citenamefont {Clajus}\ \emph {et~al.}(1995)\citenamefont {Clajus},
  \citenamefont {Albert}, \citenamefont {Bruno}, \citenamefont {Egun},
  \citenamefont {Glockle}, \citenamefont {Glombik}, \citenamefont {Gruebler},
  \citenamefont {Hautle}, \citenamefont {Kretschmer}, \citenamefont {Rauscher},
  \citenamefont {Schmelzbach}, \citenamefont {Slaus}, \citenamefont
  {Weidmann},\ and\ \citenamefont {Witala}}]{Clajus_1995}%
  \BibitemOpen
  \bibfield  {author} {\bibinfo {author} {\bibfnamefont {M.}~\bibnamefont
  {Clajus}}, \bibinfo {author} {\bibfnamefont {J.}~\bibnamefont {Albert}},
  \bibinfo {author} {\bibfnamefont {M.}~\bibnamefont {Bruno}}, \bibinfo
  {author} {\bibfnamefont {P.~M.}\ \bibnamefont {Egun}}, \bibinfo {author}
  {\bibfnamefont {W.}~\bibnamefont {Glockle}}, \bibinfo {author} {\bibfnamefont
  {A.}~\bibnamefont {Glombik}}, \bibinfo {author} {\bibfnamefont
  {W.}~\bibnamefont {Gruebler}}, \bibinfo {author} {\bibfnamefont
  {P.}~\bibnamefont {Hautle}}, \bibinfo {author} {\bibfnamefont
  {W.}~\bibnamefont {Kretschmer}}, \bibinfo {author} {\bibfnamefont
  {A.}~\bibnamefont {Rauscher}}, \bibinfo {author} {\bibfnamefont {P.~A.}\
  \bibnamefont {Schmelzbach}}, \bibinfo {author} {\bibfnamefont
  {I.}~\bibnamefont {Slaus}}, \bibinfo {author} {\bibfnamefont
  {R.}~\bibnamefont {Weidmann}},\ and\ \bibinfo {author} {\bibfnamefont
  {H.}~\bibnamefont {Witala}},\ }\bibfield  {title} {\bibinfo {title}
  {Measurement and calculation of polarization transfer coefficients in the
  reaction2h(p,p)2h at ep=22.5 {MeV}},\ }\href
  {https://doi.org/10.1088/0954-3899/21/10/010} {\bibfield  {journal} {\bibinfo
   {journal} {Journal of Physics G: Nuclear and Particle Physics}\ }\textbf
  {\bibinfo {volume} {21}},\ \bibinfo {pages} {1363} (\bibinfo {year}
  {1995})}\BibitemShut {NoStop}%
\bibitem [{\citenamefont {R\"uhl}\ \emph {et~al.}(1991)\citenamefont {R\"uhl}
  \emph {et~al.}}]{Ruhl:1991qop}%
  \BibitemOpen
  \bibfield  {author} {\bibinfo {author} {\bibfnamefont {H.}~\bibnamefont
  {R\"uhl}} \emph {et~al.},\ }\bibfield  {title} {\bibinfo {title} {{Analyzing
  power in n +d elastic scattering at 67 MeV}},\ }\href
  {https://doi.org/10.1016/0375-9474(91)90275-B} {\bibfield  {journal}
  {\bibinfo  {journal} {Nucl. Phys. A}\ }\textbf {\bibinfo {volume} {524}},\
  \bibinfo {pages} {377} (\bibinfo {year} {1991})}\BibitemShut {NoStop}%
\bibitem [{\citenamefont {Chauvin}\ \emph {et~al.}(1975)\citenamefont
  {Chauvin}, \citenamefont {Garreta},\ and\ \citenamefont
  {Fruneau}}]{Chauvin:1975fuc}%
  \BibitemOpen
  \bibfield  {author} {\bibinfo {author} {\bibfnamefont {J.}~\bibnamefont
  {Chauvin}}, \bibinfo {author} {\bibfnamefont {D.}~\bibnamefont {Garreta}},\
  and\ \bibinfo {author} {\bibfnamefont {M.}~\bibnamefont {Fruneau}},\
  }\bibfield  {title} {\bibinfo {title} {{Measurements of the spin-correlation
  coefficients C xx , C yy and S for d-p scattering at E d = 17.4, 19.5, 23.8
  and 26.1 MeV}},\ }\href {https://doi.org/10.1016/0375-9474(76)90516-9}
  {\bibfield  {journal} {\bibinfo  {journal} {Nucl. Phys. A}\ }\textbf
  {\bibinfo {volume} {247}},\ \bibinfo {pages} {335} (\bibinfo {year}
  {1975})},\ \bibinfo {note} {[Erratum: Nucl.Phys.A 262, 539--539
  (1976)]}\BibitemShut {NoStop}%
\bibitem [{\citenamefont {Bunker}\ \emph {et~al.}(1968)\citenamefont {Bunker},
  \citenamefont {Cameron}, \citenamefont {Carlson}, \citenamefont {Richardson},
  \citenamefont {Toma\v{s}}, \citenamefont {Van~Oers},\ and\ \citenamefont
  {Verba}}]{Bunker:1968omc}%
  \BibitemOpen
  \bibfield  {author} {\bibinfo {author} {\bibfnamefont {S.~N.}\ \bibnamefont
  {Bunker}}, \bibinfo {author} {\bibfnamefont {J.~M.}\ \bibnamefont {Cameron}},
  \bibinfo {author} {\bibfnamefont {R.~F.}\ \bibnamefont {Carlson}}, \bibinfo
  {author} {\bibfnamefont {J.~R.}\ \bibnamefont {Richardson}}, \bibinfo
  {author} {\bibfnamefont {P.}~\bibnamefont {Toma\v{s}}}, \bibinfo {author}
  {\bibfnamefont {W.~T.~H.}\ \bibnamefont {Van~Oers}},\ and\ \bibinfo {author}
  {\bibfnamefont {J.~W.}\ \bibnamefont {Verba}},\ }\bibfield  {title} {\bibinfo
  {title} {{Differential cross sections and polarizations in elastic p-d
  scattering at medium energies}},\ }\href
  {https://doi.org/10.1016/0375-9474(68)90418-1} {\bibfield  {journal}
  {\bibinfo  {journal} {Nucl. Phys. A}\ }\textbf {\bibinfo {volume} {113}},\
  \bibinfo {pages} {461} (\bibinfo {year} {1968})}\BibitemShut {NoStop}%
\bibitem [{\citenamefont {Wita{\l}a}\ \emph {et~al.}(1993)\citenamefont
  {Wita{\l}a}, \citenamefont {Gl{\"o}ckle}, \citenamefont {Antonuk},
  \citenamefont {Arvieux}, \citenamefont {Bachelier}, \citenamefont {Bonin},
  \citenamefont {Boudard}, \citenamefont {Cameron}, \citenamefont {Fielding},
  \citenamefont {Gar{\c c}on}, \citenamefont {Jourdan}, \citenamefont
  {Lapointe}, \citenamefont {McDonald}, \citenamefont {Pasos}, \citenamefont
  {Roy}, \citenamefont {The}, \citenamefont {Tinslay}, \citenamefont {Tornow},
  \citenamefont {Yonnet},\ and\ \citenamefont {Ziegler}}]{Witaia:1993vz}%
  \BibitemOpen
  \bibfield  {author} {\bibinfo {author} {\bibfnamefont {H.}~\bibnamefont
  {Wita{\l}a}}, \bibinfo {author} {\bibfnamefont {W.}~\bibnamefont
  {Gl{\"o}ckle}}, \bibinfo {author} {\bibfnamefont {L.~E.}\ \bibnamefont
  {Antonuk}}, \bibinfo {author} {\bibfnamefont {J.}~\bibnamefont {Arvieux}},
  \bibinfo {author} {\bibfnamefont {D.}~\bibnamefont {Bachelier}}, \bibinfo
  {author} {\bibfnamefont {B.}~\bibnamefont {Bonin}}, \bibinfo {author}
  {\bibfnamefont {A.}~\bibnamefont {Boudard}}, \bibinfo {author} {\bibfnamefont
  {J.~M.}\ \bibnamefont {Cameron}}, \bibinfo {author} {\bibfnamefont {H.~W.}\
  \bibnamefont {Fielding}}, \bibinfo {author} {\bibfnamefont {M.}~\bibnamefont
  {Gar{\c c}on}}, \bibinfo {author} {\bibfnamefont {F.}~\bibnamefont
  {Jourdan}}, \bibinfo {author} {\bibfnamefont {C.}~\bibnamefont {Lapointe}},
  \bibinfo {author} {\bibfnamefont {W.~J.}\ \bibnamefont {McDonald}}, \bibinfo
  {author} {\bibfnamefont {J.}~\bibnamefont {Pasos}}, \bibinfo {author}
  {\bibfnamefont {G.}~\bibnamefont {Roy}}, \bibinfo {author} {\bibfnamefont
  {I.}~\bibnamefont {The}}, \bibinfo {author} {\bibfnamefont {J.}~\bibnamefont
  {Tinslay}}, \bibinfo {author} {\bibfnamefont {W.}~\bibnamefont {Tornow}},
  \bibinfo {author} {\bibfnamefont {J.}~\bibnamefont {Yonnet}},\ and\ \bibinfo
  {author} {\bibfnamefont {W.}~\bibnamefont {Ziegler}},\ }\bibfield  {title}
  {\bibinfo {title} {{Complete set of deuteron analyzing powers in
  deuteron-proton elastic scattering: Measurement and realistic potential
  predictions}},\ }\href@noop {} {\bibfield  {journal} {\bibinfo  {journal}
  {Few Body Systems}\ }\textbf {\bibinfo {volume} {15}},\ \bibinfo {pages} {67}
  (\bibinfo {year} {1993})}\BibitemShut {NoStop}%
\bibitem [{\citenamefont {Sekiguchi}\ \emph {et~al.}(2004)\citenamefont
  {Sekiguchi} \emph {et~al.}}]{Sekiguchi:2004yb}%
  \BibitemOpen
  \bibfield  {author} {\bibinfo {author} {\bibfnamefont {K.}~\bibnamefont
  {Sekiguchi}} \emph {et~al.},\ }\bibfield  {title} {\bibinfo {title}
  {{Polarization transfer measurement for H-1(polarized-d, polarized-p) H-2
  elastic scattering at 135-MeV / u and three nucleon force effects}},\ }\href
  {https://doi.org/10.1103/PhysRevC.70.014001} {\bibfield  {journal} {\bibinfo
  {journal} {Phys. Rev. C}\ }\textbf {\bibinfo {volume} {70}},\ \bibinfo
  {pages} {014001} (\bibinfo {year} {2004})},\ \Eprint
  {https://arxiv.org/abs/nucl-ex/0404026} {arXiv:nucl-ex/0404026} \BibitemShut
  {NoStop}%
\bibitem [{\citenamefont {Clajus}\ \emph {et~al.}(1990)\citenamefont {Clajus}
  \emph {et~al.}}]{Clajus:1990rzf}%
  \BibitemOpen
  \bibfield  {author} {\bibinfo {author} {\bibfnamefont {M.}~\bibnamefont
  {Clajus}} \emph {et~al.},\ }\bibfield  {title} {\bibinfo {title}
  {{Investigation of the nucleon-nucleon tensor force in the three-nucleon
  system}},\ }\href {https://doi.org/10.1016/0370-2693(90)90654-O} {\bibfield
  {journal} {\bibinfo  {journal} {Phys. Lett. B}\ }\textbf {\bibinfo {volume}
  {245}},\ \bibinfo {pages} {333} (\bibinfo {year} {1990})}\BibitemShut
  {NoStop}%
\bibitem [{\citenamefont {Skibinski}\ \emph {et~al.}(2018)\citenamefont
  {Skibinski}, \citenamefont {Volkotrub}, \citenamefont {Golak}, \citenamefont
  {Topolnicki},\ and\ \citenamefont {Witala}}]{Skibinski:2018dot}%
  \BibitemOpen
  \bibfield  {author} {\bibinfo {author} {\bibfnamefont {R.}~\bibnamefont
  {Skibinski}}, \bibinfo {author} {\bibfnamefont {Y.}~\bibnamefont
  {Volkotrub}}, \bibinfo {author} {\bibfnamefont {J.}~\bibnamefont {Golak}},
  \bibinfo {author} {\bibfnamefont {K.}~\bibnamefont {Topolnicki}},\ and\
  \bibinfo {author} {\bibfnamefont {H.}~\bibnamefont {Witala}},\ }\bibfield
  {title} {\bibinfo {title} {{Theoretical uncertainties of the elastic
  nucleon-deuteron scattering observables}},\ }\href
  {https://doi.org/10.1103/PhysRevC.98.014001} {\bibfield  {journal} {\bibinfo
  {journal} {Phys. Rev. C}\ }\textbf {\bibinfo {volume} {98}},\ \bibinfo
  {pages} {014001} (\bibinfo {year} {2018})},\ \Eprint
  {https://arxiv.org/abs/1803.10345} {arXiv:1803.10345 [nucl-th]} \BibitemShut
  {NoStop}%
\bibitem [{\citenamefont {Wita\l{}a}\ \emph {et~al.}(2021)\citenamefont
  {Wita\l{}a}, \citenamefont {Golak},\ and\ \citenamefont
  {Skibi\'nski}}]{Witala:2021xqm}%
  \BibitemOpen
  \bibfield  {author} {\bibinfo {author} {\bibfnamefont {H.}~\bibnamefont
  {Wita\l{}a}}, \bibinfo {author} {\bibfnamefont {J.}~\bibnamefont {Golak}},\
  and\ \bibinfo {author} {\bibfnamefont {R.}~\bibnamefont {Skibi\'nski}},\
  }\bibfield  {title} {\bibinfo {title} {{Efficient emulator for solving
  three-nucleon continuum Faddeev equations with chiral three-nucleon force
  comprising any number of contact terms}},\ }\href
  {https://doi.org/10.1140/epja/s10050-021-00555-z} {\bibfield  {journal}
  {\bibinfo  {journal} {Eur. Phys. J. A}\ }\textbf {\bibinfo {volume} {57}},\
  \bibinfo {pages} {241} (\bibinfo {year} {2021})},\ \Eprint
  {https://arxiv.org/abs/2103.13237} {arXiv:2103.13237 [nucl-th]} \BibitemShut
  {NoStop}%
\bibitem [{\citenamefont {Frame}\ \emph {et~al.}(2018)\citenamefont {Frame},
  \citenamefont {He}, \citenamefont {Ipsen}, \citenamefont {Lee}, \citenamefont
  {Lee},\ and\ \citenamefont {Rrapaj}}]{Frame:2017fah}%
  \BibitemOpen
  \bibfield  {author} {\bibinfo {author} {\bibfnamefont {D.}~\bibnamefont
  {Frame}}, \bibinfo {author} {\bibfnamefont {R.}~\bibnamefont {He}}, \bibinfo
  {author} {\bibfnamefont {I.}~\bibnamefont {Ipsen}}, \bibinfo {author}
  {\bibfnamefont {D.}~\bibnamefont {Lee}}, \bibinfo {author} {\bibfnamefont
  {D.}~\bibnamefont {Lee}},\ and\ \bibinfo {author} {\bibfnamefont
  {E.}~\bibnamefont {Rrapaj}},\ }\bibfield  {title} {\bibinfo {title}
  {{Eigenvector continuation with subspace learning}},\ }\href
  {https://doi.org/10.1103/PhysRevLett.121.032501} {\bibfield  {journal}
  {\bibinfo  {journal} {Phys. Rev. Lett.}\ }\textbf {\bibinfo {volume} {121}},\
  \bibinfo {pages} {032501} (\bibinfo {year} {2018})},\ \Eprint
  {https://arxiv.org/abs/1711.07090} {arXiv:1711.07090 [nucl-th]} \BibitemShut
  {NoStop}%
\bibitem [{\citenamefont {Furnstahl}\ \emph {et~al.}(2020)\citenamefont
  {Furnstahl}, \citenamefont {Garcia}, \citenamefont {Millican},\ and\
  \citenamefont {Zhang}}]{Furnstahl:2020abp}%
  \BibitemOpen
  \bibfield  {author} {\bibinfo {author} {\bibfnamefont {R.~J.}\ \bibnamefont
  {Furnstahl}}, \bibinfo {author} {\bibfnamefont {A.~J.}\ \bibnamefont
  {Garcia}}, \bibinfo {author} {\bibfnamefont {P.~J.}\ \bibnamefont
  {Millican}},\ and\ \bibinfo {author} {\bibfnamefont {X.}~\bibnamefont
  {Zhang}},\ }\bibfield  {title} {\bibinfo {title} {{Efficient emulators for
  scattering using eigenvector continuation}},\ }\href
  {https://doi.org/10.1016/j.physletb.2020.135719} {\bibfield  {journal}
  {\bibinfo  {journal} {Phys. Lett. B}\ }\textbf {\bibinfo {volume} {809}},\
  \bibinfo {pages} {135719} (\bibinfo {year} {2020})},\ \Eprint
  {https://arxiv.org/abs/2007.03635} {arXiv:2007.03635 [nucl-th]} \BibitemShut
  {NoStop}%
\bibitem [{\citenamefont {Bai}\ and\ \citenamefont {Ren}(2021)}]{Bai:2021xok}%
  \BibitemOpen
  \bibfield  {author} {\bibinfo {author} {\bibfnamefont {D.}~\bibnamefont
  {Bai}}\ and\ \bibinfo {author} {\bibfnamefont {Z.}~\bibnamefont {Ren}},\
  }\bibfield  {title} {\bibinfo {title} {{Generalizing the calculable
  $R$-matrix theory and eigenvector continuation to the incoming wave boundary
  condition}},\ }\href {https://doi.org/10.1103/PhysRevC.103.014612} {\bibfield
   {journal} {\bibinfo  {journal} {Phys. Rev. C}\ }\textbf {\bibinfo {volume}
  {103}},\ \bibinfo {pages} {014612} (\bibinfo {year} {2021})},\ \Eprint
  {https://arxiv.org/abs/2101.06336} {arXiv:2101.06336 [nucl-th]} \BibitemShut
  {NoStop}%
\bibitem [{\citenamefont {Zhang}\ and\ \citenamefont
  {Furnstahl}(2021)}]{Zhang:2021jmi}%
  \BibitemOpen
  \bibfield  {author} {\bibinfo {author} {\bibfnamefont {X.}~\bibnamefont
  {Zhang}}\ and\ \bibinfo {author} {\bibfnamefont {R.~J.}\ \bibnamefont
  {Furnstahl}},\ }\href@noop {} {\bibinfo {title} {{Fast emulation of quantum
  three-body scattering}}} (\bibinfo {year} {2021}),\ \Eprint
  {https://arxiv.org/abs/2110.04269} {arXiv:2110.04269 [nucl-th]} \BibitemShut
  {NoStop}%
\bibitem [{\citenamefont {Drischler}\ \emph {et~al.}(2021)\citenamefont
  {Drischler}, \citenamefont {Quinonez}, \citenamefont {Giuliani},
  \citenamefont {Lovell},\ and\ \citenamefont {Nunes}}]{Drischler:2021qoy}%
  \BibitemOpen
  \bibfield  {author} {\bibinfo {author} {\bibfnamefont {C.}~\bibnamefont
  {Drischler}}, \bibinfo {author} {\bibfnamefont {M.}~\bibnamefont {Quinonez}},
  \bibinfo {author} {\bibfnamefont {P.}~\bibnamefont {Giuliani}}, \bibinfo
  {author} {\bibfnamefont {A.}~\bibnamefont {Lovell}},\ and\ \bibinfo {author}
  {\bibfnamefont {F.}~\bibnamefont {Nunes}},\ }\bibfield  {title} {\bibinfo
  {title} {Toward emulating nuclear reactions using eigenvector continuation},\
  }\href {https://doi.org/https://doi.org/10.1016/j.physletb.2021.136777}
  {\bibfield  {journal} {\bibinfo  {journal} {Physics Letters B}\ ,\ \bibinfo
  {pages} {136777}} (\bibinfo {year} {2021})}\BibitemShut {NoStop}%
\bibitem [{\citenamefont
  {Gl{\"o}ckle}(1983{\natexlab{b}})}]{glockle1983quantum}%
  \BibitemOpen
  \bibfield  {author} {\bibinfo {author} {\bibfnamefont {W.}~\bibnamefont
  {Gl{\"o}ckle}},\ }\href {https://books.google.se/books?id=2yM-PwAACAAJ}
  {\emph {\bibinfo {title} {The Quantum Mechanical Few-Body Problem}}},\
  Theoretical and Mathematical Physics\ (\bibinfo  {publisher} {Springer Berlin
  Heidelberg},\ \bibinfo {year} {1983})\BibitemShut {NoStop}%
\bibitem [{\citenamefont {Pomerantsev}\ \emph {et~al.}(2014)\citenamefont
  {Pomerantsev}, \citenamefont {Kukulin},\ and\ \citenamefont
  {Rubtsova}}]{Pomerantsev:2014ija}%
  \BibitemOpen
  \bibfield  {author} {\bibinfo {author} {\bibfnamefont {V.~N.}\ \bibnamefont
  {Pomerantsev}}, \bibinfo {author} {\bibfnamefont {V.~I.}\ \bibnamefont
  {Kukulin}},\ and\ \bibinfo {author} {\bibfnamefont {O.~A.}\ \bibnamefont
  {Rubtsova}},\ }\bibfield  {title} {\bibinfo {title} {{New general approach in
  few-body scattering calculations: Solving discretized Faddeev equations on a
  graphics processing unit}},\ }\href
  {https://doi.org/10.1103/PhysRevC.89.064008} {\bibfield  {journal} {\bibinfo
  {journal} {Phys. Rev. C}\ }\textbf {\bibinfo {volume} {89}},\ \bibinfo
  {pages} {064008} (\bibinfo {year} {2014})},\ \Eprint
  {https://arxiv.org/abs/1404.5253} {arXiv:1404.5253 [nucl-th]} \BibitemShut
  {NoStop}%
\bibitem [{\citenamefont {Bianchi}\ and\ \citenamefont
  {Favella}(1964)}]{bianchi1964convolution}%
  \BibitemOpen
  \bibfield  {author} {\bibinfo {author} {\bibfnamefont {L.}~\bibnamefont
  {Bianchi}}\ and\ \bibinfo {author} {\bibfnamefont {L.}~\bibnamefont
  {Favella}},\ }\bibfield  {title} {\bibinfo {title} {A convolution integral
  for the resolvent of the sum of two commuting operators},\ }\href
  {https://doi.org/10.1007/BF02750582} {\bibfield  {journal} {\bibinfo
  {journal} {Il Nuovo Cimento (1955-1965)}\ }\textbf {\bibinfo {volume} {34}},\
  \bibinfo {pages} {1825} (\bibinfo {year} {1964})}\BibitemShut {NoStop}%
\bibitem [{\citenamefont {Kukulin}\ \emph {et~al.}(2007)\citenamefont
  {Kukulin}, \citenamefont {Pomerantsev},\ and\ \citenamefont
  {Rubtsova}}]{kukulin2007wave}%
  \BibitemOpen
  \bibfield  {author} {\bibinfo {author} {\bibfnamefont {V.~I.}\ \bibnamefont
  {Kukulin}}, \bibinfo {author} {\bibfnamefont {V.~N.}\ \bibnamefont
  {Pomerantsev}},\ and\ \bibinfo {author} {\bibfnamefont {O.~A.}\ \bibnamefont
  {Rubtsova}},\ }\bibfield  {title} {\bibinfo {title} {Wave-packet continuum
  discretization method for solving the three-body scattering problem},\ }\href
  {https://doi.org/10.1007/s11232-007-0030-3} {\bibfield  {journal} {\bibinfo
  {journal} {Theoretical and Mathematical Physics}\ }\textbf {\bibinfo {volume}
  {150}},\ \bibinfo {pages} {403} (\bibinfo {year} {2007})}\BibitemShut
  {NoStop}%
\bibitem [{\citenamefont {Hoppe}\ \emph {et~al.}(2017)\citenamefont {Hoppe},
  \citenamefont {Drischler}, \citenamefont {Furnstahl}, \citenamefont
  {Hebeler},\ and\ \citenamefont {Schwenk}}]{Hoppe:2017lok}%
  \BibitemOpen
  \bibfield  {author} {\bibinfo {author} {\bibfnamefont {J.}~\bibnamefont
  {Hoppe}}, \bibinfo {author} {\bibfnamefont {C.}~\bibnamefont {Drischler}},
  \bibinfo {author} {\bibfnamefont {R.~J.}\ \bibnamefont {Furnstahl}}, \bibinfo
  {author} {\bibfnamefont {K.}~\bibnamefont {Hebeler}},\ and\ \bibinfo {author}
  {\bibfnamefont {A.}~\bibnamefont {Schwenk}},\ }\bibfield  {title} {\bibinfo
  {title} {{Weinberg eigenvalues for chiral nucleon-nucleon interactions}},\
  }\href {https://doi.org/10.1103/PhysRevC.96.054002} {\bibfield  {journal}
  {\bibinfo  {journal} {Phys. Rev. C}\ }\textbf {\bibinfo {volume} {96}},\
  \bibinfo {pages} {054002} (\bibinfo {year} {2017})},\ \Eprint
  {https://arxiv.org/abs/1707.06438} {arXiv:1707.06438 [nucl-th]} \BibitemShut
  {NoStop}%
\bibitem [{\citenamefont {Kukulin}\ \emph {et~al.}(2013)\citenamefont
  {Kukulin}, \citenamefont {Krasnopolsky},\ and\ \citenamefont
  {Hor{\'a}cek}}]{kukulin2013theory}%
  \BibitemOpen
  \bibfield  {author} {\bibinfo {author} {\bibfnamefont {V.}~\bibnamefont
  {Kukulin}}, \bibinfo {author} {\bibfnamefont {V.}~\bibnamefont
  {Krasnopolsky}},\ and\ \bibinfo {author} {\bibfnamefont {J.}~\bibnamefont
  {Hor{\'a}cek}},\ }\href {https://books.google.se/books?id=vHfoCAAAQBAJ}
  {\emph {\bibinfo {title} {Theory of Resonances: Principles and
  Applications}}},\ Reidel Texts in the Mathematical Sciences\ (\bibinfo
  {publisher} {Springer Netherlands},\ \bibinfo {year} {2013})\BibitemShut
  {NoStop}%
\bibitem [{\citenamefont {Ohlsen}(1972)}]{Ohlsen:1972zz}%
  \BibitemOpen
  \bibfield  {author} {\bibinfo {author} {\bibfnamefont {G.~G.}\ \bibnamefont
  {Ohlsen}},\ }\bibfield  {title} {\bibinfo {title} {{Polarization transfer and
  spin correlation experiments in nuclear physics}},\ }\href
  {https://doi.org/10.1088/0034-4885/35/2/305} {\bibfield  {journal} {\bibinfo
  {journal} {Rept. Prog. Phys.}\ }\textbf {\bibinfo {volume} {35}},\ \bibinfo
  {pages} {717} (\bibinfo {year} {1972})}\BibitemShut {NoStop}%
\bibitem [{\citenamefont {Seyler}(1969)}]{Seyler:1969sii}%
  \BibitemOpen
  \bibfield  {author} {\bibinfo {author} {\bibfnamefont {R.~G.}\ \bibnamefont
  {Seyler}},\ }\bibfield  {title} {\bibinfo {title} {{Polarization from
  scattering polarized spin-${\frac{1}{2}}$ on unpolarized spin-1 particles}},\
  }\href {https://doi.org/10.1016/0375-9474(69)90352-2} {\bibfield  {journal}
  {\bibinfo  {journal} {Nucl. Phys. A}\ }\textbf {\bibinfo {volume} {124}},\
  \bibinfo {pages} {253} (\bibinfo {year} {1969})}\BibitemShut {NoStop}%
\bibitem [{\citenamefont {Blatt}\ and\ \citenamefont
  {Biedenharn}(1952)}]{Blatt:1952zz}%
  \BibitemOpen
  \bibfield  {author} {\bibinfo {author} {\bibfnamefont {J.~M.}\ \bibnamefont
  {Blatt}}\ and\ \bibinfo {author} {\bibfnamefont {L.~C.}\ \bibnamefont
  {Biedenharn}},\ }\bibfield  {title} {\bibinfo {title} {{The Angular
  Distribution of Scattering and Reaction Cross Sections}},\ }\href
  {https://doi.org/10.1103/RevModPhys.24.258} {\bibfield  {journal} {\bibinfo
  {journal} {Rev. Mod. Phys.}\ }\textbf {\bibinfo {volume} {24}},\ \bibinfo
  {pages} {258} (\bibinfo {year} {1952})}\BibitemShut {NoStop}%
\bibitem [{\citenamefont {Barschall}\ and\ \citenamefont
  {Haeberli}(1971)}]{osti_4726823}%
  \BibitemOpen
  \bibfield  {author} {\bibinfo {author} {\bibfnamefont {H.~H.}\ \bibnamefont
  {Barschall}}\ and\ \bibinfo {author} {\bibfnamefont {W.}~\bibnamefont
  {Haeberli}},\ }\bibfield  {title} {\bibinfo {title} {Polarization phenomena
  in nuclear reactions},\ }\href {https://www.osti.gov/biblio/4726823}
  {\bibfield  {journal} {\bibinfo  {journal} {Proceedings of the Third
  International Symposium}\ } (\bibinfo {year} {1971})}\BibitemShut {NoStop}%
\bibitem [{\citenamefont {Huber}\ \emph {et~al.}(1995)\citenamefont {Huber},
  \citenamefont {Glockle}, \citenamefont {Golak}, \citenamefont {Witala},
  \citenamefont {Kamada}, \citenamefont {Kievsky}, \citenamefont {Rosati},\
  and\ \citenamefont {Viviani}}]{Huber:1995zza}%
  \BibitemOpen
  \bibfield  {author} {\bibinfo {author} {\bibfnamefont {D.}~\bibnamefont
  {Huber}}, \bibinfo {author} {\bibfnamefont {W.}~\bibnamefont {Glockle}},
  \bibinfo {author} {\bibfnamefont {J.}~\bibnamefont {Golak}}, \bibinfo
  {author} {\bibfnamefont {H.}~\bibnamefont {Witala}}, \bibinfo {author}
  {\bibfnamefont {H.}~\bibnamefont {Kamada}}, \bibinfo {author} {\bibfnamefont
  {A.}~\bibnamefont {Kievsky}}, \bibinfo {author} {\bibfnamefont
  {S.}~\bibnamefont {Rosati}},\ and\ \bibinfo {author} {\bibfnamefont
  {M.}~\bibnamefont {Viviani}},\ }\bibfield  {title} {\bibinfo {title}
  {{Realistic phase shift and mixing parameters for elastic neutron-deuteron
  scattering: Comparison of momentum space and configuration space methods}},\
  }\href {https://doi.org/10.1103/PhysRevC.51.1100} {\bibfield  {journal}
  {\bibinfo  {journal} {Phys. Rev. C}\ }\textbf {\bibinfo {volume} {51}},\
  \bibinfo {pages} {1100} (\bibinfo {year} {1995})}\BibitemShut {NoStop}%
\bibitem [{\citenamefont {H{\"u}ber}\ \emph {et~al.}(1995)\citenamefont
  {H{\"u}ber}, \citenamefont {Golak}, \citenamefont {Witala}, \citenamefont
  {Gl{\"o}ckle},\ and\ \citenamefont {Kamada}}]{Huber1995}%
  \BibitemOpen
  \bibfield  {author} {\bibinfo {author} {\bibfnamefont {D.}~\bibnamefont
  {H{\"u}ber}}, \bibinfo {author} {\bibfnamefont {J.}~\bibnamefont {Golak}},
  \bibinfo {author} {\bibfnamefont {H.}~\bibnamefont {Witala}}, \bibinfo
  {author} {\bibfnamefont {W.}~\bibnamefont {Gl{\"o}ckle}},\ and\ \bibinfo
  {author} {\bibfnamefont {H.}~\bibnamefont {Kamada}},\ }\bibfield  {title}
  {\bibinfo {title} {Phase shifts and mixing parameters for elastic
  neutron-deuteron scattering above breakup threshold},\ }\href
  {https://doi.org/10.1007/s006010050025} {\bibfield  {journal} {\bibinfo
  {journal} {Few-Body Systems}\ }\textbf {\bibinfo {volume} {19}},\ \bibinfo
  {pages} {175} (\bibinfo {year} {1995})}\BibitemShut {NoStop}%
\end{thebibliography}%
\end{document}